\documentclass[a4paper,10pt]{article}

\usepackage{bm}
\usepackage{verbatim}
\usepackage[colorlinks=false]{hyperref} 
\usepackage{subfigure}
\usepackage{amsmath,amsfonts,amssymb,graphics,graphicx,epsfig,color,times}
\usepackage{pgfplots}
\usepackage{verbatim}
\usepackage{epstopdf}

\def\picill#1by#2(#3)
{\vbox to #2 {\hrule width #1 height 0pt depth 0pt
\vfill\epsffile{#3}}}

\newcommand{\eq}{\begin{equation}}
\newcommand{\en}{\end{equation}}
\newcommand{\eqa}{\begin{eqnarray}}
\newcommand{\ena}{\end{eqnarray}}

\begin{document}

\setlength{\unitlength}{1mm}

\thispagestyle{empty}



 \begin{center}
{\large{\bf  Teleportation-Based Quantum Computation, Extended Temperley--Lieb Diagrammatical Approach and Yang--Baxter Equation}}
  \\[4mm]

\vspace{.5cm}
Yong Zhang ${}^{a,}$\footnote{yong\_zhang@whu.edu.cn},
Kun Zhang ${}^{a,}$\footnote{kun\_zhang@whu.edu.cn},
and Jinglong Pang ${}^{b,}$\footnote{jlpang@pku.edu.cn} \\[.5cm]

${}^a$ School of Physics and Technology, Wuhan University, P.R. China 430072

${}^b$ Department of Physics, Peking University, P.R. China 100871
\\[0.1cm]
\end{center}

\vspace{0.2cm}

\begin{center}
\parbox{13cm}{
\centerline{\small  \bf Abstract}  \noindent

This paper focuses on the study of topological features in teleportation-based quantum computation
as well as aims at presenting a detailed review on teleportaiton-based quantum computation (Gottesman
and Chuang, Nature 402, 390, 1999). In the extended Temperley--Lieb diagrammatical
approach, we clearly show that such topological features bring about the
fault-tolerant construction of both universal quantum gates and four-partite entangled states more intuitive
and simpler. Furthermore, we describe the Yang--Baxter gate by its  extended Temperley--Lieb configuration,
and then study teleportation-based quantum  circuit models using the Yang--Baxter gate.
Moreover, we discuss the relationship between the extended Temperley--Lieb diagrammatical approach and
the Yang--Baxter gate approach. With these research results,  we propose a worthwhile subject,
the extended Temperley--Lieb diagrammatical approach,  for physicists in quantum information and
quantum computation.

}

\end{center}

\vspace{.2cm}

\begin{tabbing}

\end{tabbing}

\newpage

\section{Introduction}

Quantum information and computation \cite{NC2011, Preskill97} is an interdisciplinary research field of applying fundamental principles
of quantum mechanics to information science and computer science. It represents a further development of quantum mechanics, and indeed
helps us to achieve deeper understandings on quantum physics. Quantum entanglement \cite{EPR35,Bell64} as a fundamental concept of distinguishing
classical physics from quantum physics has become a widely exploited resource in quantum information and computation. It has been experimentally
verified that, under the assistance of quantum entanglement, an unknown qubit can be transmitted  from Alice to Bob without any non-local
physical interaction between them, using the quantum information protocol called quantum teleportation \cite{BBCCJPW93,Vaidman94,BDM00,Werner01}.
Hence both quantum entanglement and quantum teleportation oblige quantum physicists to think about the relationship between Einstein's locality
and quantum non-locality  \cite{EPR35,Bell64}.

Fault-tolerant quantum computation \cite{NC2011, Preskill97} is required in practice to overcome decoherence, and
teleportation-based quantum computation \cite{GC99,Nielsen03, Leung04, ZP13, ZZ14} is a powerful approach to fault-tolerant
quantum computation in which universal quantum gate set is protected from noise using the teleportation protocol. Besides this,
teleportation-based quantum computation is an example of measurement-based quantum computation which exploits quantum measurement
as the main computing resource to determine which quantum gate is to be performed. In quantum mechanics, quantum measurement breaks
quantum coherence and is usually performed at the end of an experiment,  so measurement-based quantum computation essentially changes
our conventional viewpoint on quantum measurement and also on the standard quantum circuit model \cite{Nielsen03} in which a coherent
unitary dynamics is mainly involved.

As the title of this paper claims, we apply the extended Temperley--Lieb diagrammatical approach and the Yang--Baxter gate approach to
the reformulation of teleportation-based quantum computation. The Temperley--Lieb algebra \cite{TL71} is a well known concept in
both statistical mechanics and low dimensional topology \cite{Kauffman02}, and the Yang--Baxter equation \cite{YBE67} arises
in exactly solving both some $1+1$-dimensional quantum many-body systems and vertex models in statistics physics. The fact \cite{Kauffman02}
that a type of solutions of the Yang--Baxter equation  can be constructed using the Temperley--Lieb algebra motivates the authors to
explore teleportation-based quantum computation using the two related approaches.

The extended Temperley--Lieb diagrammatical approach \cite{Kauffman05, Zhang06} is an extension of the standard diagrammatical representation
of the Temperley--Lieb algebra, and with it, a quantum information protocol involving bipartite maximally entangled states, such as quantum
teleportation, has a very nice topological diagrammatical interpretation. The Yang--Baxter gates are nontrivial unitary solutions of the
Yang--Baxter equation,  and the Yang--Baxter gate approach \cite{Dye03, KL04, ZKG04} to quantum information and computation
is an algebraic method originally motivated by the observation that topological entanglements (like braiding configurations \cite{Kauffman02})
and quantum entanglements may have a kind of connection. Therefore, our research is expected to be interesting for physicists in quantum information
and computation, which shows that teleportation-based quantum computation admits both topological and algebraic descriptions besides its standard description
in quantum circuit model \cite{GC99,Nielsen03, Leung04}.

As we have introduced, we are going to go directly into at least three different research subjects in this paper, including quantum information
and computation, the Temperley--Lieb algebra and the Yang--Baxter equation. The guiding principle of this kind of interdisciplinary research is
that we want to explore the nature of quantum entanglement (quantum non-locality \cite{EPR35,Bell64}). As a matter of fact, nowadays, nobody is able to
state that the nature of quantum entanglement has been fully understood  \cite{NC2011, Preskill97}. In accordance with \cite{KL02}, we study the nature
of quantum entanglement by setting up a link between quantum entanglement (quantum non-locality ) and
topological entanglement (topological non-locality \cite{Kauffman02}).

Topological entanglement in the paper is represented by the extended Temperley--Lieb  diagrammatical
configuration or the Yang--Baxter gate configuration (the braiding configuration), while quantum entanglement is represented by bipartite maximally
entangled two-qubit pure  states, i.e., the Bell states  \cite{NC2011, Preskill97}.  In the extended Temperley--Lieb diagrammatical
approach \cite{Kauffman05, Zhang06}, both the Bell states and Bell measurements have the Temperley--Lieb diagrammatical configurations
on which quantum gates are allowed to move from this qubit to that qubit. In the Yang--Baxter gate approach \cite{Dye03, KL04, ZKG04}, the algebraic
formulation of teleportation-based quantum computation admits the braiding configuration on which quantum gates are permitted to move. Especially,
the relationship between such the two approaches will be clarified in this paper.

The complete scheme of teleportation-based quantum computation was proposed by Gottesman and Chuang \cite{GC99} in a very brief style, so it is necessary
firstly to make a detailed review on teleportation-based quantum computation, topics including quantum teleportation, the fault-tolerant construction of
universal quantum gate set and the fault-tolerant preparation of four-qubit entangled states. Then we present a topological diagrammatical description of
such the reviewed topics in the extended Temperley--Lieb diagrammatical approach. Afterwards, we concentrate on the Yang--Baxter gates which are the Bell
transform, a unitary basis transformation from the product basis to the Bell states, and with the help of our previous research on teleportation-based
quantum computation using the Bell transform \cite{ZZ14}, we go into the algebraic description of teleportation-based quantum computation in terms of
the Yang--Baxter gate. Finally, in view of the extended Temperley--Lieb diagrammatical representation of the Yang--Baxter gate, we set up a transparent link
between the extended Temperley--Lieb diagrammatical approach and the Yang--Baxter gate approach.

Our study in this paper is meaningful and useful in the following sense. First, as we have emphasized, we try to dig out the nature of quantum non-locality from
the viewpoint of topological non-locality. Second, the topological features in teleportation-based quantum computation make the fault-tolerant construction of
both universal quantum gate set and four-qubit entangled states more intuitive and simpler. Third, we develop the concept of teleportation operator \cite{ZZ14}
in the Yang--Baxter gate approach to catch the intrinsic characteristic  of quantum teleportation, which is capable of including the algebraic formulations of
all possible teleportaiton processes. Fourth, the methodologies underlying our research are expected to be applied to other subjects in quantum information and
computation, see Section~\ref{Concluding remarks} for concluding remarks on further research. Fifth, our research is expected to shed a light on further research
in mathematical physics, see Appendix~A.

It is obvious that what we have done in this paper is an interdisciplinary research among quantum information and computation, the low-dimensional topology and
mathematical physics such as the Yang--Baxter equation. But we aim at introducing both the Temperley--Lieb algebra and the Yang--Baxter equation to physicists in
quantum information and computation, in other words, readers can go through the entire paper without the preliminary knowledge on concepts in mathematical physics.
All the results about both the Temperley--Lieb algebra and the Yang--Baxter equation are collected in Appendix A. Furthermore, for experts in mathematical
physics, Appendix~\ref{YBE_further_research} indeed presents a list of open problems for further research which are based on our reformulation of
teleportation-based quantum computation.

This paper is organized as follows. Section 2 is a review on teleportation-based quantum computation \cite{GC99}.
Section 3 focuses on the topological diagrammatical description of teleportation-based quantum computation in the extended Temperley--Lieb diagrammatical
approach. Section 4 presents the extended Temperley--Lieb configurations of two types of Yang--Baxter gates derived in Appendix A. Section 5 describes
the Yang--Baxter gate approach to teleportation-based quantum computation.  Section 6 clarifies the relationship between the extended Temperley--Lieb
diagrammatical approach and the Yang--Baxter gate approach. To make the paper self-consistent, we present five appendices for readers' conveniences.
Appendix A reviews both the Temperley--Lieb algebra and the Yang--Baxter equation as well as presents a detailed construction on the Yang--Baxter gate
via the Temperley--Lieb algebra. Appendix B collects interesting properties of the permutation-like Yang--Baxter gates. Both Appendix C and Appendix D
present the topological diagrammatical construction of four-qubit entangled  states respectively associated with the $CNOT$ gate and $CZ$ gate \cite{NC2011}.
Appendix E is about a method of calculating the extended Temperley--Lieb diagrammatical representation of the teleportation operator.

\section{Review on teleportation-based quantum computation}

\label{section review teleportation-based quantum computation}

In this section, we make a review on teleportation-based quantum computation \cite{GC99}. The key topics include
the standard description of quantum teleportation, the fault-tolerant construction of universal quantum gate set,
and the fault-tolerant construction of four-qubit entangled states.  Meanwhile, we set up our notation and convention
for the study in the paper.

\subsection{Notation}

A single-qubit Hilbert space ${\mathcal H}_2$ has an orthonormal basis denoted by $|i\rangle$, $i=0,1$, and a single-qubit
state $|\alpha\rangle$ is given by $|\alpha\rangle= a|0\rangle +b |1\rangle$ with complex numbers $a$ and $b$.
The unit matrix $1\!\! 1_2$ and the Pauli gates $X$ and $Z$ take the form
\eq
1\!\! 1_2 =\left(\begin{array}{cc}
              1 & 0 \\
              0 & 1 \\
            \end{array}\right), \quad X=\left(\begin{array}{cc}
              0 & 1 \\
              1 & 0 \\
            \end{array} \right), \quad  Z=\left(\begin{array}{cc}
              1 & 0 \\
              0 & -1 \\
            \end{array} \right)
\en
with $Z |0\rangle=|0\rangle$ and $Z |1\rangle=-|1\rangle$, and the Pauli gate $Y$ is defined as $Y=ZX$.
A single-qubit gate is an element of the unitary group $U(2)$, and a typical single-qubit gate $W_{ij}$ in the present paper
has the form
\eq
\label{W ij}
W_{ij}=X^i Z^j, \quad i, j=0,1,
\en
satisfying $W_{ij}^T=W_{ij}^\dag$, in which the upper index $T$ denotes the matrix transpose conjugation and the upper index
$\dag$ denotes the matrix Hermitian conjugation.

A two-qubit Hilbert space ${\mathcal H}_2 \otimes {\mathcal H}_2$ has an orthonormal product basis denoted by $|ij\rangle, i,j=0,1$.
The EPR state $|\Psi\rangle$ \cite{EPR35,Bell64} takes the form
 \eq
 \label{EPR}
|\Psi\rangle =\frac 1 {\sqrt 2} (|00\rangle+|11\rangle)
 \en
and  it has  a very nice algebraic property
\eq
\label{property of EPR}
(1\!\! 1_2 \otimes U)|\Psi\rangle =(U^T \otimes 1\!\! 1_2) |\Psi\rangle
\en
with $U$ denoting any single-qubit gate. The set of the orthonormal Bell states $|\psi(ij)\rangle$ given by
\eq
\label{Bell states}
|\psi(ij)\rangle = (1\!\! 1_2 \otimes W_{ij}) |\Psi\rangle, \quad i,j=0,1
\en
is called the Bell basis \cite{NC2011, Preskill97} of the two-qubit Hilbert space ${\mathcal H}_2 \otimes {\mathcal H}_2$. Obviously, $|\Psi\rangle=|\psi(00)\rangle$.

\subsection{Quantum teleportation}

\begin{figure}
  \begin{center}
  \includegraphics[width=5cm]{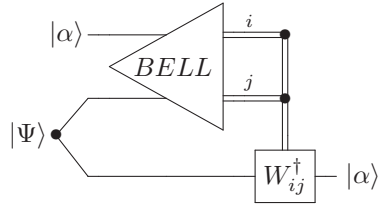}
  \end{center}
  \caption{\label{fig_tele_circuit}  Quantum circuit for quantum teleportation as a diagrammatical representation of
  the teleportation equation (\ref{tele_eq}). An unknown qubit $|\alpha\rangle$ is sent from Alice to Bob. The top two lines
  represent Alice's system and the bottom line represents Bob's system. The single-lines denote qubits and
  the double-lines denote classical bits. The triangle box represents the Bell measurement performed by Alice
  and the square  box represents the local unitary correction operator $W_{ij}^\dag$ performed by Bob to obtain
  the transmitted state $|\alpha\rangle$.}
\end{figure}

Let Alice and Bob share the EPR  state $|\Psi\rangle$, and Alice wants to transfer
an unknown quantum state $|\alpha\rangle$ to Bob. Alice and Bob prepare the quantum
state $|\alpha\rangle\otimes |\Psi\rangle$ which can be formulated as
\eq
\label{tele_eq}
 |\alpha\rangle \otimes |\Psi\rangle =\frac 1 2 \sum_{i,j=0}^1  |\psi(ij)\rangle \otimes W_{ij} |\alpha\rangle
\en
which was called the teleportation equation in \cite{Zhang06}. Then, Alice performs the Bell measurements
 $|\psi(ij)\rangle\langle\psi(ij)|\otimes 1\!\! 1_2$ on the prepared state
 $|\alpha\rangle\otimes |\Psi\rangle$,
 \eq
 (|\psi(ij)\rangle\langle\psi(ij)|\otimes 1\!\! 1_2)(|\alpha\rangle\otimes |\Psi\rangle)
 =\frac 1 2 |\psi(ij)\rangle \otimes  W_{ij} |\alpha\rangle,
 \en
and she informs Bob her measurement results labeled as $(i,j)$.
Finally, Bob applies the unitary correction operator $W_{ij}^\dag$ on his state
\eq
 (1\!\! 1_2 \otimes 1\!\! 1_2 \otimes W^\dag_{ij} ) (|\psi(ij)\rangle \otimes  W_{ij} |\alpha\rangle)
  = |\psi(ij)\rangle \otimes |\alpha\rangle
\en
to obtain the transmitted quantum state $|\alpha\rangle$. See Figure \ref{fig_tele_circuit} for
the description of the teleportation protocol in the language of quantum circuit.

Note that the teleportation protocol of Bob transmitting an unknown qubit $|\alpha\rangle$ to Alice admits another
type of the teleportation equation
\eq
\label{tele_eq_transpose}
 |\Psi\rangle \otimes  |\alpha\rangle =\frac 1 2 \sum_{i,j=0}^1  W^T_{ij} |\alpha\rangle \otimes |\psi(ij)\rangle
\en
which was called the transpose teleportation equation in \cite{ZZ14} and is to be used
in the fault-tolerant construction of two-qubit gates in Subsubsection~\ref{subsection: two-qubit}.

 \subsection{Fault-tolerant construction of universal quantum gate set}

\label{fault_tolerant_u_g_s}

Quantum gates  \cite{Barenco95b} are defined as unitary transformation matrices acting on quantum states, and the set of all $n$-qubit gates forms
a representation of the unitary group $U(2^n)$.

 \subsubsection{Universal quantum gate set}

An entangling two-qubit gate \cite{BB02} with single-qubit gates is called a universal quantum gate set which can perform universal quantum
computation in the circuit model  \cite{NC2011} of quantum computation. All single-qubit gates can be generated by both the Hadamard gate $H$
given by
\eq
\label{Hadamard gate}
H=\frac 1 {\sqrt 2} (X+Z),
\en
and the $\pi / 8$ gate \cite{BMPRV00} given by
\eq
\label{pi/8}
T=\left(\begin{array}{cc}
1 &  0\\
0 &  e^{i \frac \pi 4}\\
\end{array}\right).
\en
An entangling two-qubit gate  \cite{BB02} is defined as a two-qubit gate capable of transforming a tensor product of two single-qubit states
into an entangling two-qubit state. For examples,  the $CNOT$ gate
\eq
\label{CNOT gate}
CNOT = |0\rangle\langle 0| \otimes 1\!\! 1_2 +  |1\rangle\langle 1| \otimes X,
\en
and the $CZ$ gate
\eqa
\label{CZ gate}
CZ = |0\rangle\langle 0| \otimes 1\!\! 1_2 +  |1\rangle\langle 1| \otimes Z.
\ena
They are maximally entangling gates \cite{BG11}, namely, with single-qubit gates, they can generate the maximally entangling states such as the Bell
states (\ref{Bell states}) from the product states. The $CNOT$ gate is the well known two-qubit gate in quantum information and computation
which is the quantum analogue of the Exclusive OR gate in classical computation \cite{NC2011}. The $CZ$ gate is widely used in the one-way
quantum computation \cite{RB01}, the representative example of measurement-based quantum computation, and it is related to the $CNOT$ gate
in the way
\eq
CZ=(1\!\! 1_2 \otimes H) CNOT (1\!\! 1_2\otimes H)\equiv H_2\,\, CNOT\,\, H_2
\en
in which  the subscript of the Hadamard gate $H$ means that it is acting on the second qubit. Hence the set of the $CNOT$ gate (\ref{CNOT gate}) (or
the $CZ$ gate (\ref{CZ gate})) with single-qubit gates $H$ (\ref{Hadamard gate}) and $T$ (\ref{pi/8}) is called a universal quantum gate set
to perform universal quantum computation \cite{NC2011}.

\subsubsection{Clifford gates and fault-tolerant quantum computation}

\label{clifford_gate_fault_tolerant}

The Pauli group gates $C_1$ are generated by tensor products of the Pauli matrices $X$, $Z$ and
the identity matrix $1\!\! 1_2$ with global phase factors $\pm 1, \pm i$. Quantum gates $U$
 \cite{NC2011} are classified by
\eq
\label{C_k}
C_k \equiv \{ U | U C_{k-2} U^\dag \subseteq C_{k-1} \}
\en
where $C_2$ denotes the Clifford gates preserving
the Pauli group gates under conjugation.
Obviously, the Hadamard gate $H$ is a Clifford gate, namely $H \in C_2$, due to
\eq
 H X H =Z, \quad H Z H =X,
\en
and the $\pi / 8$ gate $T$ is not a Clifford gate, $T\in C_3$,  since
\eq
T X T^\dag =\frac {X- \sqrt{-1} Y} {\sqrt 2}, \quad T Z T^\dag = Z
\en
in which  $ ({X- \sqrt{-1} Y})/ {\sqrt 2}$ is a Clifford gate \cite{NC2011, Gottesman97}. Here is another equivalent definition of the
Clifford gate \cite{NC2011, Gottesman97}. If a quantum gate can be represented as a tensor product of the Hadamard
gate (\ref{Hadamard gate}), the phase gate
\eq
\label{phase gate}
S=\left(
    \begin{array}{cc}
      1 & 0 \\
      0 & i \\
    \end{array}
  \right),
\en
and the $CNOT$ gate (\ref{CNOT gate}), then it is a Clifford gate. Note that the phase gate $S$ is the square of
the $\pi / 8$ gate (\ref{pi/8}), $S=T^2$, and the square of the phase gate $S$ is the Pauli $Z$ gate, $S^2=Z$.

In fault-tolerant quantum computation \cite{Gottesman97}, the Pauli group gates $C_1$ and Clifford gates $C_2$ can be easily
performed in principle, but the $C_3$ gates may be difficultly realized. The teleportation-based quantum computation \cite{GC99}
is fault-tolerant quantum computation in the following sense: to perform a $C_3$ gate, it prepares a quantum state with the action
of $C_3$ gate and then applies $C_1$ or $C_2$ gates to such the quantum state using the teleportation protocol.

\subsubsection{Fault-tolerant construction of single-qubit gates}

\label{subsection: single-qubit}

\begin{figure}
 \begin{center}
  \includegraphics[width=6cm]{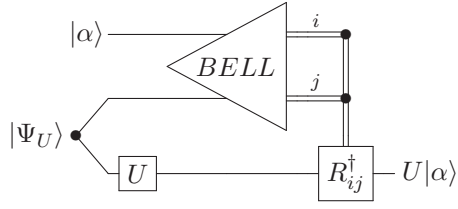}
  \end{center}
  \caption{\label{fig_tele_single_gate}  Quantum circuit for the fault-tolerant construction of the single-qubit gate $U$
 associated with the teleportation equation (\ref{tele_single}). With the such diagrammatical representation, it is obvious that
 fault-tolerantly implementing the single-qubit $U$ becomes how to fault-tolerantly perform the $R^{\dag}_{ij}$ gate and prepare
 the preliminary quantum state $|\Psi_U\rangle$ (\ref{Psi_U}).  }
\end{figure}

To perform a single-qubit gate $U\in C_k$ (\ref{C_k}) on the unknown state $|\alpha\rangle$, Alice prepares the quantum
state $|\Psi_U\rangle$ given by
\eq
\label{Psi_U}
 |\Psi_U\rangle=(1\!\! 1_2\otimes U)|\Psi\rangle
 \en
and reformulate  $|\alpha\rangle\otimes |\Psi_U\rangle$ as
\eq
\label{tele_single}
|\alpha\rangle\otimes|\Psi_U\rangle=\frac{1}{2}\sum_{i,j=0}^1|\psi(ij)\rangle \otimes R_{ij}U|\alpha\rangle,
\en
where the single-qubit gate $R_{ij}$ has the form $R_{ij}=UW_{ij}U^\dag$ with $W_{ij}$ defined in (\ref{W ij}). Then Alice makes
Bell measurements $|\psi(ij)\rangle\langle\psi(ij)|\otimes 1\!\! 1_2$ and informs Bob her measurement results labeled  by $(i,j)$. Finally,
Bob performs the unitary correction operator $R_{ij}^\dag \in C_{k-1}$ (\ref{C_k}) to attain $U |\alpha\rangle$. See Figure \ref{fig_tele_single_gate}
for the quantum circuit of fault-tolerantly performing the single-qubit gate $U$ using the teleportation protocol.

As the single-qubit gate $U$ is the Hadamard gate $H$ (\ref{Hadamard gate}), the single-qubit gate $R_{ij}$ is given by
$R_{ij}(H)=W_{ji}^T$, which is a tensor product of Pauli gates. And  as the single-qubit gate $U$ is the $\pi/8$ gate (\ref{pi/8}),
the single-qubit gate $R_{ij}$ is given by
\eq
\label{RTij}
R_{ij}(T)=(\frac {X- \sqrt{-1} Y} {\sqrt 2})^i\, Z^j,
\en
which can be easily fault-tolerantly performed, because it is a Clifford gate.
Note that the difficulty of performing a single-qubit gate $U\in C_k$ becomes how to
fault-tolerantly prepare the state $|\Psi_U\rangle$ and perform the single-qubit gate $R_{ij}^\dag \in C_{k-1}$.

\subsubsection{Fault-tolerant construction of two-qubit Clifford gates}

\label{subsection: two-qubit}

\begin{figure}
 \begin{center}
  \includegraphics[width=8cm]{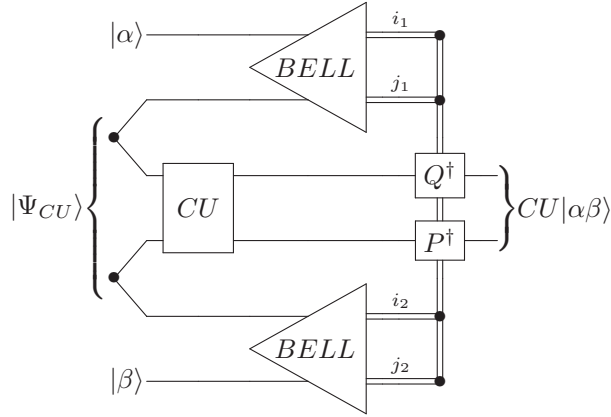}
  \end{center}
  \caption{\label{fig_tele_two_gate}  Quantum circuit for the fault-tolerant construction of a two-qubit Clifford gate $CU$,
  as a diagrammatical representation of the teleportation equation (\ref{tele_two}).  Specifically, the single-qubit gates $Q$ and $P$
  are calculated with the formula (\ref{Q P}).  }
\end{figure}

To perform a two-qubit Clifford gate $CU$ such as the $CNOT$ gate (\ref{CNOT gate}) and the $CZ$ gate (\ref{CZ gate}) on two unknown
single-qubit states $|\alpha\rangle$ and $|\beta\rangle$, we prepare a four-qubit entangled state $|\Psi_{CU}\rangle$,
\eq
\label{Psi_CU}
|\Psi_{CU}\rangle=(1\!\! 1_2\otimes CU \otimes 1\!\! 1_2)(|\Psi\rangle \otimes |\Psi\rangle),
\en
with the action of the $CU$ gate, and reformulate the prepared state $|\alpha\rangle\otimes|\Psi_{CU}\rangle\otimes|\beta\rangle$ as
\eq
\label{tele_two}
\begin{split}
& |\alpha\rangle\otimes|\Psi_{CU}\rangle\otimes|\beta\rangle\\
=&\frac{1}{4}\sum_{i_1,j_1=0}^1\sum_{i_2,j_2=0}^1(1\!\!1_4\otimes
  Q\otimes P\otimes 1\!\!1_4) (|\psi(i_1j_1)\rangle\otimes CU|\alpha\beta\rangle\otimes |\psi(i_2j_2)\rangle)
\end{split}
\en
with $1\!\! 1_4=1\!\! 1_2\otimes 1\!\! 1_2$, which is called the teleportation equation associated with the fault-tolerant
construction of the two-qubit Clifford gate $CU$.

The single-qubit gates $Q$ and $P$ in the teleportation equation (\ref{tele_two}) are calculated by
\eq
\label{Q P}
Q\otimes P=CU(W_{i_1j_1}\otimes W^T_{i_2j_2})CU^\dag.
\en
As the $CU$ gate is the $CNOT$ gate,  the gates $Q$ and $P$ are expressed as
\eq
\label{Q P for CNOT}
Q=Z^{j_2}X^{i_1}Z^{j_1},\quad P=Z^{j_2}X^{i_2}X^{i_1}.
\en
As the $CU$ gate is the $CZ$ gate, the gates $Q$ and $P$ have the form
\eq
\label{Q P for CZ}
Q=Z^{i_2}X^{i_1}Z^{j_1},\quad P=Z^{j_2}X^{i_2}Z^{i_1}.
\en
Note that the $Q$ and $P$  gates (\ref{Q P})  may be not single-qubit gates if the $CU$ gate is not a Clifford gate in
accordance with the definition of a Clifford gate \cite{NC2011, Gottesman97}.

Next, we perform the Bell measurements given by
\eq
\label{six_bell_measurement}
|\psi(i_1 j_1)\rangle\langle\psi(i_1 j_1)|\otimes 1\!\! 1_2\otimes 1\!\! 1_2 \otimes  |\psi(i_2 j_2)\rangle\langle\psi(i_2 j_2)|,
\en
and with the measurement results labeled by $(i_1,j_1)$ and $(i_2,j_2)$, we perform the unitary correction operator, $Q^\dag\otimes P^\dag$, to
obtain the exact action of the Clifford gate $CU$ on the two-qubit state $|\alpha\rangle \otimes |\beta\rangle$, namely, $CU|\alpha\beta\rangle$.

The quantum circuit for the fault-tolerant construction of the two-qubit Clifford gate $CU$ using the teleportation protocol is depicted
in Figure~\ref{fig_tele_two_gate}. Note that the two-qubit Clifford gate $CU$ we study here may be not the controlled-operation two-qubit
  gates such as the $CNOT$ gate (\ref{CNOT gate}) and the $CZ$ gate (\ref{CZ gate}).

\subsection{Construction of four-qubit entangled states}

From Subsection~\ref{fault_tolerant_u_g_s}, we already know that, the conceptual point in the fault-tolerant construction of a single-qubit quantum
gate $U$ (or a two-qubit Clifford gate $CU$) using quantum teleportation is the fault-tolerant preparation of the two-qubit entangled state (\ref{Psi_U}) (or
the four-qubit entangled state (\ref{Psi_CU})). In this subsection, we focus on the fault-tolerant preparation of the four-qubit entangled state
$|\Psi_{CU}\rangle$ (\ref{Psi_CU}) using the teleportation protocol, when the $CU$ gate is the $CNOT$ gate (\ref{CNOT gate}) and the $CZ$ gate (\ref{CZ gate})
respectively.

In Gottesman and Chuang's  original proposal \cite{GC99} of teleportation-based quantum computation, a four-qubit entangled state $|\Psi_{CU}\rangle$ is
prepared in the following steps. First, we prepare a prior entangled six-qubit state as a tensor product of a three-qubit GHZ state $|\Upsilon\rangle$
\cite{GHZ90},
\eq
\label{Upsilon}
|\Upsilon\rangle=\frac 1 {\sqrt 2} (|000\rangle+|111\rangle)
\en
and another three-qubit GHZ state $|\Upsilon\rangle$ with the local action of the Hadamard gate $H$. Second, we perform the Bell measurements on such
the six-qubit entangled state. Third, after classical communication of the Bell measurement outcomes, we perform the unitary correction operators to
obtain the four-qubit entangled state $|\Psi_{CU}\rangle$.

\subsubsection{Construction of the four-qubit entangled state $|\Psi_{CNOT}\rangle$}

\label{construction Psi CNOT}

\begin{figure}
 \begin{center}
  \includegraphics[width=8cm]{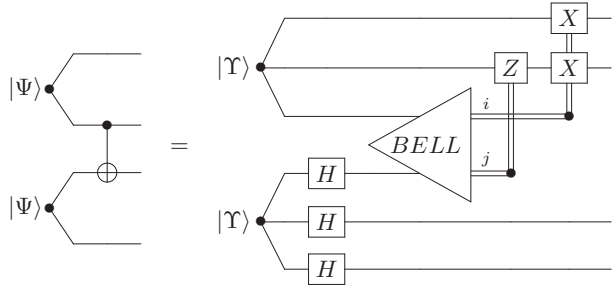}
  \end{center}
  \caption{\label{fig_psi_cnot1}  Quantum circuit for the construction of four-qubit entangled state $|\Psi_{CNOT}\rangle$ (\ref{Psi_cnot}), as a
  diagrammatical representation of the teleportation equation (\ref{tele_cnot1}). The $|\Upsilon\rangle$ state is the three-quibt GHZ state (\ref{Upsilon})
  and the single-qubit gate denoted as $H$ is the Hadamard gate (\ref{Hadamard gate}). The $CNOT$ gate (\ref{CNOT gate}) takes the conventional
  configuration in quantum circuit model \cite{NC2011}. }
\end{figure}

The four-qubit entangled state $|\Psi_{CNOT}\rangle_{1256}$ is the target state we want to construct,
\eq
\label{Psi_cnot}
|\Psi_{CNOT}\rangle_{1256}=(1\!\! 1_2 \otimes CNOT_{25} \otimes 1\!\! 1_2) (|\Psi\rangle_{12} \otimes |\Psi\rangle_{56})
\en
in which the $CNOT_{ij}$ gate requires the $i$-th qubit as the control qubit and the $j$-th qubit as
the target qubit. We prepare the six-qubit entangled state $H_4H_5H_6|\Upsilon\rangle_{123}|\Upsilon\rangle_{456}$,
which can be reformulated as
\eq
\label{tele_cnot1}
H_4H_5H_6|\Upsilon\rangle_{123}|\Upsilon\rangle_{456}=\sum_{i,j=0}^1|\psi(ij)\rangle_{34}\,\, Z_2^jX_1^iX_2^i|\Psi_{CNOT}\rangle_{1256},
\en
then make the joint Bell measurement on both the third and fourth qubits given by
\eq
\label{Bell_measurement_CNOT}
   1\!\! 1_2 \otimes 1\!\! 1_2 \otimes|\psi(ij)\rangle_{34} \langle \psi(ij) | \otimes  1\!\! 1_2  \otimes  1\!\! 1_2.
\en
With the measurement outcome $(i,j)$, we perform the unitary correction operator
\eq
 X^i\otimes  X^iZ^j\otimes 1\!\! 1_2 \otimes 1\!\! 1_2\otimes 1\!\! 1_2 \otimes 1\!\! 1_2,
\en
on the resultant quantum state to obtain the four-qubit entangled state $|\Psi_{CNOT}\rangle$ (\ref{Psi_cnot}). The quantum
circuit for the construction of the $|\Psi_{CNOT}\rangle$ state is shown in Figure~\ref{fig_psi_cnot1}.

Note that the teleportation equation (\ref{tele_cnot1}) admits an equivalent form
\eq
\label{tele_cnot2}
H_4H_5H_6|\Upsilon\rangle_{123}|\,\,\Upsilon\rangle_{456}
 =\sum_{i,j=0}^1|\psi(ij)\rangle_{34}\,\, X_5^iZ_5^jZ_6^j|\Psi_{CNOT}\rangle_{1256},
\en
so that we have the second method of constructing the $|\Psi_{CNOT}\rangle$ state (\ref{Psi_cnot}) using the teleportation protocol. These
two methods are found to be equivalent in the low dimensional topological diagrammatical approach, see Subsubsection~\ref{topo_construction_four_CNOT}.

\subsubsection{Construction of four-qubit entangled states $|\Psi_{CNOT}^\uparrow\rangle$  and $|\Psi_{CZ}\rangle$}

In Gottesman and Chuang's  original study \cite{GC99},  the fault-tolerant construction of the four-qubit entangled state $|\Psi_{CNOT}^\uparrow\rangle$  (\ref{Psi cnot down})
was presented instead of the state $|\Psi_{CNOT}\rangle$ (\ref{Psi_cnot}) that we are working on. For readers' convenience, the construction of the $|\Psi_{CNOT}^\uparrow\rangle$
state is discussed in Appendix~\ref{construction cnot down}. Besides $|\Psi_{CNOT}\rangle$ and $|\Psi_{CNOT}^\uparrow\rangle$, since the $CZ$ gate (\ref{CZ gate}) has a wide application in measurement-based quantum computation \cite{RB01}, we work out the construction of four-qubit entangled state $|\Psi_{CZ}\rangle$ (\ref{Psi_cz}) in Appendix~\ref{construction Psi CZ}.

\section{The extended Temperley--Lieb diagrammatical approach to teleportation-based quantum computation}

\label{TL diagrammatical for tele based quantum computation}

In this section\footnote{This section is an extended version of the authors' unpublished paper \cite{ZP13} in which topological
features of teleportation-based quantum computation are claimed to be associated with topological features of space-time.
}, we aim at exhibiting  topological features of teleportation-based quantum computation in the transparent style. In the
extended Temperley--Lieb diagrammatical approach \cite{Kauffman05, Zhang06}, we study the topological diagrammatical construction
of both universal quantum gate set and four-qubit entangled states in teleportation-based quantum computation \cite{GC99,Nielsen03, Leung04},
whose algebraic counterparts have been presented in Section~\ref{section review teleportation-based quantum computation}.

The extended Temperley--Lieb diagrammatical configuration \cite{Kauffman05, Zhang06} is  a kind of the extension of the standard
Temperley--Lieb configuration on which both single-qubit gates and two-qubit gates are allowed to move along the lines under certain
rules.  The definition of the Temperley--Lieb algebra and its diagrammatical representation is shown up
in Appendix~\ref{Def TL} and~\ref{Def TL and Bell states}.

\subsection{The extended Temperley--Lieb diagrammatical approach}

A single-qubit state vector $|\varphi\rangle$ is denoted by a vertical line followed by the symbol $\nabla$ on the bottom,
\eqa
\setlength{\unitlength}{0.6mm}
\begin{array}{c}
\begin{picture}(25,16)
\put(2,7){$|\varphi\rangle=$}
\put(25,4){\line(0,1){12}}
\put(23,0){\makebox(4,4){$\nabla$}}
\end{picture}
\end{array}
\ena
and the line with the symbol $\triangle$ on the top represents the covector $\langle \phi|$, so a vertical line with both symbols $\nabla$
and $\triangle$ on the boundary denotes the inner product $\langle \phi|\varphi\rangle$. A vertical line with a solid point represents a
single-qubit gate $U$ acting on the state $|\varphi\rangle$,
\eqa
\setlength{\unitlength}{0.6mm}
\begin{array}{c}
\begin{picture}(35,16)
\put(2,7){$U|\varphi\rangle=$}
\put(30,4){\line(0,1){12}}
\put(30,10){\circle*{2.}}
\put(28,0){\makebox(4,4){$\nabla$}}
\put(32,9){\tiny{$U$}}
\end{picture}
\end{array}
\ena
in which the algebraic expression is read from the right to the left and the diagrammatical representation is read from the bottom to the top.

The diagrammatical configuration of the Bell state $|\psi(ij)\rangle$ (\ref{Bell states}) is the core of the
extended Temperley--Lieb diagrammatical approach, and it is represented by a cup with a solid point
denoting the single-qubit gate $W_{ij}$ (\ref{W ij}),
\eqa
\label{TL diag Bell state}
\setlength{\unitlength}{0.6mm}
\begin{array}{c}
\begin{picture}(45,16)
\put(2,5){$|\psi(ij)\rangle=$}
\put(32,0){\line(0,1){12}}
\put(42,0){\line(0,1){12}}
\put(32,0){\line(1,0){10}}
\put(42,6){\circle*{2.}}
\put(44.,5){\tiny{$W_{ij}$}}
\end{picture}
\end{array}
\ena
so that a cup without a solid point denotes the EPR state $|\Psi\rangle$ (\ref{EPR}).
The adjoint of the Bell state, $\langle \psi(ij)|$, is represented by a cap
\eqa
\label{TL diag Bell state_cap}
\setlength{\unitlength}{0.6mm}
\begin{array}{c}
\begin{picture}(45,16)
\put(2,5){$\langle\psi(ij)|=$}
\put(32,0){\line(0,1){12}}
\put(42,0){\line(0,1){12}}
\put(32,12){\line(1,0){10}}
\put(42,6){\circle*{2.}}
\put(44.,5){\tiny{$W_{ij}^\dag$}}
\end{picture}
\end{array}
\ena
with a solid point denoting the Hermitian conjugation of $W_{ij}$. These diagrammatic states are called
a cup state or a cap state respectively. The projective measurement $|\psi(ij)\rangle\langle\psi(ij) |$
is called the Bell measurement,
\eqa
\label{TL Bell measurement}
\setlength{\unitlength}{0.6mm}
\begin{array}{c}
\begin{picture}(65,30)
\put(62,18){\line(0,1){12}}
\put(52,18){\line(0,1){12}}
\put(52,18){\line(1,0){10}}
\put(62,24){\circle*{2.}}
\put(64.,23){\tiny{$W_{ij}$}}
\put(2,14){$|\psi(ij)\rangle \langle\psi(ij)|=$}
\put(52,0){\line(0,1){12}}
\put(62,0){\line(0,1){12}}
\put(52,12){\line(1,0){10}}
\put(62,6){\circle*{2.}}
\put(64,5){\tiny{$W_{ij}^\dag$}}
\end{picture}
\end{array}
\ena
represented by a top cup state with a bottom cap state. Note that without solid points on the configuration of the Bell measurement,
such the configuration represents the basic generator of the Temperley--Lieb algebra, see
Appendix~\ref{Def TL} and~\ref{Def TL and Bell states}.

The nice property  (\ref{property of EPR}) of the EPR state has the corresponding diagrammatical expression
\eqa
\setlength{\unitlength}{0.6mm}
\begin{array}{c}
\begin{picture}(45,16)
\put(2,0){\line(0,1){12}}
\put(12,0){\line(0,1){12}}
\put(2,0){\line(1,0){10}}
\put(12,6){\circle*{2.}}
\put(14.,5){\tiny{$U$}}
\put(23.,6){$=$}
\put(32,5){\tiny{$U^T$}}
 \put(40.8,6){\circle*{2.}}
\put(40.8,0){\line(0,1){12}}
\put(50.8,0){\line(0,1){12}}
\put(40.8,0){\line(1,0){10}}
\end{picture}
\end{array}
\label{trans law: 1l}
\ena
with $U$ denoting any single-qubit gate, and the similar representation also for a cap state,
\eqa
\setlength{\unitlength}{0.6mm}
\begin{array}{c}
\begin{picture}(45,16)
\put(2,0){\line(0,1){12}}
\put(12,0){\line(0,1){12}}
\put(2,12){\line(1,0){10}}
\put(12,6){\circle*{2.}}
\put(14.,5){\tiny{$U^\dag$}}
\put(23.,6){$=$}
\put(32,5){\tiny{$U^\ast$}}
\put(40.8,6){\circle*{2.}}
\put(40.8,0){\line(0,1){12}}
\put(50.8,0){\line(0,1){12}}
\put(40.8,12){\line(1,0){10}}
\end{picture}
\end{array}
\label{trans law: 22}
\ena
with the upper index $\ast$ denoting the complex conjugation. In the diagrammatical representations (\ref{trans law: 1l}) and (\ref{trans law: 22}),
a single-qubit gate can flow from the one branch of a cup (or cap) state to its other branch with the transpose conjugation, which is beyond the
conventional utilization of the Temperley--Lieb diagrams but naturally arises in quantum computation. Note that such the operation of moving
single-qubit gates is a crucial technique in the extended Temperley--Lieb diagrammatical approach \cite{Kauffman05, Zhang06} to teleportation-based
quantum circuits.

\subsection{Topological interpretation of quantum teleportation}

In quantum teleportation, Alice and Bob prepare the quantum state $|\alpha\rangle\otimes |\Psi\rangle$, in the extended Temperley--Lieb diagrammatical language,
expressed as
\eqa
\setlength{\unitlength}{0.6mm}
\begin{array}{c}
\begin{picture}(70,16)
\put(10,7){$|\alpha\rangle\otimes |\Psi\rangle=$}
\put(45,0){\makebox(4,4){$\nabla$}}
\put(47,3.7){\line(0,1){10.3}}
\put(57,0){\line(0,1){14}}
\put(67,0){\line(0,1){14}}
\put(57,0){\line(1,0){10}}
\end{picture}
\end{array}
\ena
in which the vertical line with $\nabla$ denotes the unknown quantum state $|\alpha\rangle$  to be transmitted from Alice to Bob. Then, Alice performs
Bell measurements $|\psi(ij)\rangle\langle\psi(ij)|\otimes 1\!\! 1_2$ on the prepared state
$|\alpha\rangle\otimes |\Psi\rangle$, which has the extended Temperley--Lieb diagrammatical configuration
\eqa
\setlength{\unitlength}{0.6mm}
\begin{array}{c}
\begin{picture}(80,45)
\put(10,0){\makebox(4,4){$\nabla$}}
\put(12.,4){\line(0,1){20}}
\put(22,18){\circle*{2.}}
\put(23,17){\tiny{$W_{ij}^\dag$}}
\put(22,0){\line(0,1){24}}
\put(32,0){\line(0,1){42}}
\multiput(8,12)(1,0){30}{\line(1,0){.5}}
\put(22,0){\line(1,0){10}}
\put(12,24){\line(1,0){10}}
\put(12,30){\line(1,0){10}}
\put(12,30){\line(0,1){12}}
\put(22,30){\line(0,1){12}}
\put(22,36){\circle*{2.}}
\put(23,35){\tiny{$W_{ij}$}}
\put(40,21){\makebox(14,10){$=\frac 1 2$}}
\put(62,30){\line(1,0){10}}
\put(62,30){\line(0,1){12}}
\put(72,30){\line(0,1){12}}
\put(72,36){\circle*{2.}}
\put(73,35){\tiny{$W_{ij}$}}
\put(80,0){\makebox(4,4){$\nabla$}}
\put(82,4){\line(0,1){38}}
\put(82,21){\circle*{2.}}
\put(84,20){\tiny{$W_{ij}$}}
\end{picture}
\label{tele: tl}
\end{array}
\ena
with the single-qubit gate $W_{ij}$ defined in (\ref{W ij}).  It is obvious that this topological diagram (\ref{tele: tl}) is associated
with both the teleportation equation (\ref{tele_eq}) and the quantum circuit in Figure \ref{fig_tele_circuit}. Note that the
diagram (\ref{tele: tl}) without solid points is a standard diagrammatical representation of the Temperley--Lieb algebra  \cite{Kauffman02},
see Appendix~\ref{Def TL} and~\ref{Def TL and Bell states}.

Now let us explain the diagram (\ref{tele: tl}) in detail. On the left hand side of $=$, the diagrammatical part above the dashed line denotes
the Bell measurement, and the part under the dashed line denotes the state preparation. On the right hand side of $=$, the normalization
factor $\frac 1 2$ is contributed by the vanishing cup state and cap state according to the rules of the extend Temperley--Lieb diagrammatical
approach \cite{Kauffman05, Zhang06}. The cup state with the action of $W_{ij}$ denotes the post-measurement state which is usually neglected in
the following study for simplicity. The $W_{ij}$ gate acting on the unknown qubit $|\alpha\rangle$ is due to the operation of
moving $W_{ij}^\dag$ from the one branch of the cap state to the other branch and applying the transposition conjugation, $(W_{ij}^{\dag})^T=W_{ij}$.
Note that both classical communication and unitary correction are not shown in the
diagram (\ref{tele: tl}). In this sense, hence, the quantum information flow sending an
unknown qubit from Alice to Bob in quantum teleportation can be recognized as a result
of topological operation in the extended Temperley--Lieb diagrammatical
approach \cite{Kauffman05, Zhang06}.

The topology in the diagrammatical teleportation (\ref{tele: tl}) may be not
that obvious. Let us consider the topological configuration of the chained teleportation \cite{Childs03}
that Alice sends an unknown qubit $|\alpha\rangle$ to Bob with a sequence of standard teleportation protocols, expressed as
\eqa
\label{chain teleportation}
\setlength{\unitlength}{0.6mm}
\begin{array}{c}
\begin{picture}(70,25)
\put(10,0){\makebox(4,4){$\nabla$}}
\put(12,4){\line(0,1){20}}
\put(22,0){\line(0,1){24}}
\put(32,0){\line(0,1){24}}
\multiput(8,12)(1,0){48}{\line(1,0){.5}}
\put(22,0){\line(1,0){10}}
\put(12,24){\line(1,0){10}}
\put(32,24){\line(1,0){10}}
\put(42,0){\line(0,1){24}}
\put(42,0){\line(1,0){10}}
\put(52,0){\line(0,1){24}}
\put(57,7){\makebox(14,10){$=\frac 1 4$}}
\put(72,0){\makebox(4,4){$\nabla$}}
\put(74,4){\line(0,1){20}}
\end{picture}
\end{array}
\ena
in which the post-measurement states are neglected and only the EPR state measurements $|\Psi\rangle \langle \Psi|$ are considered.  The normalization factor $\frac 1 4$
is calculated from two vanishing cup states and two vanishing cap states. Without unitary corrections, Bob obtains the exact quantum state $|\alpha\rangle$. Hence {\em the
teleportation topology in the extended Temperley--Lieb diagrammatical approach is defined as a topological operation which straightens the configuration consisting of cap
states and cup states}. In addition, if we perform the Bell measurements $|\psi(ij)\rangle\langle\psi(ij)|$ instead of the EPR state measurements in the chained
teleportation, first of all, we have to move single-qubit gates along the path formed by top cap states with bottom cup states until boundary points of this path under
the guidance of the properties (\ref{trans law: 1l}) and  (\ref{trans law: 22}), and then apply the topological operation by straightening the relevant configuration.

 \subsection{Topological construction of universal quantum gate set}

In the authors' knowledge, a topological diagrammatical construction of universal quantum gate set \cite{NC2011,  Barenco95b} using quantum teleportation \cite{GC99}
has not been done yet in the published paper, so we study such the realization in the extended Temperley-Lieb diagrammatical approach \cite{Kauffman05,Zhang06}.

The topological diagram for the fault-tolerant construction of a single-qubit $U$ using quantum teleportation has the form
\eqa
\setlength{\unitlength}{0.7mm}
\begin{array}{c}
\begin{picture}(90,25)
\put(10,0){\makebox(4,4){$\nabla$}}
\put(12.,3.7){\line(0,1){20.3}}
\put(22,18){\circle*{1.5}}
\put(24,17){\tiny{$W_{ij}^\dag$}}
\put(22,0){\line(0,1){24}}
\put(32,0){\line(0,1){24}}
\put(32,6){\circle*{1.5}}
\put(34,5){\tiny{$U$}}
\multiput(8,12)(1,0){28}{\line(1,0){.5}}
\put(22,0){\line(1,0){10}}
\put(12,24){\line(1,0){10}}
\put(36,7){\makebox(14,10){$=\frac 1 2$}}
\put(52,0){\makebox(4,4){$\nabla$}}
\put(54.,3.7){\line(0,1){20.3}}
\put(54,12){\circle*{1.5}}
\put(56,6){\makebox(10,10){\tiny{$U W_{ij}$}}}
\put(66,7){\makebox(14,10){$=\frac 1 2$}}
\put(82,0){\makebox(4,4){$\nabla$}}
\put(84,3.7){\line(0,1){20.3}}
\put(84,18){\circle*{1.5}}
\put(86,13){\makebox(5,10){\tiny{$R_{ij}$}}}
\put(84,9){\circle*{1.5}}
\put(85,4){\makebox(5,10){\tiny{$U$}}}
\end{picture}
\label{gate: one}
\end{array}
\ena
and it is directly associated with the teleportation equation (\ref{tele_single}) and the quantum circuit model in Figure~\ref{fig_tele_single_gate},  see
Subsubsection~\ref{subsection: single-qubit}. Below the dashed line, Alice and Bob prepare the quantum state $|\alpha\rangle\otimes|\Psi_U\rangle$, and above
the dashed line, Alice performs the Bell measurements $|\psi(ij)\rangle\langle\psi(ij)|\otimes 1\!\! 1_2$. After moving the single-qubit gate $W_{ij}^\dag$
from Alice to Bob, we straighten the configuration of the top cap with the bottom cup and obtain the normalization factor $\frac 1 2$. Obviously, Bob has to
perform the local unitary correction operator $R^\dag_{ij}$ to obtain the action of the single-qubit gate $U$ on the unknown state $|\alpha\rangle$, namely,
$U|\alpha\rangle$, as is not shown up in the topological diagram~(\ref{gate: one}).

The topological diagram for the fault-tolerant construction of a two-qubit Clifford gate $CU$ using quantum teleportation is expressed as
\eqa
\setlength{\unitlength}{0.8mm}
\begin{array}{c}
\begin{picture}(150,48)
\put(14,9){\makebox(4,4){$\nabla$}}
\put(11.5,0){\makebox(9,8){$\tiny{|\alpha\rangle}$}}
\put(16,12.5){\line(0,1){32.5}}
\put(26,36){\circle*{1.5}}
\put(18,33){\makebox(6,6){\tiny{$W_{i_1j_1}^\dag$}}}
\put(26,9){\line(0,1){36}}
\multiput(12,27)(1,0){58}{\line(1,0){.5}}
\put(36,9){\line(0,1){7}}
\put(36,20){\line(0,1){25}}
\put(34,16){\line(0,1){4}}
\put(48,16){\line(0,1){4}}
\put(34,16){\line(1,0){14}}
\put(34,20){\line(1,0){14}}
\put(39,16){\makebox(4,4){\tiny{$CU$}}}
\put(26,9){\line(1,0){10}}
\put(16,45){\line(1,0){10}}
\put(46,9){\line(0,1){7}}
\put(46,20){\line(0,1){25}}
\put(46,9){\line(1,0){10}}
\put(56,9){\line(0,1){36}}
\put(56,45){\line(1,0){10}}
\put(66,12.5){\line(0,1){32.5}}
\put(66,36){\circle*{1.5}}
\put(59,34){\makebox(4,4){\tiny{$W_{i_2j_2}^\dag$}}}
\put(64,9){\makebox(4,4){$\nabla$}}
\put(61.5,0){\makebox(9,8){$\tiny{|\beta\rangle}$}}
\put(70,22){\makebox(14,10){$=~\frac 1 4$}}
\put(86,9){\makebox(4,4){$\nabla$}}
\put(82.5,0){\makebox(9,8){$\tiny{|\alpha\rangle}$}}
\put(88,29){\line(0,1){16}}
\put(88,12.5){\line(0,1){12.5}}
\put(96,9){\makebox(4,4){$\nabla$}}
\put(95.5,0){\makebox(9,8){$\tiny{|\beta\rangle}$}}
\put(86,25){\line(0,1){4}}
\put(100,25){\line(0,1){4}}
\put(86,25){\line(1,0){14}}
\put(86,29){\line(1,0){14}}
\put(91,25){\makebox(4,4){\tiny{$CU$}}}
\put(98,29){\line(0,1){16}}
\put(98,12.5){\line(0,1){12.5}}
\put(88,18){\circle*{1.5}}
\put(80,15){\makebox(6,6){\tiny{$W_{i_1j_1}$}}}
\put(98,18){\circle*{1.5}}
\put(102,16){\makebox(4,4){\tiny{$W_{i_2j_2}^T$}}}
\put(106,22){\makebox(14,10){$=~\frac 1 4$}}
\put(124,9){\makebox(4,4){$\nabla$}}
\put(120.5,0){\makebox(9,8){$\tiny{|\alpha\rangle}$}}
\put(126,29){\line(0,1){16}}
\put(126,12.5){\line(0,1){12.5}}
\put(134,9){\makebox(4,4){$\nabla$}}
\put(133.5,0){\makebox(9,8){$\tiny{|\beta\rangle}$}}
\put(124,25){\line(0,1){4}}
\put(138,25){\line(0,1){4}}
\put(124,25){\line(1,0){14}}
\put(124,29){\line(1,0){14}}
\put(129,25){\makebox(4,4){\tiny{$CU$}}}
\put(136,29){\line(0,1){16}}
\put(136,12.5){\line(0,1){12.5}}
\put(126,36){\circle*{1.5}}
\put(121,34){\makebox(4,4){\tiny{$Q$}}}
\put(136,36){\circle*{1.5}}
\put(137,34){\makebox(4,4){\tiny{$P$}}}
\end{picture}
\label{gate: two}
\end{array}
\ena
and it is associated with the teleportation equation (\ref{tele_two}) and the quantum circuit model in Figure~\ref{fig_tele_two_gate}. Below
the dashed line,  we prepare the six-qubit quantum state $|\alpha\rangle\otimes|\Psi_{CU}\rangle\otimes|\beta\rangle$, and above the dashed
line, we perform the Bell measurement (\ref{six_bell_measurement}) on such the prepared quantum state. The topological (or straightening)
operation occurs after both moving single-qubit gates $W_{i_1j_1}^\dag$ and $W_{i_2j_2}^\dag$ along the path formed by the top cup states
and bottom cup states and moving the two-qubit gate $CU$ along two vertical lines. The vanishing cap and cup states contribute the
normalization factor $\frac 1 4$.  Explicitly, the single-qubit gates $Q$ and $P$ are calculated by the formula (\ref{Q P}).

As a remark, the extended Temperley--Lieb diagrammatical approach to the fault-tolerant construction of quantum gates in teleportation-based
quantum computation appears more intuitive and simpler than other original approaches \cite{GC99,Nielsen03,Leung04}. On the topological
 representations, such as (\ref{tele: tl}), (\ref{gate: one}) and (\ref{gate: two}), one is allowed to transport an unknown
quantum state by topological operations as well as  move single-qubit or two-qubit gates along relevant configurations.

\subsection{Topological construction of four-qubit entangled states}

We combine the quantum circuit models \cite{NC2011} of the Bell states and the GHZ states with the extended Temperley--Lieb diagrammatical approach
to construct the topological diagrammatical representations of four-qubit entangled states including the $|\Psi_{CNOT}\rangle$ state (\ref{Psi_cnot}),
the $|\Psi_{CNOT}^\uparrow\rangle$ state (\ref{Psi cnot down}) and the  $|\Psi_{CZ}\rangle$ state (\ref{Psi_cz}).

\subsubsection{Diagrammatical representations of the Bell state $|\Psi\rangle$ and GHZ state $|\Upsilon\rangle$ }

The EPR state $|\Psi\rangle$ (\ref{EPR}) can be generated by the quantum circuit model in terms of the Hadamard gate $H$ and the $CNOT$ gate,
\eq
|\Psi\rangle = CNOT_{12} (H\otimes 1\!\! 1_2) |0\rangle \otimes |0\rangle
\en
with the diagrammatic representation
\eqa
\label{EPR_diag1}
\setlength{\unitlength}{0.6mm}
\begin{array}{c}
\begin{picture}(60,22)
\put(8,4){\makebox(28,10){$|\Psi\rangle=$}}
\put(32,0){\makebox(4,4){$\nabla$}}
\put(34,4){\line(0,1){20}}
\put(34,8){\circle*{2.}}
\put(34,16){\circle*{2.}}
\put(37,6){\makebox(4,4){$H$}}
\put(44,0){\makebox(4,4){$\nabla$}}
\put(46,4){\line(0,1){20}}
\put(34,16){\line(1,0){14}}
\put(46,16){\circle{4}}
\end{picture}
\end{array}
\ena
where the vertical line with $\nabla$ denotes the state $|0\rangle$. Such the configuration of the EPR state $|\Psi\rangle$ is
equivalent to the cup configuration (\ref{TL diag Bell state}) of the  $|\Psi\rangle$  state in
the extended Temperley--Lieb diagrammatical approach. The EPR state $|\Psi\rangle$ has the other equivalent quantum circuit model
\eq
|\Psi\rangle=CNOT_{21}(1\!\!1_2\otimes H)|00\rangle
\en
with the diagrammatical representation
\eqa
\label{EPR_diag2}
\setlength{\unitlength}{0.6mm}
\begin{array}{c}
\begin{picture}(60,22)
\put(8,4){\makebox(28,10){$|\Psi\rangle=$}}
\put(32,0){\makebox(4,4){$\nabla$}}
\put(44,0){\makebox(4,4){$\nabla$}}
\put(39,6){\makebox(4,4){$H$}}
\put(34,4){\line(0,1){20}}
\put(46,4){\line(0,1){20}}
\put(32,16){\line(1,0){14}}
\put(46,8){\circle*{2.}}
\put(46,16){\circle*{2.}}
\put(34,16){\circle{4}}
\end{picture}
\end{array}
\ena
which is associated with the configuration (\ref{EPR_diag1}) via the mirror symmetry.

The three-qubit GHZ state $|\Upsilon\rangle$ (\ref{Upsilon}) can be generated by the quantum circuit model in terms of
the EPR state $|\Psi\rangle$ and the $CNOT$ gate, expressed as
\eq
|\Upsilon\rangle
=(CNOT_{21}\otimes 1\!\!1_2)|0\rangle\otimes |\Psi\rangle,
\en
with the diagrammatical representation
\eqa
\setlength{\unitlength}{0.6mm}
\begin{array}{c}
\begin{picture}(36,18)
\put(0,4){\makebox(16,10){$|\Upsilon\rangle=$}}
\put(18,0){\makebox(4,4){$\nabla$}}
\put(30,0){\line(0,1){18}}
\put(40,0){\line(0,1){18}}
\put(20,4){\line(0,1){14}}
\put(30,0){\line(1,0){10}}
\put(18,12){\line(1,0){12}}
\put(20,12){\circle{4}}
\put(30,12){\circle*{2.}}
\end{picture}
\label{GHZ 1}
\end{array}
\ena
where the cup configuration (\ref{TL diag Bell state}) of the EPR state $|\Psi\rangle$ is exploited.
The $|\Upsilon\rangle$ state also allows the other equivalent quantum circuit model
\eq
|\Upsilon\rangle=(1\!\! 1_2 \otimes CNOT_{23})  |\Psi\rangle \otimes |0\rangle,
\en
with the diagrammatical representation
 \eqa
 \label{GHZ_diag2}
\setlength{\unitlength}{0.6mm}
\begin{array}{c}
\begin{picture}(36,18)
\put(2,4){\makebox(16,10){$|\Upsilon\rangle=$}}
\put(20,0){\line(0,1){18}}
\put(20,0){\line(1,0){10}}
\put(30,0){\line(0,1){18}}
\put(38,0){\makebox(4,4){$\nabla$}}
\put(40,4){\line(0,1){14}}
\put(30,12){\line(1,0){12}}
\put(40,12){\circle{4}}
\put(30,12){\circle*{2.}}
\end{picture}
\label{GHZ 2}
\end{array}
\ena
which can be obtained from the configuration (\ref{GHZ 1}) via the mirror symmetry.

Furthermore, the GHZ state $|\Upsilon\rangle$ with the local action of the Hadamard gates such as $(H\otimes H\otimes H)|\Upsilon\rangle$  has
the diagrammatical representation,
\eqa
\setlength{\unitlength}{0.6mm}
\begin{array}{c}
\begin{picture}(74,25)
\put(4,16){\makebox(4,4){\tiny{$H$}}}
\put(16,16){\makebox(4,4){\tiny{$H$}}}
\put(28,16){\makebox(4,4){\tiny{$H$}}}
\put(0,0){\makebox(4,4){$\nabla$}}
\put(14,0){\line(1,0){12}}
\put(0,10){\line(1,0){14}}
\put(2,4){\line(0,1){20}}
\put(14,0){\line(0,1){24}}
\put(26,0){\line(0,1){24}}
\put(2,18){\circle*{2.}}
\put(14,18){\circle*{2.}}
\put(26,18){\circle*{2.}}
\put(14,10){\circle*{2.}}
\put(2,10){\circle{4}}
\put(34,10){\makebox(4,4){$=$}}
\put(42,0){\makebox(4,4){$\nabla$}}
\put(46,8){\makebox(4,4){\tiny{$H$}}}
\put(56,0){\line(0,1){24}}
\put(68,0){\line(0,1){24}}
\put(44,4){\line(0,1){20}}
\put(56,0){\line(1,0){12}}
\put(44,18){\line(1,0){14}}
\put(44,18){\circle*{2.}}
\put(44,10){\circle*{2.}}
\put(56,18){\circle{4}}
\end{picture}
\label{GHZ 3}
\end{array}
\ena
where the configuration (\ref{GHZ 1}) of the GHZ state $|\Upsilon\rangle$ is used and  the formula
\eq
(H\otimes H) CNOT_{21} (H\otimes H) = CNOT_{12}
\en is applied. Under the mirror symmetry, the $(H\otimes H\otimes H)|\Upsilon\rangle$  state
has the other equivalent configuration
\eqa
\setlength{\unitlength}{0.6mm}
\begin{array}{c}
\begin{picture}(74,25)
\put(2,0){\line(0,1){24}}
\put(2,18){\circle*{2.}}
\put(4,16){\makebox(4,4){\tiny{$H$}}}
\put(2,0){\line(1,0){12}}
\put(14,0){\line(0,1){24}}
\put(14,18){\circle*{2.}}
\put(16,16){\makebox(4,4){\tiny{$H$}}}
\put(24,0){\makebox(4,4){$\nabla$}}
\put(26,4){\line(0,1){20}}
\put(26,18){\circle*{2.}}
\put(28,16){\makebox(4,4){\tiny{$H$}}}
\put(14,10){\line(1,0){14}}
\put(26,10){\circle{4}}
\put(14,10){\circle*{2.}}
\put(34,10){\makebox(4,4){$=$}}
\put(44,0){\line(0,1){24}}
\put(44,0){\line(1,0){12}}
\put(56,0){\line(0,1){24}}
\put(66,0){\makebox(4,4){$\nabla$}}
\put(68,4){\line(0,1){20}}
\put(54,18){\line(1,0){14}}
\put(68,18){\circle*{2.}}
\put(56,18){\circle{4}}
\put(68,10){\circle*{2.}}
\put(70,8){\makebox(4,4){\tiny{$H$}}}
\end{picture}
\label{GHZ 4}
\end{array}
\ena
in which the configuration (\ref{GHZ 2}) of the GHZ state $|\Upsilon\rangle$ is exploited.

\subsubsection{Topological construction of the four-qubit entangled state $|\Psi_{CNOT}\rangle$ }

\label{topo_construction_four_CNOT}

The four-qubit  entangled state $|\Psi_{CNOT}\rangle$ (\ref{Psi_cnot}) has the extended Temperley--Lieb diagrammatical configuration, expressed as
a $CNOT$ gate connecting two cup configurations,
\eqa
\setlength{\unitlength}{0.6mm}
\begin{array}{c}
\begin{picture}(55,14)
\put(-4,2){\makebox(16,10){$|\Psi_{CNOT}\rangle=$}}
\put(22,0){\line(0,1){14}}
\put(32,0){\line(0,1){14}}
\put(22,0){\line(1,0){10}}
\put(32,7){\line(1,0){12}}
\put(32,7){\circle*{2.}}
\put(42,0){\line(0,1){14}}
\put(52,0){\line(0,1){14}}
\put(42,0){\line(1,0){10}}
\put(42,7){\circle{4.}}
\end{picture}
\label{Psi_cnot_diag}
\end{array}
\ena
where the second qubit is the control qubit and the fifth qubit is the target qubit and both the third and fourth qubits are not explicitly shown up.

Following the strategy of constructing $|\Psi_{CNOT}\rangle$ (\ref{Psi_cnot}) in Subsubsection~\ref{construction Psi CNOT},
we perform the Bell measurements (\ref{Bell_measurement_CNOT}) on both the third and fourth qubits of the six-qubit state $|\Upsilon\rangle \otimes (H\otimes H \otimes H) |\Upsilon\rangle$. Using the diagrammatical representation (\ref{GHZ 1}) of the GHZ state $|\Upsilon\rangle$ and the diagrammatical representation (\ref{GHZ 3}) of the
state $(H\otimes H \otimes H)|\Upsilon\rangle$ and the diagrammatical representation (\ref{TL Bell measurement})  of  the Bell measurements (\ref{Bell_measurement_CNOT}),
we have the extended Temperley--Lieb diagrammatical configuration for the construction of the $|\Psi_{CNOT}\rangle$ state (\ref{Psi_cnot}),
\eq
\setlength{\unitlength}{0.6mm}
\begin{array}{c}
\begin{picture}(154,50)
\put(8,0){\makebox(4,4){$\nabla$}}
\put(48,8){\makebox(4,4){\tiny{$H$}}}
\put(44,0){\makebox(4,4){$\nabla$}}
\put(38,33){\makebox(6,6){\tiny{$W_{ij}^\dag$}}}
\put(75,20){\makebox(10,8){$= \frac 1 2$}}
\multiput(8,24)(1,0){65}{\line(1,0){.5}}
\put(10,4){\line(0,1){44}}
\put(22,0){\line(0,1){48}}
\put(34,0){\line(0,1){48}}
\put(46,4){\line(0,1){44}}
\put(58,0){\line(0,1){48}}
\put(70,0){\line(0,1){48}}
\put(8,18){\line(1,0){14}}
\put(22,0){\line(1,0){12}}
\put(34,48){\line(1,0){12}}
\put(46,18){\line(1,0){14}}
\put(58,0){\line(1,0){12}}
\put(46,18){\circle*{2.}}
\put(46,10){\circle*{2.}}
\put(22,18){\circle*{2.}}
\put(46,36){\circle*{2.}}
\put(10,18){\circle{4}}
\put(58,18){\circle{4}}
\put(88,0){\makebox(4,4){$\nabla$}}
\put(104,8){\makebox(4,4){\tiny{$H$}}}
\put(100,0){\makebox(4,4){$\nabla$}}
\put(105,41){\makebox(4,4){\tiny{$Z^j$}}}
\put(93,35){\makebox(4,4){\tiny{$X^i$}}}
\put(105,35){\makebox(4,4){\tiny{$X^i$}}}
\put(90,4){\line(0,1){44}}
\put(102,4){\line(0,1){44}}
\put(138,0){\line(0,1){48}}
\put(150,0){\line(0,1){48}}
\put(138,0){\line(1,0){12}}
\put(102,18){\line(1,0){38}}
\put(88,30){\line(1,0){14}}
\put(138,18){\circle{4}}
\put(90,30){\circle{4}}
\put(102,18){\circle*{2.}}
\put(102,10){\circle*{2.}}
\put(102,30){\circle*{2.}}
\put(102,36){\circle*{2.}}
\put(102,42){\circle*{2.}}
\put(90,36){\circle*{2.}}
\end{picture}
\label{fig_psi_cnot_up10}
\end{array}
\en
which is associated with the teleportation equation~(\ref{tele_cnot1}) and the quantum circuit model in Figure~\ref{fig_psi_cnot1}, and in which below
the dashed line is the prepared six-qubit state and above the dashed line is the Bell measurements.

Let us perform a series of diagrammatical operations on the diagram~(\ref{fig_psi_cnot_up10}). First,  move the single-qubit gate $W_{ij}^\dag$ from
the fourth qubit to the second qubit along the given path. Second, continue to move such the $W_{ij}^\dag$ gate across the $CNOT_{21}$ gate with
the formula
\eq
CNOT_{21}(1\!\!1_2\otimes W^\dag_{ij})CNOT_{21}=Z_2^jX_1^iX_2^i,
\en
to obtain the single-qubit gates $Z_2^jX_1^iX_2^i$ acting on the first and second qubits. Third, straighten the configuration of the bottom cup with
the top cap between the second qubit and the fourth qubit to derive the normalization factor $\frac 1 2$ and meanwhile to obtain the $CNOT_{25}$ gate
and the Hadamard gate $H_2$ respectively from the original $CNOT_{45}$ gate and the Hadamard gate $H_4$

Obviously, the $CNOT_{21}$ gate commutes with the $CNOT_{25}$ gate,  and applying such the fact on the diagram~(\ref{fig_psi_cnot_up10}) brings about
the topological diagrammatical representation
\eq
\setlength{\unitlength}{0.6mm}
\begin{array}{c}
\begin{picture}(154,50)
\put(8,0){\makebox(4,4){$\nabla$}}
\put(20,0){\makebox(4,4){$\nabla$}}
\put(24,8){\makebox(4,4){\tiny{$H$}}}
\put(25,41){\makebox(4,4){\tiny{$Z^j$}}}
\put(13,35){\makebox(4,4){\tiny{$X^i$}}}
\put(25,35){\makebox(4,4){\tiny{$X^i$}}}
\put(10,4){\line(0,1){44}}
\put(22,4){\line(0,1){44}}
\put(58,0){\line(0,1){48}}
\put(70,0){\line(0,1){48}}
\put(58,0){\line(1,0){12}}
\put(22,30){\line(1,0){38}}
\put(8,18){\line(1,0){14}}
\put(22,10){\circle*{2.}}
\put(22,18){\circle*{2.}}
\put(22,30){\circle*{2.}}
\put(22,36){\circle*{2.}}
\put(22,42){\circle*{2.}}
\put(10,36){\circle*{2.}}
\put(58,30){\circle{4}}
\put(10,18){\circle{4}}
\put(76,22){\makebox(4,4){$=$}}
\put(101,41){\makebox(4,4){\tiny{$Z^j$}}}
\put(89,35){\makebox(4,4){\tiny{$X^i$}}}
\put(101,35){\makebox(4,4){\tiny{$X^i$}}}
\put(86,0){\line(0,1){48}}
\put(98,0){\line(0,1){48}}
\put(134,0){\line(0,1){48}}
\put(146,0){\line(0,1){48}}
\put(86,0){\line(1,0){12}}
\put(98,30){\line(1,0){38}}
\put(134,0){\line(1,0){12}}
\put(98,30){\circle*{2.}}
\put(98,36){\circle*{2.}}
\put(98,42){\circle*{2.}}
\put(86,36){\circle*{2.}}
\put(134,30){\circle{4}}
\end{picture}
\label{fig_psi_cnot_up11}
\end{array}
\en
in which the diagrammatic representation (\ref{EPR_diag2}) of the EPR state $|\Psi\rangle$ is used. With both classical communication and unitary correction, therefore,
the extended Temperley--Lieb diagrammatical representation (\ref{Psi_cnot_diag}) of the four-qubit entangled state $|\Psi_{CNOT}\rangle$ can be exactly prepared
in the diagrammatical approach.

Now let us derive the teleportation equation (\ref{tele_cnot2}) in the topological diagrammatical approach. On the diagram~(\ref{fig_psi_cnot_up10}),
we replace the diagrammatical representation (\ref{GHZ 1}) of the GHZ state $|\Upsilon\rangle$   with its another diagrammatical representation (\ref{GHZ 2})
and replace the diagrammatical representation (\ref{GHZ 3}) of the $H_1H_2H_3|\Upsilon\rangle$ state with its another diagrammatical representation (\ref{GHZ 4})
so that we have the other equivalent topological diagrammatical representation,
\eq
\setlength{\unitlength}{0.6mm}
\begin{array}{c}
\begin{picture}(154,50)
\put(30,0){\makebox(4,4){$\nabla$}}
\put(66,0){\makebox(4,4){$\nabla$}}
\put(70,8){\makebox(4,4){\tiny{$H$}}}
\put(36,33){\makebox(6,6){\tiny{$W_{ij}^\dag$}}}
\multiput(6,24)(1,0){65}{\line(1,0){.5}}
\put(8,0){\line(0,1){48}}
\put(20,0){\line(0,1){48}}
\put(32,4){\line(0,1){44}}
\put(44,0){\line(0,1){48}}
\put(56,0){\line(0,1){48}}
\put(68,4){\line(0,1){44}}
\put(8,0){\line(1,0){12}}
\put(20,18){\line(1,0){14}}
\put(32,48){\line(1,0){12}}
\put(44,0){\line(1,0){12}}
\put(54,18){\line(1,0){14}}
\put(68,18){\circle*{2.}}
\put(68,10){\circle*{2.}}
\put(20,18){\circle*{2.}}
\put(44,36){\circle*{2.}}
\put(32,18){\circle{4}}
\put(56,18){\circle{4}}
\put(73,20){\makebox(10,8){$= \frac 1 2$}}
\put(139,35){\makebox(4,4){\tiny{$Z^j$}}}
\put(139,41){\makebox(4,4){\tiny{$X^i$}}}
\put(151,35){\makebox(4,4){\tiny{$Z^j$}}}
\put(88,0){\line(0,1){48}}
\put(100,0){\line(0,1){48}}
\put(136,0){\line(0,1){48}}
\put(148,0){\line(0,1){48}}
\put(88,0){\line(1,0){12}}
\put(100,30){\line(1,0){38}}
\put(136,0){\line(1,0){12}}
\put(136,30){\circle{4}}
\put(100,30){\circle*{2.}}
\put(136,36){\circle*{2.}}
\put(136,42){\circle*{2.}}
\put(148,36){\circle*{2.}}
\end{picture}
\label{fig_psi_cnot_up20}
\end{array}
\en
in which we firstly move the single-qubit gate $W_{ij}^\dag$ from the fourth qubit to the fifth qubit and then across $CNOT_{65}$ gate with the formula
\eq
CNOT_{65}(W_{ij}\otimes 1\!\!1_2)CNOT_{65}=X_5^iZ_5^jZ_6^j,
\en
to generate the single-qubit gate $X_5^i Z_5^j Z_6^j$, and secondly apply the commutative relation of the $CNOT_{25}$ gate and $CNOT_{65}$ and
the diagrammatic representation (\ref{EPR_diag2}) of the EPR state $|\Psi\rangle$. As a result, the diagram~(\ref{fig_psi_cnot_up20}) presents a kind of
proof for the teleportation equation (\ref{tele_cnot2}).

\subsubsection{Topological constructions of four-qubit entangled states $|\Psi_{CNOT}^\uparrow\rangle$ and $|\Psi_{CZ}\rangle$ }

The topological constructions of the four-qubit entangled states $|\Psi_{CNOT}^\uparrow\rangle$ (\ref{Psi cnot down})
and $|\Psi_{CZ}\rangle$ (\ref{Psi_cz}) are  shown up in Appendix~\ref{construction cnot down} and Appendix~\ref{construction Psi CZ} respectively.

\section{The Yang--Baxter gate and its extended Temperley--Lieb diagrammatical representation}

\label{section the YBG and TL}

In this section, we consider two special types of the Yang--Baxter gates derived in Appendix A as well as present their associated extended
Temperley--Lieb diagrammatical configurations, and then apply the special type II Yang--Baxter gates to teleportation-based quantum computation
in Section~\ref{YBG approach to tele based QC}. Note that a brief study on the algebraic properties of the special type I Yang--Baxter gates
is made in  Appendix B. In the first place, the extended Temperley--Lieb diagrammatical representation of a two-qubit quantum gate will be discussed.

\subsection{The extended Temperley-Lieb diagrammatical representation of a two-qubit gate}

The conventional Temperley--Lieb diagram consists of the configuration of a pair of cup and cap \cite{Kauffman02}, see Appendix~\ref{Def TL} for
the diagrammatical representation of the Temperley--Lieb algebra. As we have introduced in Section~\ref{TL diagrammatical for tele based quantum computation}, the extended Temperley--Lieb diagram permits the action of single-qubit gates, namely, it includes the following typical configuration,
\eq
\label{extended TL generator}
\setlength{\unitlength}{0.5mm}
\begin{array}{c}
\begin{picture}(65,30)
\put(62,18){\line(0,1){12}}
\put(52,18){\line(0,1){12}}
\put(52,18){\line(1,0){10}}
\put(62,24){\circle*{2.}}
\put(64,23){\tiny{$M$}}
\put(-3,14){$|\Psi_M\rangle \langle\Psi_N|=$}
\put(52,0){\line(0,1){12}}
\put(62,0){\line(0,1){12}}
\put(52,12){\line(1,0){10}}
\put(62,6){\circle*{2.}}
\put(64,5){\tiny{$N^\dag$}}
\end{picture}
\end{array}
\en
which represents the EPR state projector $|\Psi\rangle\langle\Psi|$ with the local action of single-qubit gates $M$ and $N^\dag$. Obviously, when $M=N=1\!\! 1_2$, such
the configuration (\ref{extended TL generator}) becomes the standard Temperley--Lieb configuration in  Appendix~\ref{Def TL}.

In quantum information and computation \cite{NC2011}, the two-qubit Hilbert space $\mathcal{H}_2\otimes \mathcal{H}_2$ has the two types of the orthonormal bases:
the one is denoted by the product basis $|ij\rangle$, the other is denoted by the Bell basis $|\psi(ij)\rangle$, and the unitary basis transformation matrix $T$
between these two bases satisfies $T|ij\rangle=|\psi(ij)\rangle$ and has the form,
 \eq
 \label{basis_transformation}
T=\frac 1 {\sqrt 2}\left(
    \begin{array}{cccc}
      1 & 1 & 0 & 0 \\
      0 & 0 & 1 & 1 \\
      0 & 0 & 1 & -1 \\
      1 & -1 & 0 & 0 \\
    \end{array}
  \right).
\en
In terms of the product base $|ij\rangle\langle kl|$,  a two-qubit quantum gate $G$ as a $4\times 4$ unitary matrix has an
 expression given by
\eq
G=\sum_{i,j,k,l=0}^1 G_{ij,kl}|ij\rangle\langle kl|,
\en
and in terms of the Bell base $|\psi(ij)\rangle\langle\psi(kl)|$, it has another form
\eq
\label{two-qubit gate TL representation}
\setlength{\unitlength}{0.5mm}
\begin{array}{c}
\begin{picture}(95,30)
\put(6,14){$G=\sum_{i,j,k,l=0}^1 \tilde{G}_{ij,kl}$}
\put(82,18){\line(0,1){12}}
\put(72,18){\line(0,1){12}}
\put(72,18){\line(1,0){10}}
\put(82,24){\circle*{2.}}
\put(84,23){\tiny{$X^iZ^j$}}
\put(72,0){\line(0,1){12}}
\put(82,0){\line(0,1){12}}
\put(72,12){\line(1,0){10}}
\put(82,6){\circle*{2.}}
\put(84,5){\tiny{$Z^lX^k$}}
\end{picture}
\end{array}
\en
 in the extended Temperley--Lieb diagrammatical approach. Applying the basis transformation matrix $T$ (\ref{basis_transformation}),
the coefficient matrix $\tilde{G}=(\tilde{G}_{ij,kl})$ can be obtained from
\eq
\label{trans A}
\tilde{G}=T^\dag G T.
\en

For example, the two-qubit identity gate $1\!\!1_4$ has an expansion of the Bell basis (\ref{Bell states}),
\eq
1\!\!1_4=\sum_{i,j=0}^1|\psi(ij)\rangle\langle\psi(ij)|,
\en
which represents the completeness relation defining the Bell basis, so the   two-qubit identity
gate $1\!\!1_4$ has the extended Temperley--Lieb diagrammatic representation,
\eq
\label{TL_identity}
  \setlength{\unitlength}{0.5mm}
  \begin{array}{c}
  \begin{picture}(170,30)

  \put(40,14){$=$}

  \put(22,0){\line(0,1){30}}
  \put(32,0){\line(0,1){30}}

  \put(62,18){\line(0,1){12}}
  \put(52,18){\line(0,1){12}}
  \put(52,18){\line(1,0){10}}

  \put(52,0){\line(0,1){12}}
  \put(62,0){\line(0,1){12}}
  \put(52,12){\line(1,0){10}}

  \put(70,14){$+$}

  \put(92,18){\line(0,1){12}}
  \put(82,18){\line(0,1){12}}
  \put(82,18){\line(1,0){10}}
  \put(92,24){\circle*{2.}}
  \put(94,23){\tiny{$Z$}}

  \put(82,0){\line(0,1){12}}
  \put(92,0){\line(0,1){12}}
  \put(82,12){\line(1,0){10}}
  \put(92,6){\circle*{2.}}
  \put(94,5){\tiny{$Z$}}

  \put(100,14){$+$}

  \put(122,18){\line(0,1){12}}
  \put(112,18){\line(0,1){12}}
  \put(112,18){\line(1,0){10}}
  \put(122,24){\circle*{2.}}
  \put(124,23){\tiny{$X$}}

  \put(112,0){\line(0,1){12}}
  \put(122,0){\line(0,1){12}}
  \put(112,12){\line(1,0){10}}
  \put(122,6){\circle*{2.}}
  \put(124,5){\tiny{$X$}}

  \put(130,14){$+$}

  \put(152,18){\line(0,1){12}}
  \put(142,18){\line(0,1){12}}
  \put(142,18){\line(1,0){10}}
  \put(152,24){\circle*{2.}}
  \put(154,23){\tiny{$XZ$}}

  \put(142,0){\line(0,1){12}}
  \put(152,0){\line(0,1){12}}
  \put(142,12){\line(1,0){10}}
  \put(152,6){\circle*{2.}}
  \put(154,5){\tiny{$ZX$}}
  \end{picture}
  \end{array},
  \en
where, in the conventional topological viewpoint \cite{Kauffman02},  the  two-qubit identity gate $1\!\!1_4$ is depicted as two parallel vertical lines.

For another example, let us consider the extended Temperley--Lieb diagrammatical configuration of the $CZ$ gate (\ref{CZ gate}), which is
\eq
\label{Extended_TL_CZ}
  \setlength{\unitlength}{0.5mm}
  \begin{array}{c}
  \begin{picture}(150,30)

  \put(7,14){$CZ=$}

  \put(42,18){\line(0,1){12}}
  \put(32,18){\line(0,1){12}}
  \put(32,18){\line(1,0){10}}

  \put(32,0){\line(0,1){12}}
  \put(42,0){\line(0,1){12}}
  \put(32,12){\line(1,0){10}}
  \put(42,6){\circle*{2.}}
  \put(44,5){\tiny{$Z$}}

  \put(50,14){$+$}

  \put(72,18){\line(0,1){12}}
  \put(62,18){\line(0,1){12}}
  \put(62,18){\line(1,0){10}}
  \put(72,24){\circle*{2.}}
  \put(74,23){\tiny{$Z$}}

  \put(62,0){\line(0,1){12}}
  \put(72,0){\line(0,1){12}}
  \put(62,12){\line(1,0){10}}

  \put(80,14){$+$}

  \put(102,18){\line(0,1){12}}
  \put(92,18){\line(0,1){12}}
  \put(92,18){\line(1,0){10}}
  \put(102,24){\circle*{2.}}
  \put(104,23){\tiny{$X$}}

  \put(92,0){\line(0,1){12}}
  \put(102,0){\line(0,1){12}}
  \put(92,12){\line(1,0){10}}
  \put(102,6){\circle*{2.}}
  \put(104,5){\tiny{$X$}}

  \put(110,14){$+$}

  \put(132,18){\line(0,1){12}}
  \put(122,18){\line(0,1){12}}
  \put(122,18){\line(1,0){10}}
  \put(132,24){\circle*{2.}}
  \put(134,23){\tiny{$XZ$}}

  \put(122,0){\line(0,1){12}}
  \put(132,0){\line(0,1){12}}
  \put(122,12){\line(1,0){10}}
  \put(132,6){\circle*{2.}}
  \put(134,5){\tiny{$ZX$}}
  \end{picture}
  \end{array},
  \en
where the first and second configurations from the left handside are explicitly beyond the standard Temperley--Lieb diagrammatical representation in Appendix~\ref{Def TL}.

Note that for a given two-qubit gate $G$, its extended Tempereley--Lieb diagrammatical representation is not fixed and depends on the choices of single-qubit gates acting on
the cup or cap configurations. For example, the extended Temperley--Lieb diagrammatic representation  of the $CZ$ gate (\ref{CZ gate}) can not be  (\ref{Extended_TL_CZ}) when
single-qubit gates on the configurations are not the Pauli gates.

 \subsection{Special type I Yang--Baxter gate and its extended Temperley-Lieb diagrammatical representation}

\label{type_I_YBE_TL}

The special type I Yang--Baxter gate is a typical example for the Yang--Baxter gate via the Temperley--Lieb algebra, and it is derived in Appendix~\ref{type_I_YBE}.
The main reason that we study it in this paper is that it is directly associated with the Bell states (\ref{Bell states}), and we denote the special type I Yang--Baxter gate
by $R(ij)$, which is formulated as
 \eq
 \label{R ij}
 R(i j)=1\!\!1_4-2|\psi(i j)\rangle\langle \psi(i j)|.
 \en
 As an example, for $i=j=1$, we have the Yang--Baxter gate $P \equiv R(11)$, given by
 \eq
 \label{permutation gate}
 P=\left(
         \begin{array}{cccc}
           1 & 0 & 0 & 0 \\
           0 & 0 & 1 & 0 \\
           0 & 1 & 0 & 0 \\
           0 & 0 & 0 & 1 \\
         \end{array}
       \right)
 \en
 which is the Permutation gate (or the Swap gate) $P$ in quantum computation \cite{NC2011}. Note that all of the special type I Yang--Baxter gates $R(ij)$
 are permutation-like quantum gates satisfying $R(ij)R(ij)=1\!\!1_4$, and a further research has been done in Appendix~\ref{teleportation_swapping_operator}.

The special type I  Yang--Baxter gate $R(ij)$ in the extended Temperley--Lieb diagrammatical approach is shown as
  \eq
  \setlength{\unitlength}{0.5mm}
  \begin{array}{c}
  \begin{picture}(120,30)
  \put(0,14){$R(ij)=\sum_{k,l=0}^1(1-2\delta_{i,k}\delta_{j,l})$}
  \put(102,18){\line(0,1){12}}
  \put(92,18){\line(0,1){12}}
  \put(92,18){\line(1,0){10}}
  \put(102,24){\circle*{2.}}
  \put(104,23){\tiny{$X^kZ^l$}}
  \put(92,0){\line(0,1){12}}
  \put(102,0){\line(0,1){12}}
  \put(92,12){\line(1,0){10}}
  \put(102,6){\circle*{2.}}
  \put(104,5){\tiny{$Z^lX^k$}}
  \end{picture}
  \end{array}
  \en
  where the symbol $\delta$ denotes the Kronecker delta function. For example, the Yang--Baxter gate $P=R(11)$  has the extended Temperley--Lieb
  diagrammatical representation,
  \eq
  \setlength{\unitlength}{0.5mm}
  \begin{array}{c}
  \begin{picture}(140,30)
  \put(0,14){$P=$}

  \put(32,18){\line(0,1){12}}
  \put(22,18){\line(0,1){12}}
  \put(22,18){\line(1,0){10}}
  \put(22,0){\line(0,1){12}}
  \put(32,0){\line(0,1){12}}
  \put(22,12){\line(1,0){10}}
  \put(40,14){$+$}
  \put(62,18){\line(0,1){12}}
  \put(52,18){\line(0,1){12}}
  \put(52,18){\line(1,0){10}}
  \put(62,24){\circle*{2.}}
  \put(64,23){\tiny{$X$}}
  \put(52,0){\line(0,1){12}}
  \put(62,0){\line(0,1){12}}
  \put(52,12){\line(1,0){10}}
  \put(62,6){\circle*{2.}}
  \put(64,5){\tiny{$X$}}
  \put(70,14){$+$}
  \put(92,18){\line(0,1){12}}
  \put(82,18){\line(0,1){12}}
  \put(82,18){\line(1,0){10}}
  \put(92,24){\circle*{2.}}
  \put(94,23){\tiny{$Z$}}
  \put(82,0){\line(0,1){12}}
  \put(92,0){\line(0,1){12}}
  \put(82,12){\line(1,0){10}}
  \put(92,6){\circle*{2.}}
  \put(94,5){\tiny{$Z$}}
  \put(100,14){$-$}
  \put(122,18){\line(0,1){12}}
  \put(112,18){\line(0,1){12}}
  \put(112,18){\line(1,0){10}}
  \put(122,24){\circle*{2.}}
  \put(124,23){\tiny{$XZ$}}
  \put(112,0){\line(0,1){12}}
  \put(122,0){\line(0,1){12}}
  \put(112,12){\line(1,0){10}}
  \put(122,6){\circle*{2.}}
  \put(124,5){\tiny{$ZX$}}
  \end{picture}
  \end{array}.
  \en

Note that the special type I Yang--Baxter gate $R(ij)$ will not be applied to our study on teleportation-based quantum computation in
Section~\ref{YBG approach to tele based QC}, see Appendix~\ref{teleportation_swapping_operator} for a relevant interpretation.  We
present them here because they have a simplest realization in the extended Temperley--Lieb diagrammatical approach.

 \subsection{Special type II Yang--Baxter gate and its extended Temperley-Lieb diagrammatical representation}

\label{type_II_YBE TL}

The special type II Yang--Baxter gates $B(\epsilon,\eta)$, or equivalently denoted as $B_{\epsilon,\eta}$,
with $\epsilon=\pm1$ and $\eta=\pm1$, have the form
\eq
 \label{B Bell transform}
 B(\epsilon,\eta)=\frac 1 {\sqrt 2}\left(
                                    \begin{array}{cccc}
                                      1 & 0 & 0 & \eta \\
                                      0 & 1 & \epsilon & 0 \\
                                      0 & -\epsilon & 1 & 0 \\
                                      -\eta & 0 & 0 & 1 \\
                                    \end{array}
                                  \right),
 \en
which are derived in Appendix~\ref{type_II_YBE} as examples for the Yang--Baxter gates generated by the Temperley--Lieb algebra.
These four Yang--Baxter gates $B(\epsilon,\eta)$ are related to one another  by the relations
 \eq
 \label{Four B relation}
 B(\epsilon,\eta)=B^\dag(-\epsilon,-\eta),\quad B(\epsilon,\eta)=P B(-\epsilon,\eta)P,
 \en
 where $P$ is the Permutation gate (\ref{permutation gate}), and the inverse of such the Yang--Baxter gate $B(\epsilon,\eta)$ is
 related to itself with the aid of the Pauli $Z$ gate,
 expressed as
 \eq
  \label{B and B inverse relation}
B(\epsilon,\eta)=Z_1B^\dag(\epsilon,\eta)Z_1=Z_2B^\dag(\epsilon,\eta) Z_2.
 \en

The special type II Yang--Baxter gates $B(\epsilon,\eta)$ are to be applied to our study on the reformulation of the teleportaiton-based quantum
computation in Section~\ref{YBG approach to tele based QC}, because they can generate the Bell states (\ref{Bell states}) from the product basis,
see \cite{Dye03, KL04,ZKG04}. For example, when the parameters $\epsilon=1$ and $\eta=1$, we denote the Yang--Baxter gate $B$\footnote{The Yang--Baxter
gate $B$ in this paper is denoted as the Yang--Baxter gate $B'$ in the authors' another paper \cite{ZZ14}.} as an
abbreviation of the notation $B(1,1)$,
 \eq
 \label{YBG B}
 B=\frac{1}{\sqrt 2}\left(
                      \begin{array}{cccc}
                        1 & 0 & 0 & 1 \\
                        0 & 1 & 1 & 0 \\
                        0 & -1 & 1 & 0 \\
                        -1 & 0 & 0 & 1 \\
                      \end{array}
                    \right)
 \en
 which gives rise to all four Bell states from the product basis in the way
 \eq
  \label{example B Bell transform}
  B\left(
     \begin{array}{c}
       |00\rangle \\
       |01\rangle \\
       |10\rangle \\
       |11\rangle \\
     \end{array}
   \right)=\left(
     \begin{array}{c}
       |\psi(01)\rangle \\
       |\psi(11)\rangle \\
       |\psi(10)\rangle \\
       |\psi(00)\rangle \\
     \end{array}
   \right).
 \en

 We can directly read the extended Temperley--Lieb configuration of the Yang--Baxter gate $B(\epsilon,\eta)$, shown as
   \eq
  \label{TL B Bell transform}
  \setlength{\unitlength}{0.5mm}
  \begin{array}{c}
  \begin{picture}(130,40)

  \put(-31,-2){\footnotesize{$B(\epsilon,\eta)$}}

  \put(-31,29){\line(0,1){6}}
  \put(-31,11){\line(0,-1){6}}
  \put(-13,29){\line(0,1){6}}
  \put(-13,11){\line(0,-1){6}}

  \put(-34,8){\line(0,1){24}}
  \put(-10,8){\line(0,1){24}}
  \put(-34,8){\line(1,0){24}}
  \put(-34,32){\line(1,0){24}}

  \put(-31,29){\line(1,-1){18}}
  \put(-31,11){\line(1,1){7.3}}
  \put(-13,29){\line(-1,-1){7.3}}

  \put(-3,19){$=\frac 1 {\sqrt 2}($}

  \put(32,5){\line(0,1){30}}
  \put(22,5){\line(0,1){30}}

  \put(38,19){$+\eta$}

  \put(62,23){\line(0,1){12}}
  \put(52,23){\line(0,1){12}}
  \put(52,23){\line(1,0){10}}
  \put(62,29){\circle*{2.}}
  \put(64,28){\tiny{$Z$}}

  \put(52,5){\line(0,1){12}}
  \put(62,5){\line(0,1){12}}
  \put(52,17){\line(1,0){10}}

  \put(68,19){$-\eta$}

  \put(92,23){\line(0,1){12}}
  \put(82,23){\line(0,1){12}}
  \put(82,23){\line(1,0){10}}

  \put(82,5){\line(0,1){12}}
  \put(92,5){\line(0,1){12}}
  \put(82,17){\line(1,0){10}}
  \put(92,11){\circle*{2.}}
 \put(94,10){\tiny{$Z$}}

  \put(98,19){$-\epsilon$}

  \put(122,23){\line(0,1){12}}
  \put(112,23){\line(0,1){12}}
  \put(112,23){\line(1,0){10}}
  \put(122,29){\circle*{2.}}
  \put(124,28){\tiny{$X$}}

  \put(112,5){\line(0,1){12}}
  \put(122,5){\line(0,1){12}}
  \put(112,17){\line(1,0){10}}
  \put(122,11){\circle*{2.}}
  \put(124,10){\tiny{$ZX$}}

  \put(128,19){$+\epsilon$}

  \put(152,23){\line(0,1){12}}
  \put(142,23){\line(0,1){12}}
  \put(142,23){\line(1,0){10}}
  \put(152,29){\circle*{2.}}
  \put(154,28){\tiny{$XZ$}}

  \put(142,5){\line(0,1){12}}
  \put(152,5){\line(0,1){12}}
  \put(142,17){\line(1,0){10}}
  \put(152,11){\circle*{2.}}
  \put(154,10){\tiny{$X$}}

  \put(164,19){$)$}

  \end{picture}
  \end{array}
  \en
where the over-crossing diagram represents the braiding feature \cite{Kauffman02} of the Yang--Baxter gate $B(\epsilon,\eta)$ (\ref{B Bell transform}) and the box
around it marks the feature that it can be regarded as a two-qubit quantum gate, and two parallel vertical lines represent the two-qubit identity gate (\ref{TL_identity}).
To maintain the diagrammatical representations consistent, the braiding configuration (the over-crossing configuration) has the same acting direction as the other
extended Temperley--Lieb diagrammatic representations, namely, it is read from the bottom to the top.

\section{The Yang--Baxter gate approach to teleportation-based quantum computation}

\label{YBG approach to tele based QC}

 In this section, we apply the special type II Yang--Baxter gates $B(\epsilon,\eta)$ (\ref{B Bell transform}), derived in Appendix A
 and presented in Section~\ref{section the YBG and TL} with their associated extended Temperley--Lieb diagrammatical configurations (\ref{TL B Bell transform}),
 to teleportation-based quantum computation. First of all, such the special type II Yang--Baxter gates $B(\epsilon,\eta)$  are found to be a type of examples
 for the Bell transform defined in \cite{ZZ14}, where teleportation-based quantum computation using the Bell transform has been explored in detail. Note that
 the Yang--Baxter gate approach to teleportation-based quantum computation admits an interpretation in the extended Temperley--Lieb diagrammatical approach,
 see Section~\ref{relationship_YBE_TL}.

\subsection{The Yang--Baxter gate $B(\epsilon,\eta)$ as the Bell transform and Clifford gate}

We recognize the special type II Yang--Baxter gate $B(\epsilon,\eta)$ (\ref{B Bell transform}) as the Bell transform \cite{ZZ14}, and especially
verify them as Clifford gates  \cite{NC2011, Gottesman97} in three equivalent approaches.

A type of the Bell transform \cite{ZZ14} is defined as a two-qubit quantum gate capable of generating the four Bell states (\ref{Bell states})
from the product states modulo global phase factors. For examples, the special type II Yang--Baxter gates $B(\epsilon,\eta)$,
denoted by $B_{\epsilon,\eta}$ in this section, are the Bell transform because they produce the Bell states in the way,
\eqa
&&B_{-1,1}|ij\rangle = (-1)^{i\cdot(j+1)}|\psi(i+j,j+1)\rangle,\\
&&B_{1,-1}|ij\rangle = (-1)^{i\cdot j}|\psi(i+j,j)\rangle,\\
\label{B Bell transform index}
&&B_{1,1}|ij\rangle = |\psi(i+j,i+1)\rangle,\\
\label{B inverse Bell transform index}
&&B_{-1,-1}|ij\rangle = (-1)^i|\psi(i+j,i)\rangle,
\ena
in which the addition is the binary addition and the multiplication is the logical AND operation. Obviously, the global phase factors of such the Bell states
generated by the $B_{\epsilon,\eta}$ gates are $\pm1$, so the special type II Yang--Baxter gates $B_{\epsilon,\eta}$ are Clifford gates \cite{NC2011, Gottesman97},
because the global phase factors associated with Clifford gates are allowed to be only $\pm 1$, $\pm i$ \cite{ZZ14}.

According to discussed in Subsubsection~\ref{clifford_gate_fault_tolerant}, the fact that the special type II Yang--Baxter gates $B_{\epsilon,\eta}$ are Clifford gates
is  crucial in the fault-tolerant construction of universal quantum gate set using quantum teleportation in the following subsections.
We will verify the $B_{\epsilon,\eta}$ gates as Clifford gates in another two ways. A Clifford gate can be expressed as a tensor product of elementary Clifford gates,
namely the Hadamard gate $H$ (\ref{Hadamard gate}), the phase gate $S$ (\ref{phase gate}) and the $CNOT$ gate (\ref{CNOT gate}), so we reformulate the $B_{-1,1}$ gate as
\eq
B_{-1,1}=CNOT_{21}(1\!\!1_2\otimes ZH) CNOT_{21},
\en
with $Z=S^2$. With the algebraic relations  (\ref{Four B relation}) among the $B_{\epsilon,\eta}$ gates, for examples,  the $B_{1,-1}$ gate is the inverse of the $B_{-1,1}$ gate and the $B_{1,1}$ gate is the conjugation of the $B_{-1,1}$ gate under the Permutation gate, we can decompose all the $B_{\epsilon,\eta}$ gates as
tensor products of  elementary Clifford gates.
\begin{table}
\begin{center}
\footnotesize
\begin{tabular}{c|c|c}
\hline\hline
Operation & Input & Output \\ \hline
  & $X_1$ & $X_1$ \\
  & $X_2$ & $X_1Z_2$ \\
  \raisebox{1.8ex}[0pt]{$B_{-1,1}$} & $Z_1$ & $-Y_1Y_2$ \\
  & $Z_2$ & $-X_1X_2$ \\  \hline
  & $X_1$ & $X_1$ \\
  & $X_2$ & $-X_1Z_2$ \\
  \raisebox{1.8ex}[0pt]{$B_{1,-1}$} & $Z_1$ & $Y_1Y_2$ \\
  & $Z_2$ & $X_1X_2$ \\  \hline
  & $X_1$ & $Z_1X_2$ \\
  & $X_2$ & $X_2$ \\
  \raisebox{1.8ex}[0pt]{$B_{1,1}$} & $Z_1$ & $-X_1X_2$ \\
  & $Z_2$ & $-Y_1Y_2$ \\ \hline
   & $X_1$ & $-Z_1X_2$ \\
  & $X_2$ & $X_2$ \\
  \raisebox{1.8ex}[0pt]{$B_{-1,-1}$} & $Z_1$ & $X_1X_2$ \\
  & $Z_2$ & $Y_1Y_2$ \\ \hline\hline
\end{tabular}
\caption{\label{clifford_YBG}  Transformation properties of elements of the Pauli group ${\mathcal P}_2$ under conjugation
 by the Yang--Baxter gates $B_{\epsilon,\eta}$  (\ref{B Bell transform}) where the variables $\epsilon$ and $\eta$ are
 relabeled as subscripts.   }
\end{center}
\end{table}
Besides such the decompositions, the Clifford gates $B_{\epsilon,\eta}$ are verified to preserve
the properties of Pauli group $\mathcal P_2$ generated by $X_1, X_2$ and $Z_1, Z_2$ under conjugation
 by the Yang--Baxter gates $B_{\epsilon,\eta}$  (\ref{B Bell transform}), see Table~\ref{clifford_YBG}.
 Note that the results in Table~\ref{clifford_YBG} will be exploited in the study of
the Yang--Baxter gate approach to teleportation-based quantum computation.

\subsection{Quantum teleportation circuit using the Yang--Baxter gate}

\begin{table}
\begin{center}
\footnotesize
\begin{tabular}{c|c|c|c|c|c}
\hline\hline
$\epsilon$ & $\eta$ & $p_{\epsilon,\eta}$ & $p'_{\epsilon,\eta}$ & $a_{\epsilon,\eta}$ & $b_{\epsilon,\eta}$\\ \hline
  $-1$ & $1$ & $j\cdot(j+k+l)$ & $i\cdot j+(k+l)\cdot (j+l+1)$ & & \\ \cline{1-4}
  $1$ & $-1$ & $k\cdot l+(k+l)\cdot(j+l+1)$ & $k\cdot l+(j+1)\cdot(i+k+l)$ & \raisebox{1.5ex}[0pt]{$j+l+1$} & \raisebox{1.6ex}[0pt]{$i+j+k+l$} \\ \hline
  $1$ & $1$ & $i\cdot j+(k+l)\cdot(i+k+1)$ & $i\cdot (i+k+l)$ & & \\ \cline{1-4}
  $-1$ & $-1$ & $k\cdot l+(i+1)\cdot(j+k+l)$ & $l+i\cdot(k+l)$ & \raisebox{1.5ex}[0pt]{$i+k+1$} & \raisebox{1.6ex}[0pt]{$i+j+k+l$} \\ \hline\hline
\end{tabular}
\caption{\label{index p a b}  The indices $p_{\epsilon,\eta}$, $p'_{\epsilon,\eta}$, $a_{\epsilon,\eta}$ and $b_{\epsilon,\eta}$ for both the local unitary gate
 $W_{\epsilon,\eta}$ (\ref{W epsilon}) in the teleportation equation (\ref{tele eqa B1}) and the local unitary gate $W'_{\epsilon,\eta}$ (\ref{W epsilon prime})
in the teleportation equation (\ref{tele eqa B2}). The multiplication $\cdot$ is a logical AND operation and the addition $+$ is a binary addition. }
\end{center}
\end{table}

The essential ingredient in quantum teleportation circuit model is  the teleportation operator introduced in \cite{ZZ14} which is the tensor product of
the identity operator, the Bell transform and its inverse. Here, the special type II Yang--Baxter gates $B_{\epsilon,\eta}$ are the Bell transform,
so we make use of a similar type of the teleportation operators
\eq
\label{tele operator}
(B_{\epsilon,\eta}\otimes 1\!\!1_2)(1\!\!1_2\otimes B_{\epsilon,\eta})\quad \text{or} \quad (1\!\!1_2\otimes B_{\epsilon,\eta})(B_{\epsilon,\eta}\otimes 1\!\!1_2)
\en
to act on the product state $|\alpha\rangle|kl\rangle$ or $|kl\rangle|\alpha\rangle$ respectively. They give rise to the type of the  teleportation equation
\eq
\label{tele eqa B1}
(B_{\epsilon,\eta}\otimes 1\!\!1_2)(1\!\!1_2\otimes B_{\epsilon,\eta})(|\alpha\rangle\otimes |kl\rangle)=\frac{1}{2}\sum_{i,j=0}^1|ij\rangle W_{\epsilon,\eta}|\alpha\rangle,
\en
which is associated with the teleportation equation (\ref{tele_eq}) with the local single-qubit gate
\eq
\label{W epsilon}
W_{\epsilon,\eta}=(-1)^{p_{\epsilon,\eta}}Z^{a_{\epsilon,\eta}}X^{b_{\epsilon,\eta}},
\en
to represent the protocol of transmitting an unknown qubit $|\alpha\rangle$ from Alice to Bob,
and the other type of the teleportation equation
\eq
\label{tele eqa B2}
(1\!\!1_2 \otimes B_{\epsilon,\eta})(B_{\epsilon,\eta}\otimes 1\!\!1_2)(|kl\rangle\otimes|\alpha\rangle )=\frac{1}{2}\sum_{i,j=0}^1
   W'_{\epsilon,\eta}|\alpha\rangle|ij\rangle,
\en
which is related to the teleportation equation (\ref{tele_eq_transpose}) with the local single-qubit gate
\eq
\label{W epsilon prime}
W'_{\epsilon,\eta}=(-1)^{p'_{\epsilon,\eta}}Z^{a_{\epsilon,\eta}}X^{b_{\epsilon,\eta}},
\en
to represent  the protocol of transmitting an unknown qubit $|\alpha\rangle$ from Bob to Alice, where the explicit expressions
for the indices $p_{\epsilon, \eta}$, $p'_{\epsilon, \eta}$, $a_{\epsilon,\eta}$ and $b_{\epsilon,\eta}$ in both single-qubit gates $W_{\epsilon,\eta}$ (\ref{W epsilon}) and $W'_{\epsilon,\eta}$
(\ref{W epsilon prime}) are shown up in Table~\ref{index p a b}. Note that the local single-qubit gates $W_{\epsilon,\eta}$ (\ref{W epsilon}) and $W'_{\epsilon,\eta}$
(\ref{W epsilon prime}) are almost the same except the global phase factors.

\begin{figure}
 \begin{center}
  \includegraphics[width=8cm]{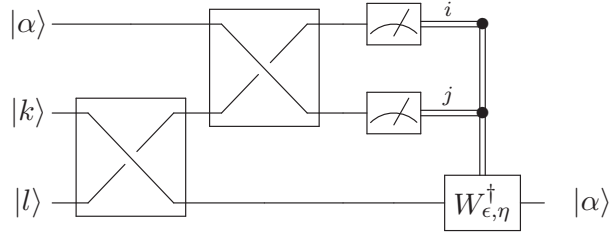}
  \end{center}
  \caption{\label{fig_tele_YBG} Quantum circuit of teleportation corresponding to the teleportation equation (\ref{tele eqa B1}).
  The Yang--Baxter gate $B(\epsilon,\eta)$ (\ref{B Bell transform}) is represented by a braiding configuration inside a two-qubit gate box.
  In accordance with the conventional rules of
  the quantum circuit diagram in Figure~\ref{fig_tele_circuit}, the braiding configuration (\ref{TL B Bell transform}) of the Yang--Baxter gate $B(\epsilon,\eta)$
  has been adjusted in the left-to-right direction from the original down-to-up direction. The left lower braiding on the product state $|kl\rangle$ is used to generate
  the prepared Bell states, and the right higher braiding followed with single-qubit state measurements works as the Bell measurements. The present quantum circuit model is essentially equivalent with the one in Figure~\ref{fig_tele_circuit}, while the superficial differences between them are due to the fact that we are using various
  of presentations  of both Bell states and Bell measurements. }
\end{figure}

Based on the above teleportation equations (\ref{tele eqa B1}) or (\ref{tele eqa B2}) using the special type II Yang--Baxter gates $B_{\epsilon,\eta}$, we continue to
perform single-qubit measurements, classical communication and local unitary corrections to complete the entire quantum teleportation protocol. For example, the
quantum teleportation circuit model associated with the teleportation equation (\ref{tele eqa B1}) is drawn in Figure~\ref{fig_tele_YBG}, where the braiding
configuration (\ref{TL B Bell transform}) of the Yang--Baxter gate $B_{\epsilon,\eta}$ has been applied. About the teleportation circuit model
in Figure~\ref{fig_tele_YBG},  the special type II Yang--Baxter gate $B(\epsilon,\eta)$ acting on the product state $|kl\rangle$ gives the prepared Bell states,
and the same Yang--Baxter gate $B(\epsilon,\eta)$ followed by two single-qubit measurements works as the Bell measurements.

 Note that the inverse of the Yang--Baxter gate $B(\epsilon,\eta)$, denoted by $B^\dag(\epsilon,\eta)$, is related to itself with the algebraic relation  (\ref{B and B inverse relation}). In other words, the inverse of the Yang--Baxter gate $B(\epsilon,\eta)$ is still the Bell transform \cite{ZZ14}.  Hence in terms of the inverse  of the Yang--Baxter
 gate $B(\epsilon,\eta)$, the teleportation operators (\ref{tele operator}) can be  reformulated  as the same type of the teleportation operators used in \cite{ZZ14} where the
 inverse of the Bell transform \cite{ZZ14} with single-qubit measurements represents the Bell measurements.

\begin{figure}
 \begin{center}
  \includegraphics[width=8cm]{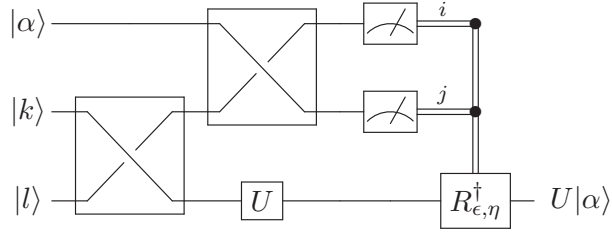}
  \end{center}
  \caption{\label{fig_single_tele_YGB} Quantum circuit for the fault-tolerant construction of a single-qubit gate $U$ in the Yang--Baxter gate approach to teleportation-based quantum computation, associated with the teleportation equation (\ref{tele one qubit gate B}), and it is equivalent with the quantum
   circuit in Figure~\ref{fig_tele_single_gate}.  }
\end{figure}

\subsection{Teleportation-based quantum computation using the Yang--Baxter gate}

 Under the guidance of the fault-tolerant construction of single-qubit gates in Subsubsection~\ref{subsection: single-qubit}, we prepare the three-qubit quantum state in the
 the Yang--Baxter gate approach, expressed as
 \eq
 (1\!\!1_2\otimes 1\!\!1_2\otimes U)(1\!\!1_2\otimes B_{\epsilon,\eta})(|\alpha\rangle\otimes|kl\rangle)
 \en
with the single-qubit gate $U$ acting on the Bell states, which are generated by the special type II Yang--Baxter gate $B(\epsilon,\eta)$ on the product basis $|kl\rangle$,
and then apply the Yang--Baxter gate $B(\epsilon,\eta)$ on the first and second qubits to derive the teleportation equation
\eq
\label{tele one qubit gate B}
(B_{\epsilon,\eta}\otimes 1\!\!1_2)(1\!\!1_2\otimes 1\!\!1_2\otimes U)(1\!\!1_2 \otimes B_{\epsilon,\eta})(|\alpha\rangle\otimes |kl\rangle)=\frac{1}{2}\sum_{i,j=0}^1|ij\rangle R_{\epsilon,\eta}U|\alpha\rangle,
\en
which is associated with the teleportation equation (\ref{tele_single}), with $R_{\epsilon,\eta}=UW_{\epsilon,\eta}U^\dag$, the single-qubit gate $W_{\epsilon,\eta}$ defined in
(\ref{W epsilon}).

For example, when the single-qubit gate $U$ is the Hadamard gate $H$ (\ref{Hadamard gate}) and $\pi/8$ gate $T$ (\ref{pi/8}), the corresponding single-qubit gates $R_{\epsilon,\eta}$ have the form respectively,
\eq
\label{local correction R for H and T}
R_{\epsilon,\eta}(H)=(-1)^{p_{\epsilon,\eta}}X^{a_{\epsilon,\eta}}Z^{b_{\epsilon,\eta}},\quad R_{\epsilon,\eta}(T)=(-1)^{p_{\epsilon,\eta}}Z^{a_{\epsilon,\eta}}\left(\frac{X-\sqrt{-1}Y}{\sqrt{2}}\right)^{b_{\epsilon,\eta}},
\en
where the indices $p_{\epsilon,\eta}$, $a_{\epsilon,\eta}$ and $b_{\epsilon,\eta}$ are those in Table~\ref{index p a b}. We can compare these results with their counterparts
obtained in Subsubsection~\ref{subsection: single-qubit}, such as (\ref{RTij}). Note that the $R_{\epsilon,\eta}(H)$ gates are the Pauli gates with the global phase factors,
whereas the $R_{\epsilon,\eta}(T)$ gates are the Clifford gates.

Combining the teleportation equation (\ref{tele one qubit gate B}) with both two single-qubit measurements and the local unitary correction operator $R^{\dag}_{\epsilon,\eta}$,
the quantum circuit model for the fault-tolerant construction of the single-qubit gate $U$ in the Yang--Baxter gate approach is presented in Figure~\ref{fig_single_tele_YGB},
which can be compared with the quantum circuit model in Figure~\ref{fig_tele_single_gate}.

\begin{figure}
 \begin{center}
  \includegraphics[width=8cm]{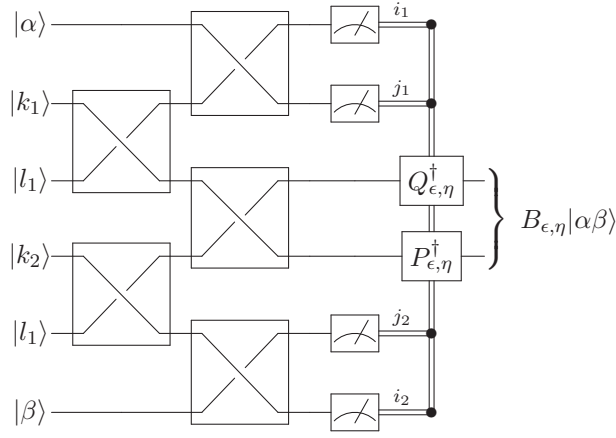}
  \end{center}
  \caption{\label{fig_two_tele_YGB}
  Quantum circuit for the fault-tolerant construction of the two-qubit gate $B_{\epsilon,\eta}$ (\ref{B Bell transform}) in the Yang--Baxter gate approach to
  teleportation-based quantum computation, associated with the teleportation equation (\ref{tele two qubit gate B}), and it is equivalent with
  the quantum circuit in Figure~\ref{fig_tele_two_gate}. Explicit expressions for single-qubit gates $Q_{\epsilon,\eta}^\dag$ and $P_{\epsilon,\eta}^\dag$
  are shown up in Table~\ref{Q and P}.}
\end{figure}

In accordance with the fault-tolerant construction of a two-qubit Clifford gate in Subsubsection \ref{subsection: two-qubit}, we perform the fault-tolerant
construction of the special type II Yang--Baxter gate $B(\epsilon,\eta)$ (\ref{B Bell transform}) in the following steps. Firstly, we prepare the four-qubit
entangled state in the Yang--Baxter gate approach as
\eq
(1\!\!1_2\otimes B_{\epsilon,\eta}\otimes 1\!\!1_2)(B_{\epsilon,\eta}\otimes B_{\epsilon,\eta})(|k_1l_1\rangle\otimes |k_2l_2\rangle).
\en
Secondly, we perform two Bell measurements respectively via two Yang--Baxter gates $B(\epsilon,\eta)$ followed by product-basis measurements,
so that we have the teleportation equation
\eqa
\label{tele two qubit gate B}
&&(B_{\epsilon,\eta}\otimes B_{\epsilon,\eta}\otimes B_{\epsilon,\eta})(1\!\!1_2\otimes B_{\epsilon,\eta}\otimes B_{\epsilon,\eta} \otimes1\!\!1_2)(|\alpha\rangle\otimes |k_1l_1\rangle\otimes |k_2l_2\rangle\otimes |\beta\rangle) \nonumber\\
&=& \frac{1}{4}\sum_{i_1,j_1=0}^1\sum_{i_2,j_2=0}^1(1\!\!1_4\otimes Q_{\epsilon,\eta}\otimes P_{\epsilon,\eta}\otimes 1\!\!1_4)(|i_1j_1\rangle\otimes B_{\epsilon,\eta}|\alpha\beta\rangle\otimes|i_2j_2\rangle),
\ena
which is related to the teleportation equation (\ref{tele_two}) and has the  quantum circuit in Figure~\ref{fig_two_tele_YGB}. The single-qubit gates $Q_{\epsilon,\eta}$ and $P_{\epsilon,\eta}$ are calculated by the formula
\eq
Q_{\epsilon,\eta}\otimes P_{\epsilon,\eta}=B_{\epsilon,\eta}(W_{\epsilon,\eta}\otimes W'_{\epsilon,\eta})B_{-\epsilon,-\eta},
\en
in which  $W_{\epsilon,\eta}$ is defined in (\ref{tele eqa B1}) and depends on the indices $i_1$, $j_1$, and $W'_{\epsilon,\eta}$ is
defined in (\ref{tele eqa B2}) and depends on the indices $i_2$, $j_2$, and the explicit expressions of $Q_{\epsilon,\eta}$ and
$P_{\epsilon,\eta}$ are listed in Table~\ref{Q and P}.

\begin{table}
\begin{center}
\footnotesize
\begin{tabular}{c|c|c|c}
\hline\hline
$\epsilon$ & $\eta$ & $Q_{\epsilon,\eta}$ & $P_{\epsilon,\eta}$ \\ \hline
  $-1$ & $1$ & $(-1)^{p_1+a_1}Y^{a_1}X^{b_1+b_2+a_2}$ & $(-1)^{p'_2+a_2}Y^{a_1}X^{a_2}Z^{b_2}$ \\ \hline
  $1$ & $-1$ & $(-1)^{p_1}Y^{a_1}X^{b_1+b_2+a_2}$ & $(-1)^{p'_2+b_2}Y^{a_1}X^{a_2}Z^{b_2}$ \\ \hline
  $1$ & $1$ & $(-1)^{p_1+a_1}X^{a_1}Z^{b_1}Y^{a_2}$ & $(-1)^{p'_2+a_2}X^{a_1+b_1}Y^{a_2}X^{b_2}$ \\ \hline
  $-1$ & $-1$ & $(-1)^{p_1+b_1}X^{a_1}Z^{b_1}Y^{a_2}$ & $(-1)^{p'_2}X^{a_1+b_1}Y^{a_2}X^{b_2}$ \\ \hline\hline
\end{tabular}
\caption{\label{Q and P} The $Q_{\epsilon,\eta}$ and $P_{\epsilon,\eta}$ gates in the teleportation equation (\ref{tele two qubit gate B}).
 For simplicity, the subscripts $\epsilon$ and $\eta$ of $p$, $p'$, $a$ and $b$, are omitted in the present table without confusion.
The parameters  $p_1$, $p_1'$, $a_1$ and $b_1$ depend on the parameters $i_1$, $j_1$, $k_1$ and $l_1$, and are calculated by the formulas
in Table~\ref{index p a b}, while the parameters  $p_2$, $p_2'$, $a_2$ and $b_2$ depend on the parameters $i_2$, $j_2$, $k_2$ and $l_2$,
and are calculated by the formulas in Table~\ref{index p a b}. }
\end{center}
\end{table}

\subsection{Special example: $\epsilon=1,\eta=1$ and $|kl\rangle=|11\rangle$}

For readers' convenience, we make a brief summary on the results  in  the Yang--Baxter gate approach to the teleportation-based quantum computation
using the Yang--Baxter gate $B$ (\ref{YBG B}), which is the $B_{\epsilon,\eta}$ gate (\ref{B Bell transform}) of $\epsilon=1,\eta=1$.
To further simplify the calculation results without losing any significant physical meaning, we choose the product state $|kl\rangle=|11\rangle$ in
the state preparation.  Note that the Yang--Baxter gate $B$ (\ref{YBG B}) acting on the state $|11\rangle$ gives rise to the EPR state (\ref{EPR}).

First, the teleportation equation associated with the teleportation operator $(B\otimes 1\!\!1_2)(1\!\!1_2 \otimes B)$ on the product state
$|\alpha\rangle\otimes |11\rangle$ has the form
\eq
\label{example tele eqa B1}
(B\otimes 1\!\!1_2)(1\!\!1_2 \otimes B)(|\alpha\rangle\otimes |11\rangle) = \frac 1 2 \sum_{i,j=0}^1 |ij\rangle W_B|\alpha\rangle,
\en
with the single-qubit gate $W_B$ given by
\eq
\label{W_B}
W_B=(-1)^{i\cdot j}Z^iX^{i+j},
\en
which is a special case of the teleportation equation (\ref{tele eqa B1}). Similarly,  the teleportation equation, as a special example
of (\ref{tele eqa B2}), is obtained to be
\eq
\label{example tele eqa B1_transpose}
(1\!\!1_2 \otimes B)(B\otimes 1\!\!1_2)(|11\rangle\otimes|\alpha\rangle ) = \frac{1}{2}\sum_{i,j=0}^1 W^T_B|\alpha\rangle|ij\rangle,
\en
where $W^T_B$ is the matrix transpose of $W_B$ (\ref{W_B}).

Second, the teleportation equation associated with the fault-tolerant construction of the single-qubit gate $U$ has the form
\eq
\label{example single qubit gate via B}
(B\otimes 1\!\!1_2)(1\!\!1_2\otimes 1\!\!1_2\otimes U)(1\!\!1_2 \otimes B)(|\alpha\rangle\otimes |11\rangle)
 =\frac{1}{2}\sum_{i,j=0}^1|ij\rangle R_B\, U|\alpha\rangle,
\en
with $R_B=UW_BU^\dag$, the $W_B$ gate defined in (\ref{W_B}), which is a special example of the teleportation equation (\ref{tele one qubit gate B}).
As the single-qubit gate $U$ is the Hadamard gate $H$ (\ref{Hadamard gate}) and the $\pi/8$ gate $T$ (\ref{pi/8}) respectively, the local
unitary gate $R_B$ has the explicit forms
\eq
R_B(H)=(-1)^{i\cdot j}X^iZ^{i+j},\quad R_B(T)=(-1)^{i\cdot j}Z^i\left(\frac{X-\sqrt{-1}Y}{\sqrt{2}}\right)^{i+j}.
\en

Third, the teleportation equation associated with the fault-tolerant construction of the Yang--Baxter gate $B$, as a special case of (\ref{tele two qubit gate B}),
has the form
\eqa
\label{example two qubit gate via B}
&&(B\otimes B\otimes B)(1\!\!1_2\otimes B\otimes B \otimes1\!\!1_2)(|\alpha\rangle\otimes |11\rangle\otimes |11\rangle\otimes |\beta\rangle) \nonumber\\
&=& \frac{1}{4}\sum_{i_1,j_1=0}^1\sum_{i_2,j_2=0}^1(1\!\!1_4\otimes Q_B\otimes P_B\otimes 1\!\!1_4)(|i_1j_1\rangle\otimes B|\alpha\beta\rangle\otimes|i_2j_2\rangle),
\ena
with the single-qubit gates $Q_B$ and $P_B$ calculated by
\eq
\label{Q P for B}
Q_B\otimes P_B=B(W_B\otimes W_B^T)B^\dag,
\en
where $W_B$ depends on the parameters $i_1$, $j_1$ and  $W_B^T$ depends on the parameters $i_2$, $j_2$. The explicit formalisms of $Q_B$ and $P_B$ are
expressed as
\eq
Q_B=(-1)^{i_1\cdot(j_1+1)}X^{i_1}Z^{i_1+j_1}Y^{i_2},\quad P_B=X^{j_1}Y^{i_2}X^{i_2+j_2}
\en
which can be derived with Table~\ref{Q and P} by picking up the third row of $\epsilon=1,\eta=1$, and setting $k_1=l_1=k_2=l_2=1$.

Note that the key reason that the results in this subsection are presented is that they are to be explained in the extended Temperley--Lieb
diagrammatical approach in Section~\ref{relationship_YBE_TL}.

\section{Relationship between the extended Temperley--Lieb  diagrammatical approach and the Yang--Baxter gate approach}

\label{relationship_YBE_TL}

So far, we study the two approaches to teleportation-based quantum computation: the one is the extended Temperley--Lieb diagrammatical approach
in Section~\ref{TL diagrammatical for tele based quantum computation}, and the other is the Yang--Baxter gate approach
in Section~\ref{YBG approach to tele based QC}. In this section, we are in a position to consider the relationship between these two approaches.
With the extended Temperley--Lieb  configuration (\ref{TL B Bell transform}) of the special type II Yang--Baxter
gate $B(\epsilon,\eta)$ (\ref{B Bell transform}) in Section~\ref{section the YBG and TL}, we are able to recast all the algebraic results
of the Yang--Baxter gate approach in Section~\ref{YBG approach to tele based QC} into the topological diagrammatical configurations, which are
found to be those in the extended Temperley--Lieb diagrammatical approach in Section~\ref{TL diagrammatical for tele based quantum computation}.
In the following discussion,  we concentrate on the Yang--Baxter gate $B$ (\ref{YBG B}) as an example for the special type II Yang--Baxter
gate $B(\epsilon,\eta)$ with $\epsilon=\eta=1$.

\subsection{The product basis and the Bell basis}

 The Yang--Baxter gate approach to teleportation-based quantum computation in Section~\ref{YBG approach to tele based QC} is based on
 the observation that the Bell states (\ref{Bell states})  can be replaced by the Yang--Baxter gate acting on the product states and
 the Bell measurements can be substituted by the Yang--Baxter gate followed with product-basis measurements. Hence, the first thing
 that we are going to do  is to describe the product basis and the Bell basis in the extended Temperley--Lieb diagrammatical approach.

 With the definition of the Bell basis (\ref{Bell states}), it is easy to formulate the two-qubit product basis in terms of the Bell basis,
 so the two-qubit product states have the extended Temperley--Lieb diagrammatical configurations respectively expressed as
 \eq
\label{product state in TL configuration 1}
\setlength{\unitlength}{0.5mm}
\begin{array}{c}
\begin{picture}(70,12)
  \put(-10,2){\line(0,1){10}}
  \put(0,2){\line(0,1){10}}
  \put(-2.1,0){\tiny{$\nabla$}}
  \put(-12.1,0){\tiny{$\nabla$}}
\put(8,4){$=\frac 1 {\sqrt2}($}
\put(42,0){\line(0,1){12}}
\put(32,0){\line(0,1){12}}
\put(32,0){\line(1,0){10}}
\put(48,4){$+$}
\put(69,0){\line(0,1){12}}
\put(59,0){\line(0,1){12}}
\put(59,0){\line(1,0){10}}
\put(69,6){\circle*{2.}}
\put(71,5){\tiny{$Z$}}
\put(75,4){$);$}

\end{picture}
\end{array}
\en
\eq
\label{product state in TL configuration 2}
\setlength{\unitlength}{0.5mm}
\begin{array}{c}
\begin{picture}(70,12)
  \put(-10,2){\line(0,1){10}}
  \put(0,2){\line(0,1){10}}
  \put(-2.1,0){\tiny{$\nabla$}}
  \put(-12.1,0){\tiny{$\nabla$}}

  \put(0,7){\circle*{2.}}
  \put(2,6){\tiny{$X$}}

\put(8,4){$=\frac 1 {\sqrt2}($}
\put(42,0){\line(0,1){12}}
\put(32,0){\line(0,1){12}}
\put(32,0){\line(1,0){10}}
\put(42,6){\circle*{2.}}
\put(44,5){\tiny{$X$}}
\put(48,4){$+$}
\put(69,0){\line(0,1){12}}
\put(59,0){\line(0,1){12}}
\put(59,0){\line(1,0){10}}
\put(69,6){\circle*{2.}}
\put(71,5){\tiny{$XZ$}}
\put(79,4){$);$}
\end{picture}
\end{array}
\en
\eq
\label{product state in TL configuration 3}
\setlength{\unitlength}{0.5mm}
\begin{array}{c}
\begin{picture}(70,12)
  \put(-10,2){\line(0,1){10}}
  \put(0,2){\line(0,1){10}}
  \put(-2.1,0){\tiny{$\nabla$}}
  \put(-12.1,0){\tiny{$\nabla$}}

  \put(-10,7){\circle*{2.}}
  \put(-8,6){\tiny{$X$}}

\put(8,4){$=\frac 1 {\sqrt2}($}
\put(42,0){\line(0,1){12}}
\put(32,0){\line(0,1){12}}
\put(32,0){\line(1,0){10}}
\put(42,6){\circle*{2.}}
\put(44,5){\tiny{$X$}}
\put(48,4){$-$}
\put(69,0){\line(0,1){12}}
\put(59,0){\line(0,1){12}}
\put(59,0){\line(1,0){10}}
\put(69,6){\circle*{2.}}
\put(71,5){\tiny{$XZ$}}
\put(79,4){$);$}

\end{picture}
\end{array}
\en
\eq
\label{product state in TL configuration 4}
\setlength{\unitlength}{0.5mm}
\begin{array}{c}
\begin{picture}(70,12)
  \put(-10,2){\line(0,1){10}}
  \put(0,2){\line(0,1){10}}
  \put(-2.1,0){\tiny{$\nabla$}}
  \put(-12.1,0){\tiny{$\nabla$}}

  \put(-10,7){\circle*{2.}}
  \put(-8,6){\tiny{$X$}}
  \put(0,7){\circle*{2.}}
  \put(2,6){\tiny{$X$}}

\put(8,4){$=\frac 1 {\sqrt2}($}
\put(42,0){\line(0,1){12}}
\put(32,0){\line(0,1){12}}
\put(32,0){\line(1,0){10}}
\put(48,4){$-$}
\put(69,0){\line(0,1){12}}
\put(59,0){\line(0,1){12}}
\put(59,0){\line(1,0){10}}
\put(69,6){\circle*{2.}}
\put(71,5){\tiny{$Z$}}
\put(75,4){$);$}

\end{picture}
\end{array}
\en
where the vertical line with $\nabla$ stands for the state $|0\rangle$  and naturally the one with the Pauli $X$ gate stands for the state $|1\rangle$.
And the adjoint of the above algebraic relations (\ref{product state in TL configuration 1})-(\ref{product state in TL configuration 4}) have the
extended Temperley--Lieb diagrammatical configurations  respectively as
\eq
\label{conjugation product state in TL configuration 1}
\setlength{\unitlength}{0.5mm}
\begin{array}{c}
\begin{picture}(70,12)
  \put(-10,0){\line(0,1){10}}
  \put(0,0){\line(0,1){10}}
  \put(-2.1,10){\tiny{$\triangle$}}
  \put(-12.1,10){\tiny{$\triangle$}}
\put(8,4){$=\frac 1 {\sqrt2}($}
\put(42,0){\line(0,1){12}}
\put(32,0){\line(0,1){12}}
\put(32,12){\line(1,0){10}}
\put(48,4){$+$}
\put(69,0){\line(0,1){12}}
\put(59,0){\line(0,1){12}}
\put(59,12){\line(1,0){10}}
\put(69,6){\circle*{2.}}
\put(71,5){\tiny{$Z$}}
\put(75,4){$);$}

\end{picture}
\end{array}
\en
\eq
\label{conjugation product state in TL configuration 2}
\setlength{\unitlength}{0.5mm}
\begin{array}{c}
\begin{picture}(70,12)
  \put(-10,0){\line(0,1){10}}
  \put(0,0){\line(0,1){10}}
  \put(-2.1,10){\tiny{$\triangle$}}
  \put(-12.1,10){\tiny{$\triangle$}}

  \put(0,5){\circle*{2.}}
  \put(2,4){\tiny{$X$}}

\put(8,4){$=\frac 1 {\sqrt2}($}
\put(42,0){\line(0,1){12}}
\put(32,0){\line(0,1){12}}
\put(32,12){\line(1,0){10}}
\put(42,6){\circle*{2.}}
\put(44,5){\tiny{$X$}}
\put(48,4){$+$}
\put(69,0){\line(0,1){12}}
\put(59,0){\line(0,1){12}}
\put(59,12){\line(1,0){10}}
\put(69,6){\circle*{2.}}
\put(71,5){\tiny{$ZX$}}
\put(79,4){$);$}

\end{picture}
\end{array}
\en
\eq
\label{conjugation product state in TL configuration 3}
\setlength{\unitlength}{0.5mm}
\begin{array}{c}
\begin{picture}(70,12)
  \put(-10,0){\line(0,1){10}}
  \put(0,0){\line(0,1){10}}
  \put(-2.1,10){\tiny{$\triangle$}}
  \put(-12.1,10){\tiny{$\triangle$}}

  \put(-10,5){\circle*{2.}}
  \put(-8,4){\tiny{$X$}}

\put(8,4){$=\frac 1 {\sqrt2}($}
\put(42,0){\line(0,1){12}}
\put(32,0){\line(0,1){12}}
\put(32,12){\line(1,0){10}}
\put(42,6){\circle*{2.}}
\put(44,5){\tiny{$X$}}
\put(48,4){$-$}
\put(69,0){\line(0,1){12}}
\put(59,0){\line(0,1){12}}
\put(59,12){\line(1,0){10}}
\put(69,6){\circle*{2.}}
\put(71,5){\tiny{$ZX$}}
\put(79,4){$);$}

\end{picture}
\end{array}
\en
\eq
\label{conjugation product state in TL configuration 4}
\setlength{\unitlength}{0.5mm}
\begin{array}{c}
\begin{picture}(70,12)
  \put(-10,0){\line(0,1){10}}
  \put(0,0){\line(0,1){10}}
  \put(-2.1,10){\tiny{$\triangle$}}
  \put(-12.1,10){\tiny{$\triangle$}}
  \put(-10,5){\circle*{2.}}
  \put(-8,4){\tiny{$X$}}
  \put(0,5){\circle*{2.}}
  \put(2,4){\tiny{$X$}}
\put(8,4){$=\frac 1 {\sqrt2}($}
\put(42,0){\line(0,1){12}}
\put(32,0){\line(0,1){12}}
\put(32,12){\line(1,0){10}}
\put(48,4){$-$}
\put(69,0){\line(0,1){12}}
\put(59,0){\line(0,1){12}}
\put(59,12){\line(1,0){10}}
\put(69,6){\circle*{2.}}
\put(71,5){\tiny{$Z$}}
\put(75,4){$).$}
\end{picture}
\end{array}
\en

Now we study the action of the Yang--Baxter gate $B$ (\ref{YBG B}) on the product state $|ij\rangle$ in the extended Temperley--Lieb diagrammatical
approach. The extended Temperley--Lieb configuration of the Yang--Baxter gate $B$ (\ref{YBG B}), has the form
\eq
\label{example TL B Bell transform}
  \setlength{\unitlength}{0.5mm}
  \begin{array}{c}
  \begin{picture}(130,40)

  \put(-25,-2){\footnotesize{$B$}}

  \put(-31,29){\line(0,1){6}}
  \put(-31,11){\line(0,-1){6}}
  \put(-13,29){\line(0,1){6}}
  \put(-13,11){\line(0,-1){6}}

  \put(-34,8){\line(0,1){24}}
  \put(-10,8){\line(0,1){24}}
  \put(-34,8){\line(1,0){24}}
  \put(-34,32){\line(1,0){24}}

  \put(-31,29){\line(1,-1){18}}
  \put(-31,11){\line(1,1){7.3}}
  \put(-13,29){\line(-1,-1){7.3}}

  \put(-3,19){$=\frac 1 {\sqrt 2}($}

  \put(32,5){\line(0,1){30}}
  \put(22,5){\line(0,1){30}}

  \put(40,19){$+$}

  \put(62,23){\line(0,1){12}}
  \put(52,23){\line(0,1){12}}
  \put(52,23){\line(1,0){10}}
  \put(62,29){\circle*{2.}}
  \put(64,28){\tiny{$Z$}}

  \put(52,5){\line(0,1){12}}
  \put(62,5){\line(0,1){12}}
  \put(52,17){\line(1,0){10}}

  \put(70,19){$-$}

  \put(92,23){\line(0,1){12}}
  \put(82,23){\line(0,1){12}}
  \put(82,23){\line(1,0){10}}

  \put(82,5){\line(0,1){12}}
  \put(92,5){\line(0,1){12}}
  \put(82,17){\line(1,0){10}}
  \put(92,11){\circle*{2.}}
 \put(94,10){\tiny{$Z$}}

  \put(100,19){$-$}

  \put(122,23){\line(0,1){12}}
  \put(112,23){\line(0,1){12}}
  \put(112,23){\line(1,0){10}}
  \put(122,29){\circle*{2.}}
  \put(124,28){\tiny{$X$}}

  \put(112,5){\line(0,1){12}}
  \put(122,5){\line(0,1){12}}
  \put(112,17){\line(1,0){10}}
  \put(122,11){\circle*{2.}}
  \put(124,10){\tiny{$ZX$}}

  \put(130,19){$+$}

  \put(152,23){\line(0,1){12}}
  \put(142,23){\line(0,1){12}}
  \put(142,23){\line(1,0){10}}
  \put(152,29){\circle*{2.}}
  \put(154,28){\tiny{$XZ$}}

  \put(142,5){\line(0,1){12}}
  \put(152,5){\line(0,1){12}}
  \put(142,17){\line(1,0){10}}
  \put(152,11){\circle*{2.}}
  \put(154,10){\tiny{$X$}}

  \put(164,19){$)$}

  \end{picture}
  \end{array}
  \en
 which is directly obtained from (\ref{TL B Bell transform}) by setting $\epsilon=1,\eta=1$. With the extended Temperley--Lieb diagrammatical
 rules \cite{Zhang06} that assign a normalization factor 1 to a loop configuration and assign a normalized trace of single-qubit gates to a loop
 with the action of associated single-qubit gates,  for example, we apply the Yang--Baxter gate $B$ on the product state $|11\rangle$, namely
 calculate $B|11\rangle$ in the diagrammatical approach,
\eq
\label{B 11 state}
\setlength{\unitlength}{0.5mm}
\begin{array}{c}
\begin{picture}(160,15)
\put(-6,4){$B|11\rangle=\frac 1 2($}
\put(42,0){\line(0,1){12}}
\put(32,0){\line(0,1){12}}
\put(32,0){\line(1,0){10}}
\put(48,4){$-$}
\put(69,0){\line(0,1){12}}
\put(59,0){\line(0,1){12}}
\put(59,0){\line(1,0){10}}
\put(69,6){\circle*{2.}}
\put(71,5){\tiny{$Z$}}
\put(76,4){$+$}
\put(96,0){\line(0,1){12}}
\put(86,0){\line(0,1){12}}
\put(86,0){\line(1,0){10}}
\put(96,6){\circle*{2.}}
\put(98,5){\tiny{$Z$}}
\put(103,4){$+$}
\put(123,0){\line(0,1){12}}
\put(113,0){\line(0,1){12}}
\put(113,0){\line(1,0){10}}
\put(128,4){$)=$}
\put(151,0){\line(0,1){12}}
\put(141,0){\line(0,1){12}}
\put(141,0){\line(1,0){10}}
\end{picture}
\end{array}
\en
 which is the cup representation of the EPR state $|\Psi\rangle$, see (\ref{TL diag Bell state}).  In the same manner, performing the Yang--Baxter gate $B$
 on the other product states, we attain the extended Temperley--Lieb configurations of the other Bell states, summarized in
\eq
\label{B ij state}
\setlength{\unitlength}{0.5mm}
\begin{array}{c}
\begin{picture}(170,15)
\put(0,4){$B(|00\rangle,|01\rangle,|10\rangle,|11\rangle)=($}
\put(95,0){\line(0,1){12}}
\put(85,0){\line(0,1){12}}
\put(85,0){\line(1,0){10}}
\put(95,6){\circle*{2.}}
\put(97,5){\tiny{$Z$}}
\put(100,1){$,$}
\put(116,0){\line(0,1){12}}
\put(106,0){\line(0,1){12}}
\put(106,0){\line(1,0){10}}
\put(116,6){\circle*{2.}}
\put(117,5){\tiny{$XZ$}}
\put(122,1){$,$}
\put(138,0){\line(0,1){12}}
\put(128,0){\line(0,1){12}}
\put(128,0){\line(1,0){10}}
\put(138,6){\circle*{2.}}
\put(140,5){\tiny{$X$}}
\put(144,1){$,$}
\put(160,0){\line(0,1){12}}
\put(150,0){\line(0,1){12}}
\put(150,0){\line(1,0){10}}
\put(165,4){$)$}
\end{picture}
\end{array}
\en
which allows a more concise expression as
\eq
\label{B kl state index}
\setlength{\unitlength}{0.5mm}
\begin{array}{c}
\begin{picture}(50,15)
\put(0,4){$B|kl\rangle=$}
\put(38,0){\line(0,1){12}}
\put(28,0){\line(0,1){12}}
\put(28,0){\line(1,0){10}}
\put(38,6){\circle*{2.}}
\put(40,5){\tiny{$V_{kl}$}}
\end{picture}
\end{array}
\en
with the single-qubit gate $V_{kl}=X^{k+l}Z^{k+1}$. Note that the diagrammatical representation (\ref{B kl state index}) is associated
with the algebraic relation~(\ref{B Bell transform index}).

 The Yang--Baxter gate $B$ followed with the product-state measurements has the algebraic form $|ij\rangle\langle ij|B$. For simplicity,
 we neglect the post-measurement state $|ij\rangle$, and only calculate $\langle ij|B$ in the extended Temperley--Lieb diagrammatical
 approach. For example, the quantum state $\langle00|B$ is calculated in the way
\eq
\label{11 B state}
\setlength{\unitlength}{0.5mm}
\begin{array}{c}
\begin{picture}(160,15)
\put(-6,4){$\langle00|B=\frac 1 2($}
\put(42,0){\line(0,1){12}}
\put(32,0){\line(0,1){12}}
\put(32,12){\line(1,0){10}}
\put(48,4){$+$}
\put(69,0){\line(0,1){12}}
\put(59,0){\line(0,1){12}}
\put(59,12){\line(1,0){10}}
\put(69,6){\circle*{2.}}
\put(71,5){\tiny{$Z$}}
\put(76,4){$-$}
\put(96,0){\line(0,1){12}}
\put(86,0){\line(0,1){12}}
\put(86,12){\line(1,0){10}}
\put(96,6){\circle*{2.}}
\put(98,5){\tiny{$Z$}}
\put(103,4){$+$}
\put(123,0){\line(0,1){12}}
\put(113,0){\line(0,1){12}}
\put(113,12){\line(1,0){10}}
\put(128,4){$)=$}
\put(151,0){\line(0,1){12}}
\put(141,0){\line(0,1){12}}
\put(141,12){\line(1,0){10}}
\end{picture}
\end{array}
\en
 which is the cap representation of the EPR state $|\Psi\rangle$, see (\ref{TL diag Bell state_cap}). After calculating the
 other cases, it turns out that  the state  $\langle ij|B$ has the representation,
\eq
\label{ij B state index}
\setlength{\unitlength}{0.5mm}
\begin{array}{c}
\begin{picture}(50,15)
\put(0,4){$\langle ij|B=$}
\put(38,0){\line(0,1){12}}
\put(28,0){\line(0,1){12}}
\put(28,12){\line(1,0){10}}
\put(38,6){\circle*{2.}}
\put(40,5){\tiny{$U_{ij}$}}
\end{picture}
\end{array}
\en
with the single-qubit gate $U_{ij}=(-1)^iZ^iX^{i+j}$. Note that the diagrammatical representation (\ref{ij B state index}) is associated with
the adjoint of the algebraic relation~(\ref{B inverse Bell transform index}) in which the Bell transform $B_{-1,-1}$ is the inverse of the Bell
transform $B$.

\subsection{Teleportation-based quantum computation}

Let us derive the extended Temperley--Lieb configurations in Section~\ref{TL diagrammatical for tele based quantum computation} respectively from
 the Yang--Baxter gate approach to teleportation-based quantum computation in Section~\ref{YBG approach to tele based QC}.

The topological representation of quantum teleportation, such as (\ref{tele: tl}), can be regarded as  the diagrammatical representation of
a matrix element of the teleportation equation~(\ref{example tele eqa B1}) using the Yang--Baxter gate $B$. With the cup state (\ref{B 11 state}) generated
by $B|11\rangle$ and the cap state (\ref{ij B state index}) generated by $\langle ij|B$, the teleportation operator $(B\otimes 1\!\!1_2)(1\!\!1\otimes B)$
has the matrix element as
\eq
\label{example teleportation operator as teleportation}
\setlength{\unitlength}{0.7mm}
\begin{array}{c}
\begin{picture}(140,25)
\put(-8,11){$(\langle ij|\otimes 1\!\! 1_2)(B\otimes 1\!\!1_2)(1\!\!1_2\otimes B)|\alpha\rangle\otimes|11\rangle=$}
\put(96,0){\line(0,1){24}}
\put(86,3.5){\line(0,1){20.5}}
\put(86,24){\line(1,0){10}}
\put(84,0){\makebox(4,4){$\nabla$}}
\put(96,0){\line(1,0){10}}
\put(106,0){\line(0,1){24}}
\put(96,18){\circle*{1.5}}
\put(97.5,17){\tiny{$U_{ij}$}}
\multiput(81,12)(1,0){30}{\line(1,0){.5}}
\put(115,11){$=\frac 1 2$}
\put(130,3.5){\line(0,1){20.5}}
\put(128,0){\makebox(4,4){$\nabla$}}
\put(130,13){\circle*{1.5}}
\put(132.5,12){\tiny{$U_{ij}^T$}}
\end{picture}
\end{array}
\en
in which the diagrammatical part below the dashed line denotes the prepared state $|\alpha\rangle\otimes|\Psi\rangle$ and the part above the dashed line
denotes the Bell measurement labeled by the single-qubit gate $U_{ij}$ in (\ref{ij B state index}). The transpose of $U_{ij}$ is
the local unitary gate $W_B$ (\ref{W_B}), namely, $U_{ij}^T=W_B$. Note that the post-measurement state $B^\dag|ij\rangle$ is neglected for simplicity. As
another example, the topological configuration  (\ref{chain teleportation}) of the chained teleportation is obtained in the Yang--Baxter
gate approach,
\eqa
\label{chain teleportation B}
\setlength{\unitlength}{0.6mm}
\begin{array}{c}
\begin{picture}(170,25)
\put(-22,11){($_{1234}\langle0000|\otimes 1\!\! 1_2)B_{12}B_{34}B_{23}B_{45}|\alpha\rangle_1\otimes|1111\rangle_{2345}=$}
\put(110,0){\makebox(4,4){$\nabla$}}
\put(112,4){\line(0,1){20}}
\put(122,0){\line(0,1){24}}
\put(132,0){\line(0,1){24}}
\multiput(108,12)(1,0){48}{\line(1,0){.5}}
\put(122,0){\line(1,0){10}}
\put(112,24){\line(1,0){10}}
\put(132,24){\line(1,0){10}}
\put(142,0){\line(0,1){24}}
\put(142,0){\line(1,0){10}}
\put(152,0){\line(0,1){24}}
\put(158,7){\makebox(14,10){$=\frac 1 4$}}
\put(172,0){\makebox(4,4){$\nabla$}}
\put(174,4){\line(0,1){20}}
\end{picture}
\end{array}
\ena
where $B|11\rangle$ denotes  the cup state (\ref{B 11 state}) and  $\langle00|B$ denotes the cap state (\ref{11 B state}) and the symbol $B_{ij}$ means
that the Yang--Baxter gate $B$ is acting on both the $i$-th and $j$-th qubits.

The topological configuration (\ref{gate: one}) for the fault-tolerant construction of the single-qubit gate $U$ in teleportation-based
quantum computation is associated with the matrix element of the teleportation equation~(\ref{example single qubit gate via B}),
\eq
\label{example single qubit gate as teleportation}
\setlength{\unitlength}{0.7mm}
\begin{array}{c}
\begin{picture}(145,25)
\put(-20,11){$(\langle ij|\otimes 1\!\! 1_2)(B\otimes 1\!\!1_2)(1\!\!1_2\otimes1\!\!1_2\otimes U)(1\!\!1\otimes B)|\alpha\rangle\otimes|11\rangle=$}
\put(116,0){\line(0,1){24}}
\put(106,3.5){\line(0,1){20.5}}
\put(106,24){\line(1,0){10}}
\put(104,0){\makebox(4,4){$\nabla$}}
\put(116,0){\line(1,0){10}}
\put(126,0){\line(0,1){24}}
\put(116,18){\circle*{1.5}}
\put(117.5,17){\tiny{$W_B^T$}}
\put(126,6){\circle*{1.5}}
\put(127.5,5){\tiny{$U$}}
\multiput(101,12)(1,0){30}{\line(1,0){.5}}
\put(134,11){$=\frac 1 2$}
\put(149,3.5){\line(0,1){20.5}}
\put(147,0){\makebox(4,4){$\nabla$}}
\put(149,18){\circle*{1.5}}
\put(151.5,17){\tiny{$R_B$}}
\put(149,8){\circle*{1.5}}
\put(151.5,7){\tiny{$U$}}
\end{picture}
\end{array}
\en
with $R_B=UW_BU^\dag$, where the single-qubit gate $U_{ij}$ in (\ref{ij B state index}) is denoted as the $W_B^T$ gate to avoid a possible notational confusion. Furthermore,
the algebraic counterpart of the topological configuration (\ref{gate: two}) of the fault-tolerant construction of the Yang--Baxter gate $B$ which is a two-qubit Clifford
gate, is the matrix element of  the teleportation equation (\ref{example two qubit gate via B}), expressed as
\eqa
\label{example two qubit gate as teleportation}
\setlength{\unitlength}{0.7mm}
\begin{array}{c}
\begin{picture}(150,60)
\put(80,50){\makebox(9,8){$(\tiny{\langle i_1j_1|\otimes 1\!\!1_2\otimes 1\!\!1_2\otimes \langle i_2j_2|)(B\otimes B\otimes B)(1\!\!1_2\otimes B\otimes B \otimes1\!\!1_2)(|\alpha\rangle\otimes |11\rangle\otimes |11\rangle\otimes |\beta\rangle)}$}}
\put(0,22){\makebox(14,10){$=$}}
\put(14,9){\makebox(4,4){$\nabla$}}
\put(16,12.5){\line(0,1){32.5}}
\put(26,36){\circle*{1.5}}
\put(18,33){\makebox(6,6){\tiny{$W^T_B$}}}
\put(26,9){\line(0,1){36}}
\multiput(12,27)(1,0){58}{\line(1,0){.5}}
\put(36,9){\line(0,1){7}}
\put(36,20){\line(0,1){25}}
\put(34,16){\line(0,1){4}}
\put(48,16){\line(0,1){4}}
\put(34,16){\line(1,0){14}}
\put(34,20){\line(1,0){14}}
\put(39,16){\makebox(4,4){\tiny{$B$}}}
\put(26,9){\line(1,0){10}}
\put(16,45){\line(1,0){10}}
\put(46,9){\line(0,1){7}}
\put(46,20){\line(0,1){25}}
\put(46,9){\line(1,0){10}}
\put(56,9){\line(0,1){36}}
\put(56,45){\line(1,0){10}}
\put(66,12.5){\line(0,1){32.5}}
\put(66,36){\circle*{1.5}}
\put(59,34){\makebox(4,4){\tiny{$W^T_B$}}}
\put(64,9){\makebox(4,4){$\nabla$}}

\put(70,22){\makebox(14,10){$=~\frac 1 4$}}
\put(86,9){\makebox(4,4){$\nabla$}}

\put(88,29){\line(0,1){16}}
\put(88,12.5){\line(0,1){12.5}}
\put(96,9){\makebox(4,4){$\nabla$}}

\put(86,25){\line(0,1){4}}
\put(100,25){\line(0,1){4}}
\put(86,25){\line(1,0){14}}
\put(86,29){\line(1,0){14}}
\put(91,25){\makebox(4,4){\tiny{$B$}}}
\put(98,29){\line(0,1){16}}
\put(98,12.5){\line(0,1){12.5}}
\put(88,18){\circle*{1.5}}
\put(80,15){\makebox(6,6){\tiny{$W_B$}}}
\put(98,18){\circle*{1.5}}
\put(101,16){\makebox(4,4){\tiny{$W_B^T$}}}
\put(106,22){\makebox(14,10){$=~\frac 1 4$}}
\put(124,9){\makebox(4,4){$\nabla$}}

\put(126,29){\line(0,1){16}}
\put(126,12.5){\line(0,1){12.5}}
\put(134,9){\makebox(4,4){$\nabla$}}

\put(124,25){\line(0,1){4}}
\put(138,25){\line(0,1){4}}
\put(124,25){\line(1,0){14}}
\put(124,29){\line(1,0){14}}
\put(129,25){\makebox(4,4){\tiny{$B$}}}
\put(136,29){\line(0,1){16}}
\put(136,12.5){\line(0,1){12.5}}
\put(126,36){\circle*{1.5}}
\put(119,34){\makebox(4,4){\tiny{$Q_B$}}}
\put(136,36){\circle*{1.5}}
\put(139,34){\makebox(4,4){\tiny{$P_B$}}}
\put(133.5,0){\makebox(9,8){\footnotesize{$|\beta\rangle$}}}
\put(120.5,0){\makebox(9,8){\footnotesize{$|\alpha\rangle$}}}
\put(95.5,0){\makebox(9,8){\footnotesize{$|\beta\rangle$}}}
\put(82.5,0){\makebox(9,8){\footnotesize{$|\alpha\rangle$}}}
\put(61.5,0){\makebox(9,8){\footnotesize{$|\beta\rangle$}}}
\put(11.5,0){\makebox(9,8){\footnotesize{$|\alpha\rangle$}}}
\end{picture}
\label{gate: two2}
\end{array}
\ena
where the single-qubit gate $W_B^T$ acting on the second qubit depends on the indices $i_1$, $j_1$ and
the single-qubit gate $W_B^T$ on the sixth qubit depends on the indices $i_2$, $j_2$ and the single-qubit
gates $Q_B$ and $P_B$ are derived by the formula~(\ref{Q P for B}).

\subsection{The teleportation operator and the teleportation equation}

\label{The teleportation operator and the teleportation equation}


As shown up in Section~\ref{YBG approach to tele based QC},  the teleportation operator (\ref{tele operator}) plays the key role in the Yang--Baxter gate approach
to teleportation-based quantum computation, and it is accompanied with both the product state preparation and the product state measurement to perform the teleportation protocol. In this subsection, we study the diagrammatical representation of the teleportation operator $(B\otimes 1\!\!1_2)(1\!\!1_2\otimes B)$ by combining the configuration of the product basis with the extended Temperley--Lieb diagrammatical approach, and from such the diagrammatical representation, we can easily derive the teleportation equation of the type~(\ref{example tele eqa B1}).

With the extended Temperley--Lieb configuration (\ref{TL_identity}) of the two-qubit identity gate, the Yang--Baxter gate $B$ (\ref{YBG B})
has the configuration
\eq
\label{example TL B Bell transform expansion}
\setlength{\unitlength}{0.5mm}
\begin{array}{c}
\begin{picture}(20,40)
\put(-115,-2){\footnotesize{$B$}}

  \put(-121,29){\line(0,1){6}}
  \put(-121,11){\line(0,-1){6}}
  \put(-103,29){\line(0,1){6}}
  \put(-103,11){\line(0,-1){6}}

  \put(-124,8){\line(0,1){24}}
  \put(-100,8){\line(0,1){24}}
  \put(-124,8){\line(1,0){24}}
  \put(-124,32){\line(1,0){24}}

  \put(-121,29){\line(1,-1){18}}
  \put(-121,11){\line(1,1){7.3}}
  \put(-103,29){\line(-1,-1){7.3}}

\put(-93,19){$=\frac 1 {\sqrt 2}($}
\put(-58,23){\line(0,1){12}}
\put(-68,23){\line(0,1){12}}
\put(-68,23){\line(1,0){10}}
\put(-68,5){\line(0,1){12}}
\put(-58,5){\line(0,1){12}}
\put(-68,17){\line(1,0){10}}
\put(-53,19){$+$}
\put(-32,23){\line(0,1){12}}
\put(-42,23){\line(0,1){12}}
\put(-42,23){\line(1,0){10}}
\put(-32,29){\circle*{2.}}
\put(-30,28){\tiny{$Z$}}
\put(-42,5){\line(0,1){12}}
\put(-32,5){\line(0,1){12}}
\put(-42,17){\line(1,0){10}}
\put(-32,11){\circle*{2.}}
\put(-30,10){\tiny{$Z$}}
\put(-26,19){$+$}
\put(-6,23){\line(0,1){12}}
\put(-16,23){\line(0,1){12}}
\put(-16,23){\line(1,0){10}}
\put(-6,29){\circle*{2.}}
\put(-4,28){\tiny{$X$}}
\put(-16,5){\line(0,1){12}}
\put(-6,5){\line(0,1){12}}
\put(-16,17){\line(1,0){10}}
\put(-6,11){\circle*{2.}}
\put(-4,10){\tiny{$X$}}
\put(0,19){$+$}
\put(20,23){\line(0,1){12}}
\put(10,23){\line(0,1){12}}
\put(10,23){\line(1,0){10}}
\put(20,29){\circle*{2.}}
\put(22,28){\tiny{$XZ$}}
\put(10,5){\line(0,1){12}}
\put(20,5){\line(0,1){12}}
\put(10,17){\line(1,0){10}}
\put(20,11){\circle*{2.}}
\put(22,10){\tiny{$ZX$}}
\put(24,19){$+$}
\put(44,23){\line(0,1){12}}
\put(34,23){\line(0,1){12}}
\put(34,23){\line(1,0){10}}
\put(44,29){\circle*{2.}}
\put(46,28){\tiny{$Z$}}
\put(34,5){\line(0,1){12}}
\put(44,5){\line(0,1){12}}
\put(34,17){\line(1,0){10}}
\put(48,19){$-$}
\put(68,23){\line(0,1){12}}
\put(58,23){\line(0,1){12}}
\put(58,23){\line(1,0){10}}
\put(58,5){\line(0,1){12}}
\put(68,5){\line(0,1){12}}
\put(58,17){\line(1,0){10}}
\put(68,11){\circle*{2.}}
\put(70,10){\tiny{$Z$}}
\put(72,19){$+$}
\put(92,23){\line(0,1){12}}
\put(82,23){\line(0,1){12}}
\put(82,23){\line(1,0){10}}
\put(92,29){\circle*{2.}}
\put(94,28){\tiny{$X$}}
\put(82,5){\line(0,1){12}}
\put(92,5){\line(0,1){12}}
\put(82,17){\line(1,0){10}}
\put(92,11){\circle*{2.}}
\put(94,10){\tiny{$ZX$}}
\put(96,19){$-$}
\put(116,23){\line(0,1){12}}
\put(106,23){\line(0,1){12}}
\put(106,23){\line(1,0){10}}
\put(116,29){\circle*{2.}}
\put(118,28){\tiny{$XZ$}}
\put(106,5){\line(0,1){12}}
\put(116,5){\line(0,1){12}}
\put(106,17){\line(1,0){10}}
\put(116,11){\circle*{2.}}
\put(118,10){\tiny{$X$}}
\put(124,19){$)$}
\end{picture}
\end{array}
\en
which is equivalent to a special case of the extended Temperley--Lieb configuration (\ref{TL B general}). Applying the algebraic relations (\ref{conjugation product state in TL configuration 1})-(\ref{conjugation product state in TL configuration 4}) on the configuration~(\ref{example TL B Bell transform expansion}),
the Yang--Baxter gate $B$ (\ref{YBG B}) has its compact diagrammatical representation
\eq
\label{diagram B Bell transform}
\setlength{\unitlength}{0.5mm}
\begin{array}{c}
\begin{picture}(210,40)

\put(15,-2){\footnotesize{$B$}}

  \put(9,29){\line(0,1){6}}
  \put(9,11){\line(0,-1){6}}
  \put(27,29){\line(0,1){6}}
  \put(27,11){\line(0,-1){6}}

  \put(6,8){\line(0,1){24}}
  \put(30,8){\line(0,1){24}}
  \put(6,8){\line(1,0){24}}
  \put(6,32){\line(1,0){24}}

  \put(9,29){\line(1,-1){18}}
  \put(9,11){\line(1,1){7.3}}
  \put(27,29){\line(-1,-1){7.3}}

\put(37,19){$=$}

\put(62,23){\line(0,1){12}}
\put(52,23){\line(0,1){12}}
\put(52,23){\line(1,0){10}}
\put(62,29){\circle*{2.}}
\put(64,28){\tiny{$Z$}}

\put(52,5){\line(0,1){10}}
\put(62,5){\line(0,1){10}}

\put(49.9,15){\tiny{$\triangle$}}
\put(59.9,15){\tiny{$\triangle$}}

\put(70,19){$+$}

\put(92,23){\line(0,1){12}}
\put(82,23){\line(0,1){12}}
\put(82,23){\line(1,0){10}}
\put(92,29){\circle*{2.}}
\put(94,28){\tiny{$XZ$}}

\put(82,5){\line(0,1){10}}
\put(92,5){\line(0,1){10}}
\put(92,11){\circle*{2.}}
\put(94,10){\tiny{$X$}}

\put(79.9,15){\tiny{$\triangle$}}
\put(89.9,15){\tiny{$\triangle$}}

\put(100,19){$+$}

\put(122,23){\line(0,1){12}}
\put(112,23){\line(0,1){12}}
\put(112,23){\line(1,0){10}}
\put(122,29){\circle*{2.}}
\put(124,28){\tiny{$X$}}

\put(112,5){\line(0,1){10}}
\put(122,5){\line(0,1){10}}
\put(112,11){\circle*{2.}}
\put(114,10){\tiny{$X$}}

\put(109.9,15){\tiny{$\triangle$}}
\put(119.9,15){\tiny{$\triangle$}}

\put(130,19){$+$}

\put(152,23){\line(0,1){12}}
\put(142,23){\line(0,1){12}}
\put(142,23){\line(1,0){10}}

\put(142,5){\line(0,1){10}}
\put(152,5){\line(0,1){10}}
\put(142,11){\circle*{2.}}
\put(144,10){\tiny{$X$}}
\put(152,11){\circle*{2.}}
\put(154,10){\tiny{$X$}}

\put(139.9,15){\tiny{$\triangle$}}
\put(149.9,15){\tiny{$\triangle$}}

\put(160,19){$=\sum_{k,l=0}^1$}

\put(204,23){\line(0,1){12}}
\put(194,23){\line(0,1){12}}
\put(194,23){\line(1,0){10}}
\put(204,29){\circle*{2.}}
\put(206,28){\tiny{$V_{kl}$}}

\put(194,5){\line(0,1){10}}
\put(204,5){\line(0,1){10}}
\put(204,11){\circle*{2.}}
\put(205,10){\tiny{$X^l$}}
\put(194,11){\circle*{2.}}
\put(195,10){\tiny{$X^k$}}

\put(191.9,15){\tiny{$\triangle$}}
\put(201.9,15){\tiny{$\triangle$}}

\end{picture}
\end{array}
\en
where the single-qubit gate $V_{kl}$ is defined in~(\ref{B kl state index}), and the associated algebraic formulation of such the
configuration~(\ref{diagram B Bell transform}) is expressed as
\eq
\label{algebraic B Bell transform}
B=\sum_{k,l=0}^1 |\Psi_{V_{kl}}\rangle\langle kl|=\sum_{k,l=0}^1|\psi(k+l,k+1)\rangle\langle kl|,
\en
with $|\Psi_{V_{kl}}\rangle$ defined in (\ref{Psi_U}),  which is obviously the defining relation~(\ref{B Bell transform index})
of the Bell transform $B$.

On the other hand, applying the algebraic relations (\ref{product state in TL configuration 1})-(\ref{product state in TL configuration 4}) on
the configuration (\ref{example TL B Bell transform expansion})  of the Yang--Baxter gate $B$ gives rise to its another compact configuration
\eq
\label{diagram B inverse Bell transform}
\setlength{\unitlength}{0.5mm}
\begin{array}{c}
\begin{picture}(210,40)

\put(15,-2){\footnotesize{$B$}}

  \put(9,29){\line(0,1){6}}
  \put(9,11){\line(0,-1){6}}
  \put(27,29){\line(0,1){6}}
  \put(27,11){\line(0,-1){6}}

  \put(6,8){\line(0,1){24}}
  \put(30,8){\line(0,1){24}}
  \put(6,8){\line(1,0){24}}
  \put(6,32){\line(1,0){24}}

  \put(9,29){\line(1,-1){18}}
  \put(9,11){\line(1,1){7.3}}
  \put(27,29){\line(-1,-1){7.3}}

\put(37,19){$=$}

\put(62,25){\line(0,1){10}}
\put(52,25){\line(0,1){10}}

\put(49.9,23){\tiny{$\nabla$}}
\put(59.9,23){\tiny{$\nabla$}}

\put(52,5){\line(0,1){12}}
\put(62,5){\line(0,1){12}}
\put(52,17){\line(1,0){10}}

\put(70,19){$+$}

\put(92,25){\line(0,1){10}}
\put(82,25){\line(0,1){10}}
\put(92,29){\circle*{2.}}
\put(94,28){\tiny{$X$}}

\put(79.9,23){\tiny{$\nabla$}}
\put(89.9,23){\tiny{$\nabla$}}

\put(82,5){\line(0,1){12}}
\put(92,5){\line(0,1){12}}
\put(82,17){\line(1,0){10}}
\put(92,11){\circle*{2.}}
\put(94,10){\tiny{$X$}}

\put(100,19){$-$}

\put(122,25){\line(0,1){10}}
\put(112,25){\line(0,1){10}}
\put(112,29){\circle*{2.}}
\put(114,28){\tiny{$X$}}

\put(109.9,23){\tiny{$\nabla$}}
\put(119.9,23){\tiny{$\nabla$}}

\put(112,5){\line(0,1){12}}
\put(122,5){\line(0,1){12}}
\put(112,17){\line(1,0){10}}
\put(122,11){\circle*{2.}}
\put(124,10){\tiny{$ZX$}}

\put(130,19){$-$}

\put(152,25){\line(0,1){10}}
\put(142,25){\line(0,1){10}}
\put(152,29){\circle*{2.}}
\put(154,28){\tiny{$X$}}
\put(142,29){\circle*{2.}}
\put(144,28){\tiny{$X$}}

\put(139.9,23){\tiny{$\nabla$}}
\put(149.9,23){\tiny{$\nabla$}}

\put(142,5){\line(0,1){12}}
\put(152,5){\line(0,1){12}}
\put(142,17){\line(1,0){10}}
\put(152,11){\circle*{2.}}
\put(154,10){\tiny{$Z$}}

\put(160,19){$=\sum_{i,j=0}^1$}

\put(204,25){\line(0,1){10}}
\put(194,25){\line(0,1){10}}
\put(204,29){\circle*{2.}}
\put(205,28){\tiny{$X^j$}}
\put(194,29){\circle*{2.}}
\put(195,28){\tiny{$X^i$}}

\put(192.2,23){\tiny{$\nabla$}}
\put(202.2,23){\tiny{$\nabla$}}

\put(194,5){\line(0,1){12}}
\put(204,5){\line(0,1){12}}
\put(194,17){\line(1,0){10}}
\put(203.9,11){\circle*{2.}}
\put(205.9,10){\tiny{$U_{ij}$}}

\end{picture}
\end{array}
\en
where the single-qubit gate $U_{ij}$ is defined in (\ref{ij B state index}), and the associated algebraic expression is
\eq
\label{algebraic B inverse Bell transform}
B=\sum_{i,j=0}^1|ij\rangle\langle\Psi_{U_{ij}}|=\sum_{i,j=0}^1(-1)^j|ij\rangle\langle\psi(i+j,i)|,
\en
with $|\Psi_{U_{ij}}\rangle$ defined in (\ref{Psi_U}), from which the Yang--Baxter gate $B$ can be also viewed as the inverse of
the Bell transform from  the Bell basis to the product basis. Note that in \cite{ZZ14} the inverse of the Bell transform  with
the product basis measurement is regarded as the Bell measurement, hence the Yang--Baxter gate $B$ acted by the product state
can be viewed as the Bell measurement.

In the algebraic approach, the teleportation operator $(B\otimes 1\!\!1_2)(1\!\!1_2\otimes B)$ has the form
\eq
(B\otimes 1\!\!1_2)(1\!\!1_2\otimes B)=\sum_{i,j,k,l=0}^1(|ij\rangle\langle\Psi_{U_{ij}}|\otimes 1\!\!1_2)(1\!\!1_2\otimes
 |\Psi_{V_{kl}}\rangle\langle kl| )
\en
where the left Yang--Baxter gate $B$ represents the inverse of the Bell transform (\ref{algebraic B inverse Bell transform}) with the configuration (\ref{diagram B inverse Bell transform}) and the right one denotes the Bell transform (\ref{algebraic B Bell transform}) with the configuration (\ref{diagram B Bell transform}).
Therefore, such the teleportation operator is calculated in the diagrammatical approach,
\eq
\label{TL for teleportation operator in Bell transform}
  \setlength{\unitlength}{0.5mm}
  \begin{array}{c}
  \begin{picture}(30,65)


  \put(-66,25){\line(0,1){6}}
  \put(-66,7){\line(0,-1){6}}
  \put(-48,25){\line(0,1){6}}
  \put(-48,7){\line(0,-1){6}}

  \put(-69,4){\line(0,1){24}}
  \put(-45,4){\line(0,1){24}}
  \put(-69,4){\line(1,0){24}}
  \put(-69,28){\line(1,0){24}}

  \put(-66,25){\line(1,-1){18}}
  \put(-66,7){\line(1,1){7.3}}
  \put(-48,25){\line(-1,-1){7.3}}


  \put(-84,55){\line(0,1){6}}
  \put(-84,37){\line(0,-1){6}}
  \put(-66,55){\line(0,1){6}}
  \put(-66,37){\line(0,-1){6}}

  \put(-87,34){\line(0,1){24}}
  \put(-63,34){\line(0,1){24}}
  \put(-87,34){\line(1,0){24}}
  \put(-87,58){\line(1,0){24}}

  \put(-84,55){\line(1,-1){18}}
  \put(-84,37){\line(1,1){7.3}}
  \put(-66,55){\line(-1,-1){7.3}}


  \put(-48,31){\line(0,1){30}}
  \put(-84,31){\line(0,-1){30}}


  \put(-40,29){$=$}


  \put(-30,29){$\sum_{i,j,k,l=0}^1$}

  \put(14.5,48){\line(0,1){12}}
  \put(2,48){\line(0,1){12}}
  \put(14.5,54){\circle*{2.}}
  \put(15.5,53){\tiny{$X^j$}}
  \put(2,54){\circle*{2.}}
  \put(3,53){\tiny{$X^i$}}

  \put(-.2,45){{\scriptsize$\nabla$}}
  \put(12.3,45){\scriptsize{$\nabla$}}

  \put(14.5,30){\circle*{2.}}
  \put(16,29){\tiny{$U_{ij}$}}

  \put(14.5,0){\line(0,1){12}}
  \put(27,0){\line(0,1){12}}
  \put(27,6){\circle*{2.}}
  \put(28,5){\tiny{$X^l$}}
  \put(14.5,6){\circle*{2.}}
  \put(15.5,5){\tiny{$X^k$}}

  \put(12.3,12){\tiny{$\triangle$}}
  \put(24.8,12){\tiny{$\triangle$}}

  \put(2,0){\line(0,1){37.5}}
  \put(2,37.5){\line(1,0){12.5}}
  \put(14.5,22.5){\line(0,1){15}}
  \put(27,22.5){\line(0,1){37.5}}
  \put(14.5,22.5){\line(1,0){12.5}}

  \put(27,30){\circle*{2.}}
  \put(28.5,29){\tiny{$V_{kl}$}}

  \put(40,29){$=\sum_{i,j,k,l=0}^1$}

  \put(94.5,48){\line(0,1){12}}
  \put(82,48){\line(0,1){12}}
  \put(94.5,54){\circle*{2.}}
  \put(95.5,53){\tiny{$X^j$}}
  \put(82,54){\circle*{2.}}
  \put(83,53){\tiny{$X^i$}}

  \put(79.8,45){{\scriptsize$\nabla$}}
  \put(92.3,45){\scriptsize{$\nabla$}}


  \put(94.5,0){\line(0,1){12}}
  \put(107,0){\line(0,1){12}}
  \put(107,6){\circle*{2.}}
  \put(108,5){\tiny{$X^l$}}
  \put(94.5,6){\circle*{2.}}
  \put(95.5,5){\tiny{$X^k$}}

  \put(92.3,12){\tiny{$\triangle$}}
  \put(104.8,12){\tiny{$\triangle$}}

  \put(82,0){\line(0,1){37.5}}
  \put(82,37.5){\line(1,0){12.5}}
  \put(94.5,22.5){\line(0,1){15}}
  \put(107,22.5){\line(0,1){37.5}}
  \put(94.5,22.5){\line(1,0){12.5}}

  \put(107,54){\circle*{2.}}
 \put(108.5,53){\tiny{$W_{1,1}$}}

  \end{picture}
  \end{array}
  \en
where the single-qubit gate $W_{1,1}=V_{kl}U^T_{ij}$ expressed as
\eq
\label{W 1 1}
W_{1,1} = (-1)^{i\cdot j+(k+l)\cdot(i+k+1)}Z^{i+k+1}X^{i+j+k+l},
\en
can be also calculated from the formula~(\ref{W epsilon}) by setting $\epsilon=\eta=1$. See Table~\ref{Table W 1 1} for the explicit
expressions of the single-qubit gate $W_{1,1}$.

\begin{table}
\begin{center}
\footnotesize
\begin{tabular}{c|c|c|c|c}
\hline\hline
   & $i=0$, $j=0$ & $i=0$, $j=1$ & $i=1$, $j=0$ & $i=1$, $j=1$ \\ \hline
  $k=0$, $l=0$ & $Z$  & $-XZ$ & $X$ & $-1\!\!1_2$ \\ \hline
  $k=0$, $l=1$ & $XZ$ & $-Z$ & $1\!\!1_2$ & $-X$  \\ \hline
  $k=1$, $l=0$ & $X$ & $1\!\!1_2$ & $-Z$ & $XZ$ \\ \hline
  $k=1$, $l=1$ & $1\!\!1_2$ & $X$ & $-XZ$ & $-Z$  \\ \hline\hline
\end{tabular}
\caption{\label{Table W 1 1}  The single-qubit gate $W_{1,1}$ (\ref{W 1 1}) in  both the algebraic expansion (\ref{algebraic tl tele operator product state sum}) of
 the  teleportation operator $(B\otimes 1\!\!1_2)(1\!\!1_2\otimes B)$ and the teleportation equation (\ref{example tele eqa B}).}
\end{center}
\end{table}

Furthermore, the algebraic counterpart of the diagram~(\ref{TL for teleportation operator in Bell transform}) can be rewritten  as
\eq
\label{algebraic tl tele operator product state sum}
(B\otimes 1\!\!1_2)(1\!\!1_2\otimes B)=\sum_{i,j,k,l=0}^1(|ij\rangle, W_{1,1}, \langle kl|),
\en
with the algebraic notation $(|ij\rangle, W_{1,1}, \langle kl|)$ of the associated diagrammatical term,  which gives rise to
the teleportation equation
\eq
\label{example tele eqa B}
(B\otimes 1\!\!1_2)(1\!\!1_2\otimes B)|\alpha\rangle\otimes |kl\rangle=\frac 1 2\sum_{i,j=0}^1|ij\rangle W_{1,1}|\alpha\rangle,
\en
with the teleportation equation~(\ref{example tele eqa B1}) as its special example of $k=l=1$. In this sense, each one of the total 16 teleportation
configurations in the diagram~(\ref{TL for teleportation operator in Bell transform}) can be extracted by applying the product basis measurement on
the teleportation operator $(B\otimes 1\!\!1_2)(1\!\!1_2\otimes B)$. In addition, Appendix E presents another equivalent method of deriving the
algebraic structure (\ref{algebraic tl tele operator product state sum}) of the teleportation operator $(B\otimes 1\!\!1_2)(1\!\!1_2\otimes B)$.

\begin{figure}
 \begin{center}
  \includegraphics[width=13cm]{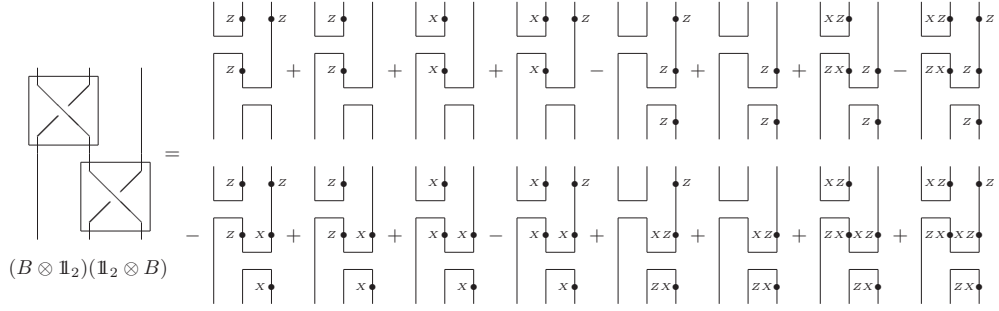}
  \end{center}
  \caption{\label{fig_tl_tele_operator_Bell_measurement} The extended Temperley--Lieb configuration of the teleportation operator $(B\otimes 1\!\!1_2)(1\!\!1_2\otimes B)$
  from the viewpoint of Bell measurements. The result is derived by directly applying the extended Temperley--Lieb configuration~(\ref{example TL B Bell transform expansion})
  of the Yang--Baxter gate $B$ (\ref{YBG B}). The vertical braiding configuration can be regarded as the one obtained by clockwise rotating the horizontal braiding
  configuration in the quantum circuit in Figure~\ref{fig_tele_YBG} by 90 degrees.        }
\end{figure}

\subsection{The teleportation operator using Bell measurements  }

  The extended Temperley--Lieb diagrammatical representation of the teleportation operator $(B\otimes 1\!\!1_2)(1\!\!1_2\otimes B)$
  can be derived in a rather straightforward diagrammatical approach using the extended Temperley--Lieb configuration~(\ref{example TL B Bell transform expansion})
  of the Yang--Baxter gate $B$, and the result is shown up in Figure~\ref{fig_tl_tele_operator_Bell_measurement}, which can be explicitly regarded as a kind of
  linear combination of 16 typical quantum teleportation processes using Bell measurements.

\begin{figure}
 \begin{center}
  \includegraphics[width=13cm]{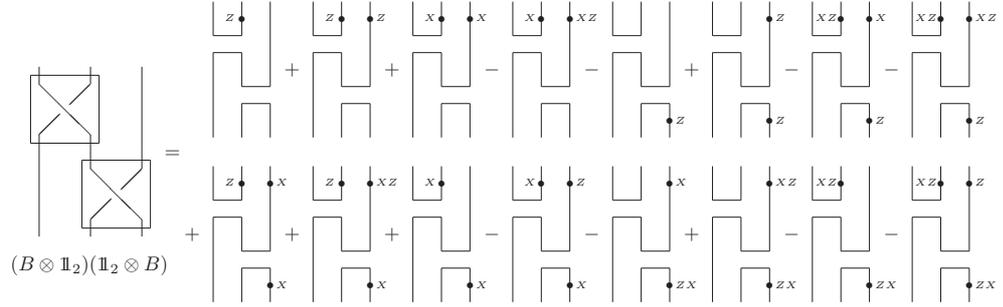}
  \end{center}
  \caption{\label{fig_tl_tele_operator}  The extended Temperley--Lieb configuration of the teleportation operator $(B\otimes 1\!\!1_2)(1\!\!1_2\otimes B)$ as
  a simplified version of the configuration in Figure~\ref{fig_tl_tele_operator_Bell_measurement}.      }
\end{figure}

   Although each one of these 16 diagrammatical  teleportation terms in Figure~\ref{fig_tl_tele_operator_Bell_measurement} can be
  interpreted in the viewpoint of quantum teleportation, a linear combination of these diagrammatical terms can not be usually viewed as quantum teleportation.
  The reason is that the teleportation operator with the product basis measurement (instead of Bell measurements) gives rise to quantum teleportation,
  as discussed in Subsection~\ref{The teleportation operator and the teleportation equation}. Note that the simplified version of
  Figure~\ref{fig_tl_tele_operator_Bell_measurement} is shown up in Figure~\ref{fig_tl_tele_operator}.

  Appendix~\ref{algebraic_study_fig_tl_tele_operator} presents an algebraic method of deriving the extended Temperley--Lieb configuration
  in Figure~\ref{fig_tl_tele_operator}, and it is obvious that  the topological configuration in Figure~\ref{fig_tl_tele_operator_Bell_measurement} can
  be easily obtained from the configuration in Figure~\ref{fig_tl_tele_operator}.
  It is worthwhile pointing out that such the extended Temperley--Lieb configurations in both the diagram~(\ref{TL for teleportation operator in Bell transform})
  and Figure~\ref{fig_tl_tele_operator} come naturally from our topological and algebraic reformulations
  of teleportation-based quantum computation, whereas they are indeed unexpected if we consider the standard low dimensional
  topology \cite{Kauffman02}. Interested readers are invited to refer to Appendix A.

\section{Concluding remarks}

\label{Concluding remarks}

In this paper, we reformulate teleportation-based quantum computation \cite{GC99,Nielsen03,Leung04} in both the extended Temperley--Lieb diagrammatical
approach \cite{Kauffman05,Zhang06} and the Yang--Baxter gate approach \cite{Dye03, KL04, ZKG04}. Such the two approaches can be respectively regarded as
the topological aspect and algebraic aspect of a unified approach. On the other hand, through our research, the Yang--Baxter gate
configuration (the braiding configuration) admits an equivalent description of a set of the extended Temperley--Lieb diagrammatical configurations,
so we finally propose the extended Temperley--Lieb diagrammatical approach as an interesting topic for physicists in quantum information and computation.
Our results show that the fact that quantum entanglement (or quantum non-locality) admits a kind of interpretation of topological
entanglement (topological non-locality) takes the responsibility for topological
features in teleportation-based quantum computation. Such topological features regard teleportation-based quantum circuit models as the two-dimensional
topological deformations of the extended Temperley--Lieb diagrammatical configurations, and they greatly simplify the algebraic analysis in
teleportation-based quantum computation.
About further research, since teleportation-based quantum computation  is an example for measurement-based quantum computation which includes
 the one-way quantum computation \cite{RB01, BFN08} as another example, so we expect that the one-way quantum computation \cite{RB01} can be also
 understood from both the extended Temperley--Lieb diagrammatical approach and the Yang--Baxter gate approach. Furthermore, if we consider the categorical
 description  \cite{Coecke04, AC04, Abramsky09} of quantum teleportation, it is no doubt to obtain new insights on both our research results in this paper
 and categorical quantum information and computation.  In addition, further research problems in mathematical physics are collected in
 Appendix~\ref{YBE_further_research}.

\section*{Acknowledgements}

 Yong Zhang is supported by the starting grant--273732 of Wuhan University.

\appendix

\section{The Yang--Baxter gate via the Temperley--Lieb algebra and its extended Temperley--Lieb diagrammatical representation}

Recent years, non-trivial unitary solutions of the Yang--Baxter equation \cite{YBE67}, called the Yang--Baxter gates \cite{Dye03, KL04, ZKG04},
have been proposed in the study of quantum information and computation \cite{NC2011, Preskill97}. Since the central topic of the present paper is
about the application of  the extended Temperley--Lieb diagrammatical approach to teleportation-based quantum
computation~\cite{GC99,Nielsen03, Leung04}, we study the construction of the Yang--Baxter gates using the Temperley--Lieb
algebra \cite{TL71, Kauffman02} in detail.

\begin{figure}
 \begin{center}
  \includegraphics[width=11cm]{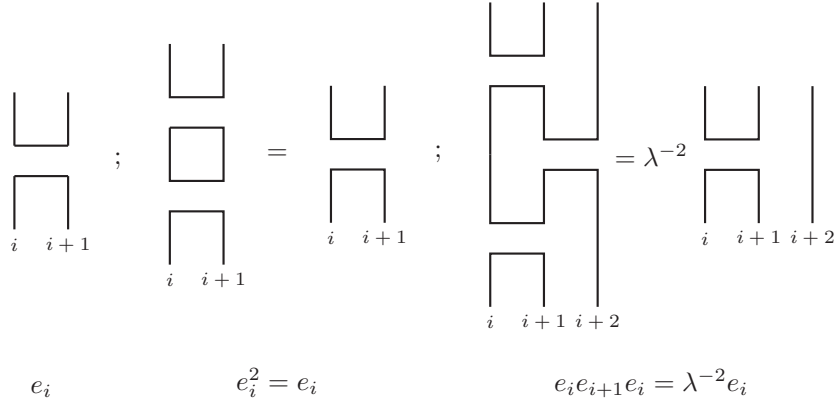}
  \end{center}
  \caption{\label{fig_tl}  Diagrammatical representation of the algebraic relations defining the Temperley--Lieb algebra (\ref{TL}). The leftside diagram denotes
   the type of the idempotent $e_i$ (\ref{T_generator}) acting on the $i$-th and $i+1$-th Hilbert spaces, where the vertical lines denoting an identity action on
   the 1-th, 2-th, $\cdots$, $i-1$-th, $i+2$-th, $\cdots$, $n$-th Hilbert spaces are neglected. The middle diagram denotes $e_i^2 =e_i$ in which the box (or the loop)
   represents the normalization factor 1 and can be thus omitted. The rightside diagram denotes $e_i e_{i+1}e_i=\lambda^{-2} e_i$, where the factor
   $\lambda^{-2}$ is contributed by all the vanishing cups and caps in the topological straightening operation. }
\end{figure}

\subsection{The Temperley--Lieb algebra and its diagrammatical representation}

\label{Def TL}

The Temperley--Lieb algebra \cite{TL71, Kauffman02}, denoted by $TL_n(\lambda)$, is generated by idempotents $e_i$ which satisfy the algebraic
relations:
\eqa
\label{TL}
e_i^2 & =& e_i, \quad i=1, \cdots, n-1; \nonumber\\
e_ie_{i\pm1}e_i & =& \lambda^{-2} e_i; \nonumber\\
e_ie_j & = & e_j e_i,\quad  i, j=1, \cdots, n-1, \left|i-j\right|>1;
\ena
where  $i \pm 1$ is a positive integer between 1 and $n-1$, and $\lambda$ is called the loop parameter, a non-vanishing complex number. In this section,
we study the Temperley--Lieb algebra $TL_n(\lambda)$ with the following type of idempotents:
\eq
\label{T_generator}
e_i = 1\!\!1^{\otimes(i-1)}\otimes T \otimes  1\!\!1^{\otimes (n-i-1)},\quad i=1,2,\cdots,n-1,
\en
where  $1\!\!1$ is the identity operator on the Hilbert space $V$ and the generator $T$ is a linear mapping on $V\otimes V$ to be specified
in a given circumstance.

There is a well known diagrammatical representation to catch the essential algebraic properties of the Temperley--Lieb algebra $TL_n(\lambda)$,
see \cite{Kauffman02}.  In Figure~\ref{fig_tl}, the generator $e_i$ is depicted as a pair
of cup and cap on the sites of $i$ and $i+1$. The algebraic relations (\ref{TL}) defining the Temperley--Lieb algebra are read from the
right to the left, while the corresponding diagrammatical expressions are read from the bottom to the top.
The points on the same site (or in the same vertical row) are connected; the lines connecting points on different sites can be straightened
via topological diagrammatical deformations, and two pairs of vanishing cups and caps contribute the normalization factor $\lambda^{-2}$.


\subsection{The extended Temperley--Lieb diagrammatical representation for the Temperley--Lieb algebra generated by Bell states (\ref{Bell states})}

\label{Def TL and Bell states}

The fact that the EPR state (\ref{EPR}), a maximal bipartite entangled state, generates a representation of
the Temperley--Lieb algebra $TL_n(\lambda)$, has been discussed in  \cite{Zhang06}, and here we make a brief sketch.
The projector of the EPR state $|\Psi\rangle$  has the form
\eq
|\Psi\rangle\langle\Psi|=\frac{1}{d} \sum_{i,j=0}^{d-1}|ii\rangle\langle j j|,
\en
with $d=2$, the dimension of the Hilbert space $V$. With the $T$ generator (\ref{T_generator}) given by $T=|\Psi\rangle\langle\Psi|$,
the idempotents $e_i$ of the Temperley--Lieb algebra $TL_n(\lambda)$ are defined as
\eq
e_i= 1\!\!1_2^{\otimes(i-1)}\otimes|\Psi\rangle\langle\Psi|\otimes  1\!\!1_2^{\otimes (n-i-1)},\quad i=1,2,\cdots,n-1
\en
which naturally give rise to $e_i^2=e_i$ and $e_i e_j =e_j e_i, \quad |i-j|>1$. We calculate
\eq
e_1e_2e_1|\alpha\beta\gamma\rangle=\frac 1 d \sum_{l=0}^{d-1} e_1e_2|ll\gamma\rangle\delta_{\alpha\beta}=\frac{1}{d^3}\sum_{l=0}^{d-1} |nn\gamma\rangle\delta_{\alpha\beta}=\frac{1}{d^2}e_1|\alpha\beta\gamma\rangle,\\
\en
to determine the loop parameter $\lambda=d=2$.

Similarly, a representation of the Temperley--Lieb algebra  $TL_n(\lambda)$ can be generated by the Bell states $|\psi(ij)\rangle$ (\ref{Bell states}),
and the loop parameter is still $\lambda=d=2$, see Subsubsection~\ref{type_I_YBE}. When the EPR state $|\Psi\rangle$ is described by the cup
configuration and single-qubit gates are by solid points on the cup, the diagrammatical representation for the Temperley--Lieb algebra
$TL_n(\lambda)$ generated by $|\psi(ij)\rangle\langle \psi(ij) |$ is the extended Temperley--Lieb diagrammatical
configuration in \cite{Zhang06}.

The key reason that we propose such the extended Temperley--Lieb configuration is that it is capable of describing both
quantum teleportation and teleportation-based quantum computation \cite{GC99,Nielsen03,Leung04,ZP13,ZZ14}. Such the diagrammatical representation
is beyond the standard Temperley--Lieb diagrammatical representation in Figure~\ref{fig_tl},  because the action of single-qubit and two-qubit quantum
gates are not allowed by the defining algebraic relations of the Temperley--Lieb algebra (\ref{TL}) in general.

\subsection{The Yang--Baxter equation and the Yang--Baxter gate}

\label{Def YBE}

\begin{figure}
 \begin{center}
  \includegraphics[width=12cm]{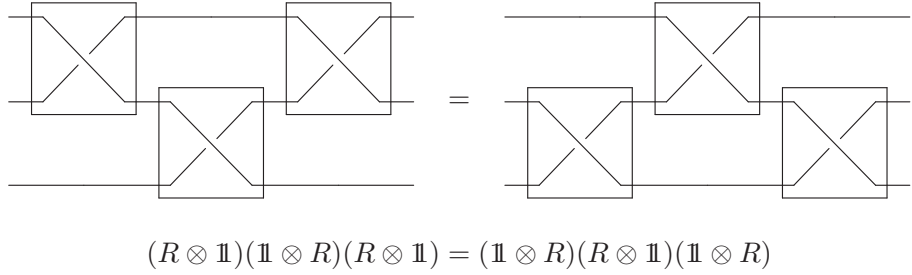}
  \end{center}
  \caption{\label{fig_YBE} Diagrammatical representation of the Yang--Baxter equation (\ref{YBE}).  Such the diagram without the boxes is the standard
  diagrammatical representation of the Yang--Baxter equation  (\ref{YBE}) in which a solution of the Yang--Baxter equation is denoted as an over-crossing
  braiding configuration. On the other hand,  as the dimension of the Hilbert space $V$ is $d=2$, such the diagram is a quantum circuit model in terms
  of the Yang--Baxter gate which is a two-qubit gate represented by a box with two single-qubit lines. }
\end{figure}

The Yang--Baxter equation \cite{YBE67} has the form
\eq
\label{YBE}
(R\otimes 1\!\!1)(1\!\!1\otimes R)(R\otimes 1\!\!1)=(1\!\!1\otimes R)(R\otimes 1\!\!1)(1\!\!1\otimes R)
\en
where $R$ is a linear operator on $V\otimes V$. Note that it is also called the constant
Yang--Baxter equation in the literature \cite{YBE67}, because it has no explicit dependence on parameters.

It is well known that a solution of the
Yang--Baxter equation (\ref{YBE}) naturally yields a representation of the braid group ${\cal B}_n$ in low dimensional topology \cite{Kauffman02}.
The braid group ${\cal B}_n$ has generators $\sigma_1$, $\cdots$, $\sigma_{n-1}$, and the defining algebraic relations are
\eqa
\sigma_i\sigma_{i+1}\sigma_i &=& \sigma_{i+1}\sigma_i\sigma_{i+1}, \quad 1\leq i<n;\nonumber\\
\sigma_i\sigma_j &=& \sigma_j\sigma_i,\quad |i-j|>1.
\ena
The braid group representation generated by a solution of the Yang--Baxter equation (\ref{YBE}) usually takes the form
\eq
\sigma_i= 1\!\!1^{\otimes(i-1)}\otimes R \otimes  1\!\!1^{\otimes (n-i-1)},\quad i=1,2,\cdots,n-1.
\en

The Yang--Baxter gates \cite{Dye03, KL04} are defined as unitary solutions of the Yang--Baxter equation \cite{YBE67} and satisfy the unitarity
condition given by
\eq
\label{unitarity}
R\, R^\dag=R^\dag R=1\!\!1.
\en
In low dimensional topology \cite{Kauffman02},  a solution of the Yang--Baxter equation is denoted as an over-crossing vertex.
In quantum information and computation \cite{NC2011}, a two-qubit quantum gate is represented as a box acting with two single-qubit lines.
Hence, Figure~\ref{fig_YBE} as the  diagrammatical representation of the Yang--Baxter equation (with the two-dimensional identity operator)
can be understood both in the viewpoint of low dimensional topology and in the viewpoint of quantum circuit model.

\subsection{The Yang--Baxter gate via the Temperley--Lieb algebra}

\label{YBG via TL}

In this subsection, we present a detailed calculation on how to construct a type of the Yang--Baxter gates
under the guidance of the state model construction of a  type of solutions of the Yang--Baxter equation \cite{Kauffman02}.
We find that the Yang--Baxter gate imposes the strict constraint condition on the loop parameter $\lambda$ in
the Temperley--Lieb algebra (\ref{TL}), whereas there is no such constraint on the loop parameter $\lambda$ in
the state model construction of a type of solutions of the Yang--Baxter equation \cite{Kauffman02}.

In view of the state model in knot theory \cite{Kauffman02},  we suppose that the Yang--Baxter gate $R$ has the form,
\eq
\label{construct YBG}
R=a 1\!\!1+b T,
\en
where $a$ and $b$ are non-vanishing complex constants, and $T$ is the type of generator in the construction (\ref{T_generator})
of the Temperley--Lieb algebra idempotents. Next, we specify the coefficients $a$ and $b$ with the constraints of both
the Yang--Baxter equation (\ref{YBE}) and  the unitary property (\ref{unitarity}) of quantum gates.

Substituting the Yang--Baxter gate (\ref{construct YBG}) into the Yang--Baxter equation (\ref{YBE}), after some algebra, we derive an equation
of parameters $a$ and $b$,
\eq
\label{YBE constrain}
a^2+a b+\frac {b^2} {\lambda^2}=0.
\en
Substituting the Yang--Baxter gate (\ref{construct YBG}) into the unitarity condition (\ref{unitarity}), we have another constraint equations of
parameters $a$ and $b$,
\eq
\label{unitary constrain}
\left\{
\begin{array}{l}
|a|^2=1;\\
a^*b+ab^*+|b|^2=0.
\end{array}\right.
\en

Now, let us take the complex conjugation of the equation (\ref{YBE constrain}) and then multiply the resultant equation with the
equation (\ref{YBE constrain}), we obtain the equation
\eq
|a|^4+|a|^2(a^*b+ab^*+|b|^2)=\frac {|b|^4} {|\lambda|^4},
\en
with the help of the unitarity constraint conditions (\ref{unitary constrain}), which can be simplified into the equation
\eq
|b|^2=|\lambda|^2.
\en

With the above constraint conditions on  the coefficients $a$ and $b$,  we assume
\eq
a=e^{i\mu},\quad b=|\lambda|e^{i \nu},
\en
with real parameters $\mu$ and $\nu$, which give rise to the ratio of $a$ and $b$,
\eq
\label{alpha/beta}
\frac a b=\frac 1 {|\lambda|}e^{i(\mu-\nu)}=\frac 1 {|\lambda|}\cos (\mu-\nu)+\frac i {|\lambda|}\sin(\mu-\nu).
\en
Thus it is much better directly to consider the equations of the parameter $a/b$.
Reformulate the Yang--Baxter equation constraint condition (\ref{YBE constrain}) and the unitarity constraint
condition (\ref{unitary constrain}) in terms  of the parameter $a/b$ respectively, and we have
\eq
\label{YBE constrains II}
\left(\frac a b + \frac 1 2\right)^2=\frac 1 4 -\frac 1 {\lambda^2},
\en
and
\eq
\label{unitary constrains II}
\frac a b+\frac {a^*}{b^*}=-1.
\en

Clearly, there exists a constraint on the loop parameter $\lambda$, because the above three equations (\ref{alpha/beta}), (\ref{YBE constrains II}) and
(\ref{unitary constrains II}) have to be consistent with one another. Combine the equation  (\ref{alpha/beta}) with the equation (\ref{unitary constrains II}),
we have
\eq
\label{lambda less 2}
\frac 1 {|\lambda|}\cos (\mu-\nu)=-\frac 1 2,
\en
which gives the constraint for $\lambda$, that is $0<|\lambda|\leq2$. Similarly, with the relations (\ref{YBE constrains II}) and (\ref{lambda less 2}),
we have
\eq
\frac {\sin^2(\mu-\nu)} {|\lambda|^2}=\frac 1 {\lambda^2}-\frac 1 4,
\en
which determines the loop parameter $\lambda$ as a real number, $\lambda\in \mathbb{R}$, because $\sin^2(\mu-\nu)$ is a real number. Therefore,
we are allowed to choose $0<\lambda\leq 2$ for convenience, since only $\lambda^2$ appears in the defining algebraic relations of the Temperley--Lieb
algebra (\ref{TL}).

Obviously,  the Yang--Baxter gate (\ref{construct YBG}) modulo a phase factor is still a unitary solution of both the Yang--Baxter equation (\ref{YBE})
and the unitarity constraint equation (\ref{unitarity}), thus we set the coefficient $a$ as 1 and then multiply the Yang--Baxter gate with the globe
phase factor $e^{i\mu}$. Finally, with the coefficients $a$ and $b$ given by
\eq
\label{alpha_beta_eq}
a= e^{i\mu}, \quad b = -\frac \lambda 2 (\lambda \pm i\sqrt{4-\lambda^2}) e^{i\mu},
\en
we have  the Yang--Baxter gate via the Temperley--Lieb algebra, expressed as
\eq
R=e^{i\mu} (1\!\!1-\frac \lambda 2 (\lambda \pm i\sqrt{4-\lambda^2}) T).
\en
To be clarified again, the range of the loop parameter $\lambda$, $0<\lambda\leq 2$\footnote{The case for the loop parameter $\lambda=2$ has been discussed in \cite{Dye03}.}, is determined by both the Yang--Baxter equation (\ref{YBE})
and the unitarity constraint equation (\ref{unitarity}), so it is independent of the dimension of the chosen Hilbert space $V$.

\subsection{Examples for the Yang--Baxter gate via the Temperley--Lieb algebra}

\label{two types YBG}

We present typical examples for the Yang--Baxter gate via the Temperley--Lieb algebra, and they are
respectively about $\lambda=2$ and $\lambda=\sqrt{2}$. They are to be exploited
in this paper because the dimension of the associated Hilbert space is $d=2$, the dimension of a qubit \cite{NC2011,Preskill97}.

\subsubsection{Type I Yang--Baxter gate for $\lambda=2$ and its extended Temperley--Lib diagrammatical representation}

\label{type_I_YBE}

For the case of $\lambda=2$, we solve the equations (\ref{alpha_beta_eq}) and obtain the Yang--Baxter gate as
\eq
R_I(\pm 1, \tau)=1\!\!1-2 T_I(\pm 1, \tau)
\en
modulo a global phase factor $e^{i\mu}$, with a parameter $\tau$ to be determined,  which is called the type I
Yang--Baxter gate in this paper. In the
literature, see \cite{NXZG2011} (and its quoted earlier references), a suitable representation for $T_I$ in the
two-qubit Hilbert space, has the form
\eq
\label{T_I_matrix}
T_{I}(1,\tau)= \frac 1 2\left(
               \begin{array}{cccc}
                 1 & 0 & 0 & \tau \\
                 0 & 0 & 0 & 0 \\
                 0 & 0 & 0 & 0 \\
                 \tau^{-1} & 0 & 0 & 1 \\
               \end{array}
             \right) \quad \textrm{or} \quad
T_{I}(-1,\tau)=\frac 1 2\left(
               \begin{array}{cccc}
                 0 & 0 & 0 & 0 \\
                 0 & 1 & \tau & 0 \\
                 0 & \tau^{-1} & 1 & 0 \\
                 0 & 0 & 0 & 0 \\
               \end{array}
             \right),
\en
where $\tau$ is a complex number with norm 1, so the type I Yang--Baxter gate $R$ has the form
\eq
\label{R_I_matrix}
R_I(1, \tau) = \left(
          \begin{array}{cccc}
            0 & 0 & 0 & -\tau \\
            0 & 1 & 0  & 0 \\
            0 & 0  & 1 & 0 \\
            -\tau^{-1} & 0 & 0 & 0 \\
          \end{array}
        \right)
\quad \textrm{or} \quad
 R_I(-1, \tau)= \left(
          \begin{array}{cccc}
            1 & 0 & 0 & 0 \\
            0 & 0 & -\tau & 0 \\
            0 & -\tau^{-1} & 0 & 0 \\
            0 & 0 & 0 & 1 \\
          \end{array}
        \right).
\en

The $T_{I}$ matrix (\ref{T_I_matrix}) can be respectively written as a projector generated by the EPR state (\ref{EPR}) with the local action of
the single-qubit gate,
\eqa
T_{I}(1,\tau) &=& (1\!\!1_2\otimes L)|\psi(00)\rangle\langle \psi(00)|(1\!\!1_2\otimes L^\dag)\\
T_{I}(-1,\tau) &=& (1\!\!1_2\otimes XL)|\psi(00)\rangle\langle \psi(00)|(1\!\!1_2\otimes L^\dag X),
\ena
with the single-qubit gate $L$ given by
\eq
L=\left(
    \begin{array}{cc}
      1 & 0 \\
      0 & \tau^{-1} \\
    \end{array}
  \right)
\en
where $L=1\!\!1_2$ for $\tau=1$ and $L=Z$ for $\tau=-1$. So the type I Yang--Baxter gate $R_I(1, \tau)$ (\ref{R_I_matrix})
has the extended Temperley--Lieb diagrammatical representation,
  \eq
  \setlength{\unitlength}{0.5mm}
  \begin{array}{c}
  \begin{picture}(200,40)

  \put(16,22){$R_I(1,\tau)=$}

  \put(62,26){\line(0,1){12}}
  \put(52,26){\line(0,1){12}}
  \put(52,26){\line(1,0){10}}

  \put(52,8){\line(0,1){12}}
  \put(62,8){\line(0,1){12}}
  \put(52,20){\line(1,0){10}}

  \put(70,22){$+$}

  \put(92,26){\line(0,1){12}}
  \put(82,26){\line(0,1){12}}
  \put(82,26){\line(1,0){10}}
  \put(92,32){\circle*{2.}}
  \put(94,31){\tiny{$Z$}}

  \put(82,8){\line(0,1){12}}
  \put(92,8){\line(0,1){12}}
  \put(82,20){\line(1,0){10}}
  \put(92,14){\circle*{2.}}
  \put(94,13){\tiny{$Z$}}

  \put(100,22){$+$}

  \put(122,26){\line(0,1){12}}
  \put(112,26){\line(0,1){12}}
  \put(112,26){\line(1,0){10}}
  \put(122,32){\circle*{2.}}
  \put(124,31){\tiny{$X$}}

  \put(112,8){\line(0,1){12}}
  \put(122,8){\line(0,1){12}}
  \put(112,20){\line(1,0){10}}
  \put(122,14){\circle*{2.}}
  \put(124,13){\tiny{$X$}}

  \put(130,22){$+$}

  \put(152,26){\line(0,1){12}}
  \put(142,26){\line(0,1){12}}
  \put(142,26){\line(1,0){10}}
  \put(152,32){\circle*{2.}}
  \put(154,31){\tiny{$XZ$}}

  \put(142,8){\line(0,1){12}}
  \put(152,8){\line(0,1){12}}
  \put(142,20){\line(1,0){10}}
  \put(152,14){\circle*{2.}}
  \put(154,13){\tiny{$ZX$}}

  \put(160,22){$-2$}

  \put(182,26){\line(0,1){12}}
  \put(172,26){\line(0,1){12}}
  \put(172,26){\line(1,0){10}}
  \put(182,32){\circle*{2.}}
  \put(184,31){\tiny{$L$}}

  \put(182,8){\line(0,1){12}}
  \put(172,8){\line(0,1){12}}
  \put(172,20){\line(1,0){10}}
  \put(182,14){\circle*{2.}}
  \put(184,13){\tiny{$L^\dag$}}
  \end{picture}
  \end{array},
  \en
and the type I Yang--Baxter gate $R_I(-1, \tau)$ (\ref{R_I_matrix}) has the other extended Temperley--Lieb
diagrammatical representation,
    \eq
  \setlength{\unitlength}{0.5mm}
  \begin{array}{c}
  \begin{picture}(200,40)

  \put(11,22){$R_I(-1,\tau)=$}

  \put(62,26){\line(0,1){12}}
  \put(52,26){\line(0,1){12}}
  \put(52,26){\line(1,0){10}}

  \put(52,8){\line(0,1){12}}
  \put(62,8){\line(0,1){12}}
  \put(52,20){\line(1,0){10}}

  \put(70,22){$+$}

  \put(92,26){\line(0,1){12}}
  \put(82,26){\line(0,1){12}}
  \put(82,26){\line(1,0){10}}
  \put(92,32){\circle*{2.}}
  \put(94,31){\tiny{$Z$}}

  \put(82,8){\line(0,1){12}}
  \put(92,8){\line(0,1){12}}
  \put(82,20){\line(1,0){10}}
  \put(92,14){\circle*{2.}}
  \put(94,13){\tiny{$Z$}}

  \put(100,22){$+$}

  \put(122,26){\line(0,1){12}}
  \put(112,26){\line(0,1){12}}
  \put(112,26){\line(1,0){10}}
  \put(122,32){\circle*{2.}}
  \put(124,31){\tiny{$X$}}

  \put(112,8){\line(0,1){12}}
  \put(122,8){\line(0,1){12}}
  \put(112,20){\line(1,0){10}}
  \put(122,14){\circle*{2.}}
  \put(124,13){\tiny{$X$}}

  \put(130,22){$+$}

  \put(152,26){\line(0,1){12}}
  \put(142,26){\line(0,1){12}}
  \put(142,26){\line(1,0){10}}
  \put(152,32){\circle*{2.}}
  \put(154,31){\tiny{$XZ$}}

  \put(142,8){\line(0,1){12}}
  \put(152,8){\line(0,1){12}}
  \put(142,20){\line(1,0){10}}
  \put(152,14){\circle*{2.}}
  \put(154,13){\tiny{$ZX$}}

  \put(160,22){$-2$}

  \put(182,26){\line(0,1){12}}
  \put(172,26){\line(0,1){12}}
  \put(172,26){\line(1,0){10}}
  \put(182,32){\circle*{2.}}
  \put(184,31){\tiny{$XL$}}

  \put(182,8){\line(0,1){12}}
  \put(172,8){\line(0,1){12}}
  \put(172,20){\line(1,0){10}}
  \put(182,14){\circle*{2.}}
  \put(184,13){\tiny{$L^\dag X$}}
  \end{picture}
  \end{array}.
  \en

In this paper, we only consider the special type I Yang--Baxter gates for the cases $\tau=\pm1$, which
are denoted by the Yang--Baxter gates $R(ij)$ (\ref{R ij}),
\eqa
 && R(00) = R_{I}(1,1), \quad R(01)= R_{I}(1,-1), \nonumber\\
 && R(10) = R_{I}(-1,1), \quad R(11)= R_{I}(-1,-1),
\ena
and the associated $T_I$ matrices (\ref{T_I_matrix}) are respectively related to four projectors of the
Bell states $|\psi(ij)\rangle$ (\ref{Bell states}),
\eqa
 && |\psi(00)\rangle \langle \psi(00)| = T_{I}(1,1), \quad |\psi(01)\rangle \langle \psi(01)|= T_{I}(1,-1), \nonumber\\
 && |\psi(10)\rangle \langle \psi(10)| = T_{I}(-1,1), \quad |\psi(11)\rangle \langle \psi(11)|= T_{I}(-1,-1),
\ena
which have been used in Subsection~\ref{Def TL and Bell states}. The extended Temperley--Lieb diagrammatical
representation for the Yang--Baxter gate $R(ij)$  is presented in Subsection~\ref{type_I_YBE_TL}.

\subsubsection{Type II Yang--Baxter gate for $\lambda=\sqrt{2}$ and its extended Temperley--Lieb diagrammatical representation}

\label{type_II_YBE}

For the case of $\lambda=\sqrt 2$, we solve the equations (\ref{alpha_beta_eq}) and have the solutions,
\eq
a=e^{i\mu}, \quad b=-\sqrt{2} e^{\pm i \frac \pi 4} e^{i\mu}
\en
and with the suitable phase factors $e^{i\mu}$, we have two kinds of the type II Yang--Baxter gates,
\eq
R_{II}(+)=e^{i \frac \pi 4} 1\!\!1-i \sqrt 2 T_{II}, \quad R_{II}(-)=e^{-i \frac \pi 4} 1\!\!1+i \sqrt 2 T_{II},
\en
where $\mu=\pi/4$ in $R_{II}(+)$ and  $\mu=-\pi/4$ in $R_{II}(-)$.

We exploit the representation of $T_{II}$ in the two-qubit Hilbert space, which is shown in \cite{NXZG2011}
(and its quoted earlier references),
\eq
\label{T_II_matrix}
T_{II}(\epsilon,\varphi)=\frac 1 2\left(
               \begin{array}{cccc}
                 1 & 0 & 0 & -ie^{-i\varphi} \\
                 0 & 1 & -i\epsilon & 0 \\
                 0 & i\epsilon & 1 & 0 \\
                 ie^{i\varphi} & 0 & 0 & 1 \\
               \end{array}
             \right),
\en
and the associated Yang--Baxter gates $R_{II}(+)$ and  $R_{II}(-)$ can have a unified formalism,
\eqa
\label{B via TL}
R(\epsilon,\varphi)&=&e^{-i\frac \pi 4}1\!\!1_4+i\sqrt 2T_{II} \nonumber\\
 &=&\frac 1 {\sqrt 2}\left(
      \begin{array}{cccc}
        1 & 0 & 0 & e^{-i\varphi} \\
        0 & 1 & \epsilon & 0 \\
        0 & -\epsilon & 1 & 0 \\
        -e^{i\varphi} & 0 & 0 & 1 \\
      \end{array}
    \right),
\ena
with $\epsilon=\pm1$. About the application of the type II Yang--Baxter gate $R(\epsilon,\varphi)$ to quantum information and computation,
interested readers are invited to refer to \cite{ZKG04}.

The $T_{II}(\epsilon,\varphi)$ matrix (\ref{T_II_matrix}) can be formulated as a sum of two projectors, each of which is generated
by the Bell states with the local action of single-qubit gates,
\eq
T_{II}(\epsilon,\varphi)=|\psi_\varphi(00)\rangle\langle\psi_\varphi(00)|+|\psi_\epsilon(10)\rangle\langle\psi_\epsilon(10)|,
\en
where $|\psi_\varphi(00)\rangle$ and $|\psi_\epsilon(10)\rangle$ are respectively given by
\eqa
 &&|\psi_\varphi(00)\rangle =\frac 1 {\sqrt 2}(1\!\!1_2\otimes M)|\psi(00)\rangle, \nonumber \\
 &&|\psi_\epsilon(10)\rangle =\frac 1 {\sqrt 2}(1\!\!1_2\otimes X N)|\psi(00)\rangle,
\ena
with the single-qubit gates $M$ and $N$ given by
\eq
M=\left(
    \begin{array}{cc}
      1 & 0 \\
      0 & ie^{i\varphi} \\
    \end{array}
  \right),\quad N=\left(
    \begin{array}{cc}
      1 & 0 \\
      0 & i\epsilon \\
    \end{array}
  \right).
\en
Consequently,  the type II Yang--Baxter gate $R(\epsilon,\varphi)$ (\ref{B via TL}) has the extended Temperley--Lieb configuration,
    \eq
\label{TL B general}
\setlength{\unitlength}{0.5mm}
\begin{picture}(180,40)

  \put(-28,-2){\footnotesize{$R(\epsilon,\varphi)$}}

  \put(-28,29){\line(0,1){6}}
  \put(-28,11){\line(0,-1){6}}
  \put(-10,29){\line(0,1){6}}
  \put(-10,11){\line(0,-1){6}}

  \put(-31,8){\line(0,1){24}}
  \put(-07,8){\line(0,1){24}}
  \put(-31,8){\line(1,0){24}}
  \put(-31,32){\line(1,0){24}}

  \put(-28,29){\line(1,-1){18}}
  \put(-28,11){\line(1,1){7.3}}
  \put(-10,29){\line(-1,-1){7.3}}

\put(-00,19){$=e^{-i\frac \pi 4}($}
\put(42,23){\line(0,1){12}}
\put(32,23){\line(0,1){12}}
\put(32,23){\line(1,0){10}}
\put(32,5){\line(0,1){12}}
\put(42,5){\line(0,1){12}}
\put(32,17){\line(1,0){10}}
\put(47,19){$+$}
\put(68,23){\line(0,1){12}}
\put(58,23){\line(0,1){12}}
\put(58,23){\line(1,0){10}}
\put(68,29){\circle*{2.}}
\put(70,28){\tiny{$Z$}}
\put(58,5){\line(0,1){12}}
\put(68,5){\line(0,1){12}}
\put(58,17){\line(1,0){10}}
\put(68,11){\circle*{2.}}
\put(70,10){\tiny{$Z$}}
\put(74,19){$+$}
\put(94,23){\line(0,1){12}}
\put(84,23){\line(0,1){12}}
\put(84,23){\line(1,0){10}}
\put(94,29){\circle*{2.}}
\put(96,28){\tiny{$X$}}
\put(84,5){\line(0,1){12}}
\put(94,5){\line(0,1){12}}
\put(84,17){\line(1,0){10}}
\put(94,11){\circle*{2.}}
\put(96,10){\tiny{$X$}}
\put(100,19){$+$}
\put(120,23){\line(0,1){12}}
\put(110,23){\line(0,1){12}}
\put(110,23){\line(1,0){10}}
\put(120,29){\circle*{2.}}
\put(122,28){\tiny{$XZ$}}
\put(110,5){\line(0,1){12}}
\put(120,5){\line(0,1){12}}
\put(110,17){\line(1,0){10}}
\put(120,11){\circle*{2.}}
\put(122,10){\tiny{$ZX$}}
\put(130,19){$)$}
\put(134,19){$+i\sqrt 2($}
\put(168,23){\line(0,1){12}}
\put(158,23){\line(0,1){12}}
\put(158,23){\line(1,0){10}}
\put(168,29){\circle*{2.}}
\put(170,28){\tiny{$M$}}
\put(158,5){\line(0,1){12}}
\put(168,5){\line(0,1){12}}
\put(158,17){\line(1,0){10}}
\put(168,11){\circle*{2.}}
\put(170,10){\tiny{$M^\dag$}}
\put(174,19){$+$}
\put(194,23){\line(0,1){12}}
\put(184,23){\line(0,1){12}}
\put(184,23){\line(1,0){10}}
\put(194,29){\circle*{2.}}
\put(196,28){\tiny{$XN$}}
\put(194,5){\line(0,1){12}}
\put(184,5){\line(0,1){12}}
\put(184,17){\line(1,0){10}}
\put(194,11){\circle*{2.}}
\put(196,10){\tiny{$N^\dag X$}}
\put(202,19){$)$}
\end{picture}
\en
where the extended Temperley--Lieb configuration (\ref{TL_identity}) of the two-qubit identity gate is exploited.

In this paper, we consider the special type II Yang--Baxter gates, denoted by the $B(\epsilon,\eta)$ gate (\ref{B Bell transform}), satisfying
\eq
B(\epsilon, 1)= R(\epsilon, 0), \quad B(\epsilon, -1)= R(\epsilon, \pi),
\en
which are the Bell transform introduced in \cite{ZZ14}, and apply them to the study of teleportation-based quantum computation in
Section~\ref{YBG approach to tele based QC}.

Note that the same quantum gate allows various of the extended Temperley--Lieb diagrammatical representations, and the reason is that single-qubit
gates acting on such the configuration can be chosen in purpose. For example, the extended Temperley--Lieb configuration  (\ref{TL B Bell transform})
 for the special type II Yang--Baxter gate $B(\epsilon,\eta)$ (\ref{B Bell transform}) is essentially equivalent to that one obtained from the extended
 Temperley--Lieb configuration (\ref{TL B general}) of the type II Yang--Baxter gate $R(\epsilon, \varphi)$ (\ref{B via TL}) by taking
 $\varphi$ as $0$ or $\pi$, but they indeed look different in form.

\subsubsection{Type III Yang--Baxter gate: $\lambda=\sqrt{3}$}

\label{type_III_YBE}

For the case of  $\lambda=\sqrt 3$, we solve the equations (\ref{alpha_beta_eq}) and obtain the third type of the Yang--Baxter gate as
\eq
R_{III}=1\!\!1-\sqrt 3\,\, e^{\pm i\frac \pi 6}\,\,T_{III}
\en
modulo a global phase factor $e^{i\mu}$. In the literature  \cite{NXZG2011}, two kinds of the $T_{III}$ matrices have been constructed with
the associate dimension of the Hilbert space, $d=3$, which are not directly related to the main topic of this paper so are not presented here.

\subsection{Remarks on further research}

\label{YBE_further_research}

With our research results in this paper,  we suggest two interesting configurations for both physicists in quantum information and
computation \cite{NC2011}  and mathematicians in low dimensional topology \cite{Kauffman02}: the one is the extended Temperley--Lieb diagrammatical
configuration, and the other is the extended braiding configuration (the braiding gate is the Yang--Baxter gate) with the action of quantum gates.
Here we propose five open  problems for interested readers as follows.

The first problem is how to construct more nontrivial examples for the Yang--Baxter gates via the Temperley--Lieb algebra,
besides the above three types of examples: $\lambda=2$ in Subsubsection~\ref{type_I_YBE}, $\lambda=\sqrt{2}$ in Subsubsection~\ref{type_II_YBE} and
$\lambda=\sqrt{3}$ in Subsubsection~\ref{type_III_YBE}.  The papers in \cite{NXZG2011} and its quoted earlier references may be helpful for solving this
problem in a systematic approach.   Note that $0< \lambda \le 2$ is derived in the present paper.

The second problem is to study the algebraic formulation of the extended Temperley--Lieb algebraic approach. The reason for it is that the diagrammatical
representation of both quantum teleportation and teleportation-based quantum computation is the extended Temperley--Lieb configuration.  Obviously, the
Temperley--Lieb algebra (\ref{TL}) is to be a special example of such the algebraic structure, if it exists. Interested readers are invited to refer
to \cite{Zhang06, Velez09} for some insights on this problem.

The third problem is how to construct solutions of the Yang--Baxter equation directly in the extended Temperley--Lieb diagrammatical approach. Note that
all of our examples for the Yang--Baxter gates in this paper are constructed using the Temperley--Lieb algebra, although they admit the extended Temperley--Lieb
representations. So it is meaningful to study solutions of the Yang--Baxter equation which are not associated with the Tempereley--Lieb algebra but
have the extended Temperley--Lieb diagrammatical representation. For example, we can construct the Yang--Baxter gates using the Birman-Wenzl-Murakami
algebra \cite{WXSLZ15}  instead of the Temperley--Lieb algebra.

The fourth problem is to study the algebraic structure underlying the braiding configuration with the local actions of single-qubit gates. It is an interesting
question, because our results clearly show that this kind of configuration is capable of describing teleportation-based quantum computation. If such the algebraic
structure exists, it will be directly related to the algebraic formulation of the extended Temperley--Lieb configuration.

The fifth problem is to study a possibility of constructing a new type of the state model \cite{Kauffman02} (or a new type of knot invariants) using the
extended Temperley--Lieb diagrammatical configuration of the braiding gate (the Yang--Baxter gate). Such the research may bring a new way of
thinking about the relationship between topological entanglement and quantum entanglement \cite{KL02}.

As a conclusion of this section, we want to emphasize the last thing that quantum information and computation indeed offers both mathematicians and mathematical
physicists a lot of interesting problems to solve. For example, the above problems are just motivated by our study on both topological and algebraic descriptions
of teleportation-based quantum computation.

\section{The special type I Yang--Baxter gates (\ref{R ij}) are permutation-like quantum gates}

\label{teleportation_swapping_operator}

In this paper, we do not apply the special type I Yang--Baxter gates (\ref{R ij}) to teleportation-based quantum computation,  and for readers' convenience,
a brief study on the algebraic properties of the special type I Yang--Baxter gates (\ref{R ij}) is performed in the following.

The special type I Yang--Baxter gate $R(ij)$ (\ref{R ij}) has the matrix form
  \eqa
   \label{R ij matrix}
   R(ij)&=&\sum_{k,l=0}^1(1-2\delta_{i,k}\delta_{j,l})|\psi(kl)\rangle\langle \psi(kl)|\\
   &=&\left(
   \begin{array}{cccc}
     1-\delta_{i,0} & 0 & 0 & (-1)^{j+1}\delta_{i,0} \\
     0 & 1-\delta_{i,1} & (-1)^{j+1}\delta_{i,1} & 0 \\
     0 & (-1)^{j+1}\delta_{i,1} & 1-\delta_{i,1} & 0 \\
     (-1)^{j+1}\delta_{i,0} & 0 & 0 & 1-\delta_{i,0} \\
   \end{array}
   \right),
   \ena
   where the symbol $\delta$ denotes the Kronecker delta function, so each $R(ij)$ gate has four non-vanishing matrix entries, for example,
   the Yang--Baxter gate $R(11)$ is the Permutation gate $P$ (\ref{permutation gate}).

 With the relation (\ref{Bell states}) that the four Bell states are transformed to one another by local Pauli gates,
 the special type I Yang--Baxter gates $R(ij)$ can be rewritten as the Permutation gate $P$ (\ref{permutation gate}) with the local action of Pauli gates,
 \eq
 \label{R ij as permuataion gate with Pauli gate}
 R(ij)=Z_1^{j+1} X_1^{i+1} P X_1^{i+1}Z_1^{j+1}=X_2^{i+1}Z_2^{j+1} P Z_2^{j+1}X_2^{i+1},
 \en
 where the subscripts 1 and 2 denote the qubit sites that the associated Pauli gates are acting on.  Since the entangling power  \cite{BG11} of both the Permutation
 gate $P$ and local single-qubit gates is zero, then the entangling power of the special type I Yang--Baxter gates $R(ij)$ is zero. In other words, the two-qubit
 quantum gates $R(ij)$ with the action of arbitrary single-qubit gates can not perform universal quantum computation \cite{NC2011}, which is the key reason that
 we do not apply the special  type I Yang--Baxter gate to teleportation-based quantum computation in this paper.

 In terms of the special type I Yang--Baxter gate, we define the  teleportation swapping operator \cite{Zhang06}, denoted by $S(ij)$,
 \eqa
  S(ij) &=& R(ij)_{23}R(ij)_{12}R(ij)_{23}=R(ij)_{12}R(ij)_{23}R(ij)_{12}  \nonumber \\
  &=& Z_2^{j+1} X_2^{i+1} P_{23}P_{12}P_{23}X_2^{i+1}Z_2^{j+1},
  \ena
 where $R(ij)_{23}=1\!\!1_2\otimes R(ij)$ and $R(ij)_{12}= R(ij)\otimes 1\!\!1_2$. The teleportation swapping operator $S(ij)$ acting on the  product state of
 arbitrary three qubit states $|\lambda\rangle\otimes|\mu\rangle\otimes|\nu\rangle$, gives rise to the result
 \eq
  S(ij) (|\lambda\rangle\otimes|\mu\rangle\otimes|\nu\rangle) = |\nu\rangle\otimes|\mu\rangle\otimes|\lambda\rangle,
  \en
  where the first qubit $|\lambda\rangle$ and the third qubit $|\nu\rangle$ have been swapped with each other.

\section{Topological construction of the four-qubit state $|\Psi_{CNOT}^\uparrow\rangle$ }

\label{construction cnot down}

Compared with the four-qubit entangled state $|\Psi_{CNOT}\rangle$ (\ref{Psi_cnot}),  the other
four-qubit entangled state $|\Psi_{CNOT}^\uparrow\rangle$ takes the form
\eq
\label{Psi cnot down}
|\Psi_{CNOT}^\uparrow\rangle_{1256}=(1\!\! 1_2 \otimes CNOT_{52} \otimes 1\!\! 1_2)|\Psi\rangle_{12} \otimes |\Psi\rangle_{56},
\en
with the extended Temperley--Lieb diagrammatical representation
\eqa
\setlength{\unitlength}{0.6mm}
\begin{array}{c}
\begin{picture}(55,15)
\put(-2,2){\makebox(13,10){$|\Psi_{CNOT}^\uparrow\rangle=$}}
\put(22,0){\line(0,1){14}}
\put(32,0){\line(0,1){14}}
\put(22,0){\line(1,0){10}}
\put(30,7){\line(1,0){12}}
\put(32,7){\circle{4.}}
\put(42,0){\line(0,1){14}}
\put(52,0){\line(0,1){14}}
\put(42,0){\line(1,0){10}}
\put(42,7){\circle*{2.}}
\end{picture}
\label{Psi_cnot down_diag}
\end{array}
\ena
in which the $CNOT_{52}$ gate has the fifth qubit as the control qubit and the second qubit as the target qubit.

\begin{figure}
 \begin{center}
  \includegraphics[width=8cm]{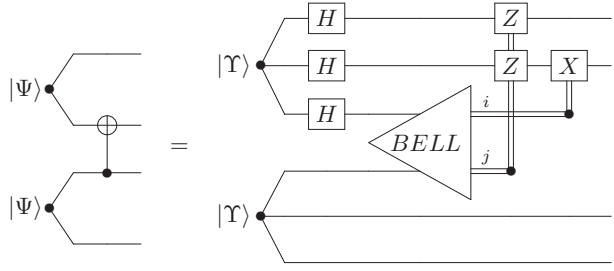}
  \end{center}
  \caption{\label{fig_psi_cnot_down1} Quantum circuit for the construction of the four-qubit entangled state $|\Psi_{CNOT}^\uparrow\rangle$ (\ref{Psi cnot down}), as
  the diagrammatical representation of the teleportation equation (\ref{tele cnot down1}). After performing the Bell measurements on the third and fourth qubits of
  the six-qubit state $H_1H_2H_3|\Upsilon\rangle_{123}|\Upsilon\rangle_{456}$, according to the measurement results, we perform the local unitary correction gate
  $X_2^iZ_1^jZ_2^j$ to complete the construction.  }
\end{figure}

In order to construct the $|\Psi_{CNOT}^\uparrow\rangle$ state, we prepare the six-qubit state $H_1H_2H_3|\Upsilon\rangle_{123}|\Upsilon\rangle_{456}$
and perform the Bell measurements on its third and fourth qubits, and such the procedure gives rise to the teleportation equation
\eq
\label{tele cnot down1}
H_1H_2H_3|\Upsilon\rangle_{123}|\Upsilon\rangle_{456}=\sum_{i,j=0}^1|\psi(ij)\rangle Z_1^jZ_2^jX_2^i|\Psi_{CNOT}^\uparrow\rangle_{1256},
\en
which is associated with the quantum circuit model in Figure~\ref{fig_psi_cnot_down1}.

In the extended Temperley--Lieb diagrammatical approach, using the diagrammatical representation (\ref{GHZ 3}) of the $H_1H_2H_3|\Upsilon\rangle_{123}$ state and
the diagrammatical representation (\ref{GHZ 1}) of the  $|\Upsilon\rangle_{456}$ state and the diagrammatical representation (\ref{TL Bell measurement}) of the
Bell measurement, we have the topological diagrammatical representation of the teleportation equation~(\ref{tele cnot down1}),
 \eq
\setlength{\unitlength}{0.6mm}
\begin{array}{c}
\begin{picture}(154,50)
\put(8,0){\makebox(4,4){$\nabla$}}
\put(12,8){\makebox(4,4){\tiny{$H$}}}
\put(44,0){\makebox(4,4){$\nabla$}}
\put(38,33){\makebox(6,6){\tiny{$W_{ij}^\dag$}}}
\put(75,20){\makebox(10,8){$= \frac 1 2$}}
\multiput(8,24)(1,0){65}{\line(1,0){.5}}
\put(10,4){\line(0,1){44}}
\put(22,0){\line(0,1){48}}
\put(34,0){\line(0,1){48}}
\put(46,4){\line(0,1){44}}
\put(58,0){\line(0,1){48}}
\put(70,0){\line(0,1){48}}
\put(10,18){\line(1,0){14}}
\put(22,0){\line(1,0){12}}
\put(34,48){\line(1,0){12}}
\put(44,18){\line(1,0){14}}
\put(58,0){\line(1,0){12}}
\put(58,18){\circle*{2.}}
\put(10,10){\circle*{2.}}
\put(10,18){\circle*{2.}}
\put(46,36){\circle*{2.}}
\put(22,18){\circle{4}}
\put(46,18){\circle{4}}
\put(105,40){\makebox(4,4){\tiny{$Z^j$}}}
\put(93,40){\makebox(4,4){\tiny{$Z^j$}}}
\put(105,34){\makebox(4,4){\tiny{$X^i$}}}
\put(90,0){\line(0,1){48}}
\put(102,0){\line(0,1){48}}
\put(138,0){\line(0,1){48}}
\put(150,0){\line(0,1){48}}
\put(90,0){\line(1,0){12}}
\put(100,30){\line(1,0){38}}
\put(138,0){\line(1,0){12}}
\put(138,30){\circle*{2.}}
\put(102,36){\circle*{2.}}
\put(102,42){\circle*{2.}}
\put(90,42){\circle*{2.}}
\put(102,30){\circle{4}}
\end{picture}
\label{fig_psi_cnot_down10}
\end{array}
\en
in which we move the single-qubit gate $W_{ij}^\dag$ from the fourth qubit to the second qubit and across the $CNOT_{12}$ gate via the formula
\eq
CNOT_{12}(1\!\!1_2\otimes W^\dag_{ij})CNOT_{12}=Z_1^jZ_2^jX_2^i;
\en then interchange the $CNOT_{12}$ gate and the $CNOT_{52}$ gate, and finally apply the diagrammatical representation (\ref{EPR_diag1}) of the EPR
state $|\Psi\rangle$.

Furthermore, on the diagram~(\ref{fig_psi_cnot_down10}),  we replace the configuration of the $H_1H_2H_3|\Upsilon\rangle_{123}$ state with its another
equivalent configuration (\ref{GHZ 4}) and replace the configuration  of the $|\Upsilon\rangle_{456}$ state with its another equivalent configuration
(\ref{GHZ 2}) to obtain the other diagram,
\eq
\setlength{\unitlength}{0.6mm}
\begin{array}{c}
\begin{picture}(154,50)
\multiput(6,24)(1,0){65}{\line(1,0){.5}}
\put(8,0){\line(0,1){48}}
\put(8,0){\line(1,0){12}}
\put(20,0){\line(0,1){48}}
\put(30,0){\makebox(4,4){$\nabla$}}
\put(32,4){\line(0,1){44}}
\put(18,18){\line(1,0){14}}
\put(32,18){\circle*{2.}}
\put(20,18){\circle{4}}
\put(32,10){\circle*{2.}}
\put(34,8){\makebox(4,4){\tiny{$H$}}}
\put(32,48){\line(1,0){12}}
\put(44,0){\line(0,1){48}}
\put(44,0){\line(1,0){12}}
\put(56,0){\line(0,1){48}}
\put(66,0){\makebox(4,4){$\nabla$}}
\put(68,4){\line(0,1){44}}
\put(56,18){\line(1,0){14}}
\put(68,18){\circle{4}}
\put(56,18){\circle*{2.}}
\put(44,36){\circle*{2.}}
\put(36,33){\makebox(6,6){\tiny{$W_{ij}^\dag$}}}
\put(73,20){\makebox(10,8){$= \frac 1 2$}}
\put(88,0){\line(0,1){48}}
\put(88,0){\line(1,0){12}}
\put(100,0){\line(0,1){48}}
\put(98,30){\line(1,0){38}}
\put(100,30){\circle{4}}
\put(136,0){\line(0,1){48}}
\put(136,0){\line(1,0){12}}
\put(148,0){\line(0,1){48}}
\put(136,30){\circle*{2.}}
\put(136,36){\circle*{2.}}
\put(139,34){\makebox(4,4){\tiny{$Z^j$}}}
\put(136,42){\circle*{2.}}
\put(139,40){\makebox(4,4){\tiny{$X^i$}}}
\put(148,42){\circle*{2.}}
\put(151,40){\makebox(4,4){\tiny{$X^i$}}}
\end{picture}
\label{fig_psi_cnot_down20}
\end{array}
\en
in which moving the single-qubit gate $W_{ij}^\dag$ across the $CNOT_{56}$ gate involves the formula
\eq
CNOT_{56}(W_{ij}\otimes1\!\!1_2 )CNOT_{56}=X_5^iX_6^iZ_5^j.
\en
With the topological diagrammatical representation~(\ref{fig_psi_cnot_down20}), we derive its algebraic counterpart as
\eq
\label{tele cnot down2}
H_1H_2H_3 \, |\Upsilon\rangle_{123} |\Upsilon\rangle_{456}=\sum_{i,j=0}^1|\psi(i j)\rangle X_5^iX_6^iZ_5^j
 |\Psi_{CNOT}^\uparrow\rangle_{1256},
\en
which is equivalent to the teleportation equation~(\ref{tele cnot down1}).

\section{Topological construction of the four-qubit state $|\Psi_{CZ}\rangle$}

\label{construction Psi CZ}

The four-qubit entangled state $|\Psi_{CZ}\rangle$  used to perform $CZ$ gate (\ref{CZ gate}) in teleportation-based
quantum computation, takes the form
\eq
\label{Psi_cz}
|\Psi_{CZ}\rangle_{1256}=(1\!\! 1_2 \otimes CZ_{25} \otimes 1\!\! 1_2)( |\Psi\rangle_{12} \otimes |\Psi\rangle_{56}),
\en
with the extended Temperley--Lieb diagrammatical form
\eqa
\setlength{\unitlength}{0.6mm}
\begin{array}{c}
\begin{picture}(55,15)
\put(0,2){\makebox(16,10){$|\Psi_{CZ}\rangle=$}}
\put(22,0){\line(0,1){14}}
\put(32,0){\line(0,1){14}}
\put(22,0){\line(1,0){10}}
\put(32,7){\line(1,0){10}}
\put(32,7){\circle*{2.}}
\put(42,0){\line(0,1){14}}
\put(52,0){\line(0,1){14}}
\put(42,0){\line(1,0){10}}
\put(42,7){\circle*{2.}}
\end{picture}
\label{Psi_cz_diag}
\end{array}
\ena
in which two solid points linked by the horizontal line represents the $CZ$ gate and the symmetrical diagrammatical notation of the $CZ$ gate
indicates $CZ_{25}=CZ_{52}$.

\begin{figure}
 \begin{center}
  \includegraphics[width=8cm]{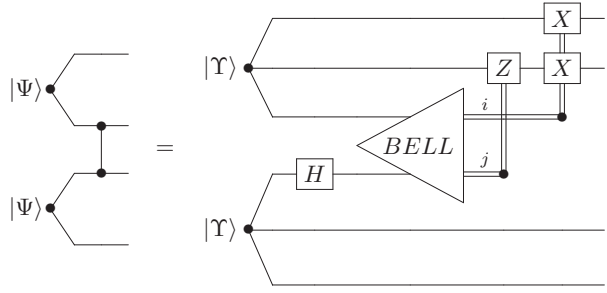}
  \end{center}
  \caption{\label{fig_psi_cz1}  Quantum circuit for the construction of the four-qubit entangled state $|\Psi_{CZ}\rangle$ (\ref{Psi_cz}) as the
  diagrammatical representation of the teleportation equation~(\ref{tele_cz1}).
 }
\end{figure}

With the algebraic relation between the $|\Psi_{CZ}\rangle$ state and the other four-qubit entangled state $|\Psi_{CNOT}\rangle$ (\ref{Psi_cnot})
given by
\eq
|\Psi_{CZ}\rangle_{1256}= H_5\, H_6 \, |\Psi_{CNOT}\rangle_{1256},
\en
we obtain the  teleportation equation for the construction of the $|\Psi_{CZ}\rangle$ state directly from
the teleportation equation (\ref{tele_cnot1}) for the construction of the $|\Psi_{CNOT}\rangle$ state,
\eq
\label{tele_cz1}
H_4|\Upsilon\rangle_{123}|\Upsilon\rangle_{456}=\sum_{i,j=0}^1|\psi(ij)\rangle_{34}\,\, Z_2^jX_1^iX_2^i|\Psi_{CZ}\rangle_{1256},
\en
which has the associated quantum circuit model in Figure~\ref{fig_psi_cz1}.

Using the diagrammatical representation~(\ref{GHZ 1}) of the three-qubit GHZ state $|\Upsilon\rangle$ and the diagrammatical
representation (\ref{TL Bell measurement}) of the Bell measurement, we have the topological diagrammatical representation of the
teleportation equation~(\ref{tele_cz1}),
\eq
\setlength{\unitlength}{0.6mm}
\begin{array}{c}
\begin{picture}(154,50)
\put(8,0){\makebox(4,4){$\nabla$}}
\put(48,16){\makebox(4,4){\tiny{$H$}}}
\put(44,0){\makebox(4,4){$\nabla$}}
\put(38,33){\makebox(6,6){\tiny{$W_{ij}^\dag$}}}
\put(75,20){\makebox(10,8){$= \frac 1 2$}}
\multiput(8,24)(1,0){65}{\line(1,0){.5}}
\put(10,4){\line(0,1){44}}
\put(22,0){\line(0,1){48}}
\put(34,0){\line(0,1){48}}
\put(46,4){\line(0,1){44}}
\put(58,0){\line(0,1){48}}
\put(70,0){\line(0,1){48}}
\put(8,18){\line(1,0){14}}
\put(22,0){\line(1,0){12}}
\put(34,48){\line(1,0){12}}
\put(44,10){\line(1,0){14}}
\put(58,0){\line(1,0){12}}
\put(58,10){\circle*{2.}}
\put(46,18){\circle*{2.}}
\put(22,18){\circle*{2.}}
\put(46,36){\circle*{2.}}
\put(10,18){\circle{4}}
\put(46,10){\circle{4}}
\put(105,40){\makebox(4,4){\tiny{$Z^j$}}}
\put(93,34){\makebox(4,4){\tiny{$X^i$}}}
\put(105,34){\makebox(4,4){\tiny{$X^i$}}}
\put(90,0){\line(0,1){48}}
\put(102,0){\line(0,1){48}}
\put(138,0){\line(0,1){48}}
\put(150,0){\line(0,1){48}}
\put(90,0){\line(1,0){12}}
\put(102,30){\line(1,0){36}}
\put(138,0){\line(1,0){12}}
\put(138,30){\circle*{2.}}
\put(102,36){\circle*{2.}}
\put(102,42){\circle*{2.}}
\put(90,36){\circle*{2.}}
\put(102,30){\circle*{2.}}
\end{picture}
\label{fig_psi_cz10}
\end{array}
\en
in which we firstly move the single-qubit gate $W_{ij}^\dag$ gate from the fourth qubit to the second qubit and across the $CNOT_{21}$ gate via the formula
\eq
CNOT_{21}(1\!\!1_2\otimes W^\dag_{ij})CNOT_{21}=Z_2^jX_1^iX_2^i;
\en
secondly perform the topological straightening operation to derive the normalization factor $\frac 1 2$ after moving the Hadamard gate from the fourth qubit
to the second qubit; thirdly apply the relation $H_2CNOT_{52}H_2=CZ_{25}$ between the $CNOT$ gate and the $CZ$ gate and then interchange the $CNOT_{21}$ gate
with the $CZ_{25}$ gate; and finally use the diagrammatical representation (\ref{EPR_diag2}) of the EPR state $|\Psi\rangle$.

On the other hand, with the diagrammatical representation (\ref{GHZ 2}) of the three-qubit GHZ state $|\Upsilon\rangle$, the diagram~(\ref{fig_psi_cz10})
has the other equivalent representation
\eq
\setlength{\unitlength}{0.6mm}
\begin{array}{c}
\begin{picture}(154,50)
\put(30,0){\makebox(4,4){$\nabla$}}
\put(66,0){\makebox(4,4){$\nabla$}}
\put(46,16){\makebox(4,4){\tiny{$H$}}}
\put(36,33){\makebox(6,6){\tiny{$W_{ij}^\dag$}}}
\multiput(6,24)(1,0){65}{\line(1,0){.5}}
\put(8,0){\line(0,1){48}}
\put(20,0){\line(0,1){48}}
\put(32,4){\line(0,1){44}}
\put(44,0){\line(0,1){48}}
\put(56,0){\line(0,1){48}}
\put(68,4){\line(0,1){44}}
\put(8,0){\line(1,0){12}}
\put(20,18){\line(1,0){14}}
\put(32,48){\line(1,0){12}}
\put(44,0){\line(1,0){12}}
\put(56,18){\line(1,0){14}}
\put(56,18){\circle*{2.}}
\put(44,18){\circle*{2.}}
\put(20,18){\circle*{2.}}
\put(44,36){\circle*{2.}}
\put(32,18){\circle{4}}
\put(68,18){\circle{4}}
\put(73,20){\makebox(10,8){$= \frac 1 2$}}
\put(139,34){\makebox(4,4){\tiny{$X^j$}}}
\put(139,40){\makebox(4,4){\tiny{$Z^i$}}}
\put(151,34){\makebox(4,4){\tiny{$X^j$}}}
\put(88,0){\line(0,1){48}}
\put(100,0){\line(0,1){48}}
\put(136,0){\line(0,1){48}}
\put(148,0){\line(0,1){48}}
\put(88,0){\line(1,0){12}}
\put(100,30){\line(1,0){36}}
\put(136,0){\line(1,0){12}}
\put(136,30){\circle*{2.}}
\put(100,30){\circle*{2.}}
\put(136,36){\circle*{2.}}
\put(136,42){\circle*{2.}}
\put(148,36){\circle*{2.}}
\end{picture}
\label{fig_psi_cz20}
\end{array}
\en
in which the single-qubit gate $W^\dag_{ij}H$ gate is moved from the fourth qubit to the fifth qubit and across the
$CNOT_{56}$ gate via the formula
\eq
CNOT_{56}(H\otimes 1\!\!1_2)(W_{ij}\otimes 1\!\!1_2)(H\otimes 1\!\!1_2)CNOT_{56} =Z_5^iX_5^jX_6^j,
\en
then the algebraic relation $H_5CNOT_{25}H_5=CZ_{25}$ is used to derive the $CZ_{25}$ gate, and the $CZ_{25}$ gate and the $CNOT_{56}$ gate
are interchanged to apply the diagrammatical representation (\ref{EPR_diag1}) of the EPR state $|\Psi\rangle$. Such the topological
diagrammatical representation~(\ref{fig_psi_cz20}) admits the form of the teleportation equation
\eq
\label{tele_cz2}
H_4|\Upsilon\rangle_{123}|\Upsilon\rangle_{456}=\sum_{i,j=0}^1|\psi(ij)\rangle_{34}\,\, Z_5^iX_5^jX_6^j|\Psi_{CZ}\rangle_{1256},
\en
which is obviously the other equivalent form of the teleportation equation~(\ref{tele_cz1}).

\section{An algebraic method of deriving both the extended Temperley--Lieb configuration in Figure~\ref{fig_tl_tele_operator} and the
algebraic expansion~(\ref{algebraic tl tele operator product state sum}) of the teleportation operator $(B\otimes 1\!\!1_2)(1\!\!1_2\otimes B)$}

\label{algebraic_study_fig_tl_tele_operator}

Let us make an  algebraic study of how to derive the extended Temperley--Lieb configuration in Figure~\ref{fig_tl_tele_operator}, which
is originally derived in the diagrammatical approach. First of all, we introduce the algebraic notation for the typical extended Temperley--Lieb
configurations in teleportation-based quantum computation,
\eqa
 ((U_1,U_2),1\!\!1_2) &\equiv&  |\Psi_{U_1}\rangle \langle \Psi_{U_2}| \otimes 1\!\! 1_2   \nonumber\\
  (1\!\!1_2,(U_3,U_4)) &\equiv&  1\!\! 1_2  \otimes |\Psi_{U_3}\rangle \langle \Psi_{U_4}| \nonumber\\
  (U_1,U_3 U_2^T,U_4)  &\equiv&  (|\Psi_{U_1}\rangle \langle \Psi|  \otimes U_3 U_2^T) (1\!\! 1_2  \otimes |\Psi\rangle \langle\Psi_{U_4}|)
\ena
with $|\Psi_{U}\rangle$  defined in (\ref{Psi_U}), and we have the formula as
   \eq
  \label{multiplication rule}
  ((U_1,U_2),1\!\!1_2)(1\!\!1_2,(U_3,U_4))=(U_1,U_3 U_2^T,U_4)
  \en
   to characterize the diagrammatical representation,
  \eq
\label{TL for teleportation operator}
  \setlength{\unitlength}{0.5mm}
  \begin{array}{c}
  \begin{picture}(60,50)

  \put(12,36){\line(0,1){12}}
  \put(2,36){\line(0,1){12}}
  \put(2,36){\line(1,0){10}}
  \put(12,42){\circle*{2.}}
  \put(4,41){\tiny{$U_1$}}

  \put(12,24){\circle*{2.}}
  \put(4,23){\tiny{$U_2$}}

  \put(12,0){\line(0,1){12}}
  \put(22,0){\line(0,1){12}}
  \put(12,12){\line(1,0){10}}
  \put(22,6){\circle*{2.}}
  \put(24,6){\tiny{$U_4$}}

  \put(2,0){\line(0,1){30}}
  \put(2,30){\line(1,0){10}}
  \put(12,18){\line(0,1){12}}
  \put(22,18){\line(0,1){30}}
  \put(12,18){\line(1,0){10}}
  \put(22,24){\circle*{2.}}
  \put(24,23){\tiny{$U_3$}}

  \put(33,22){$=$}

  \put(57,36){\line(0,1){12}}
  \put(47,36){\line(0,1){12}}
  \put(47,36){\line(1,0){10}}
  \put(57,42){\circle*{2.}}
  \put(49,41){\tiny{$U_1$}}

  \put(57,0){\line(0,1){12}}
  \put(67,0){\line(0,1){12}}
  \put(57,12){\line(1,0){10}}
  \put(67,6){\circle*{2.}}
  \put(69,5){\tiny{$U_4$}}

  \put(47,0){\line(0,1){30}}
  \put(47,30){\line(1,0){10}}
  \put(57,18){\line(0,1){12}}
  \put(67,18){\line(0,1){30}}
  \put(57,18){\line(1,0){10}}
  \put(67,42){\circle*{2.}}
  \put(69,41){\tiny{$U_3 U_2^T$}}
  \end{picture}
  \end{array}\nonumber
  \en
 in which $U_2^T$  is obtained by moving $U_2$ from the second qubit to the third qubit.  Applying the
 formula~(\ref{multiplication rule}),  the teleportation operator $(B\otimes 1\!\!1_2)(1\!\!1_2\otimes B)$ can be reformulated as
  \eq
  \label{algebraic tl tele operator}
  \begin{split}
  &(B\otimes 1\!\!1_2)(1\!\!1_2\otimes B)\\
  =&(Z,1\!\!1_2,1\!\!1_2)+(Z,Z,1\!\!1_2)+(X,X,1\!\!1_2)-(X,XZ,1\!\!1_2)\\
  -&(1\!\!1_2,1\!\!1_2,Z)+(1\!\!1_2,Z,Z)-(XZ,X,Z)-(XZ,XZ,Z)\\
  +&(Z,X,X)+(Z,XZ,X)+(X,1\!\!1_2,X)-(X,Z,X)\\
  -&(1\!\!1_2,X,ZX)+(1\!\!1_2,XZ,ZX)-(XZ,1\!\!1_2,ZX)-(XZ,Z,ZX),
  \end{split}
  \en
 each term of which corresponds to the relevant diagrammatical term in Figure~\ref{fig_tl_tele_operator}.

With the result~(\ref{algebraic tl tele operator}), furthermore,  we are able to derive the algebraic
expansion (\ref{algebraic tl tele operator product state sum}) of the teleportation operator $(B\otimes 1\!\!1_2)(1\!\!1_2\otimes B)$.
Before calculation, we introduce the notation
\eq
(|ij\rangle, W_{1,1},\langle kl| )\equiv(|ij\rangle \langle \Psi | \otimes W_{1,1} ) (1\!\! 1_2 \otimes |\Psi\rangle \langle kl|),
\en
which has already appeared in~(\ref{algebraic tl tele operator product state sum}), and with it, for example, we verify the formula
\eq
\label{product_basis_tele_config}
(Z,1\!\!1_2,1\!\!1_2)-(1\!\!1_2,1\!\!1_2, Z)
=(|00\rangle,1\!\!1_2,\langle 11|)-(|11\rangle,1\!\!1_2,\langle 00|),
\en
characterized by the topological configuration
 \eq
  \setlength{\unitlength}{0.5mm}
  \begin{array}{c}
  \begin{picture}(140,50)


  \put(12,36){\line(0,1){12}}
  \put(2,36){\line(0,1){12}}
  \put(2,36){\line(1,0){10}}
  \put(12,42){\circle*{2.}}
  \put(6,41){\tiny{$Z$}}

  \put(12,0){\line(0,1){12}}
  \put(22,0){\line(0,1){12}}
  \put(12,12){\line(1,0){10}}

  \put(2,0){\line(0,1){30}}
  \put(2,30){\line(1,0){10}}
  \put(12,18){\line(0,1){12}}
  \put(22,18){\line(0,1){30}}
  \put(12,18){\line(1,0){10}}

  \put(27,22){$-$}

  \put(47,36){\line(0,1){12}}
  \put(37,36){\line(0,1){12}}
  \put(37,36){\line(1,0){10}}

  \put(47,0){\line(0,1){12}}
  \put(57,0){\line(0,1){12}}
  \put(47,12){\line(1,0){10}}
  \put(57,6){\circle*{2.}}
  \put(59,5){\tiny{$Z$}}

  \put(37,0){\line(0,1){30}}
  \put(37,30){\line(1,0){10}}
  \put(47,18){\line(0,1){12}}
  \put(57,18){\line(0,1){30}}
  \put(47,18){\line(1,0){10}}

  \put(62,22){$=$}

 \put(82,38){\line(0,1){10}}
 \put(72,38){\line(0,1){10}}
 \put(69.9,36){\tiny{$\nabla$}}
 \put(79.9,36){\tiny{$\nabla$}}

  \put(82,0){\line(0,1){10}}
  \put(92,0){\line(0,1){10}}
  \put(89.9,10){\tiny{$\triangle$}}
  \put(79.9,10){\tiny{$\triangle$}}
   \put(92,6){\circle*{2.}}
  \put(94,5){\tiny{$X$}}
  \put(82,6){\circle*{2.}}
  \put(84,5){\tiny{$X$}}

  \put(72,0){\line(0,1){30}}
  \put(72,30){\line(1,0){10}}
  \put(82,18){\line(0,1){12}}
  \put(92,18){\line(0,1){30}}
  \put(82,18){\line(1,0){10}}

  \put(97,22){$-$}

  \put(117,38){\line(0,1){10}}
  \put(107,38){\line(0,1){10}}
  \put(114.9,36){\tiny{$\nabla$}}
  \put(104.9,36){\tiny{$\nabla$}}

  \put(117,42){\circle*{2.}}
  \put(119,41){\tiny{$X$}}

  \put(107,42){\circle*{2.}}
  \put(109,41){\tiny{$X$}}

  \put(117,0){\line(0,1){10}}
  \put(127,0){\line(0,1){10}}
  \put(114.9,10){\tiny{$\triangle$}}
  \put(124.9,10){\tiny{$\triangle$}}

  \put(107,0){\line(0,1){30}}
  \put(107,30){\line(1,0){10}}
  \put(117,18){\line(0,1){12}}
  \put(127,18){\line(0,1){30}}
  \put(117,18){\line(1,0){10}}


  \end{picture}
  \end{array}
  \en
where the vertical line with $\nabla$ stands for the $|0\rangle$ state. With the technique underlying the formula~(\ref{product_basis_tele_config}),
we rewrite the algebraic  expression (\ref{algebraic tl tele operator}) into
\eq
\label{algebraic tl tele operator product state}
  \begin{split}
  &(B\otimes 1\!\!1_2)(1\!\!1_2\otimes B)\\
  =&(|00\rangle,Z,\langle 00|)-(|01\rangle,XZ,\langle 00|)+(|10\rangle,X,\langle 00|)-(|11\rangle,1\!\!1_2,\langle 00|)\\
  +&(|00\rangle,XZ,\langle 01|)-(|01\rangle,Z,\langle 01|)+(|10\rangle,1\!\!1_2,\langle 01|)-(|11\rangle,X,\langle 01|)\\
  +&(|00\rangle,X,\langle 10|)+(|01\rangle,1\!\!1_2,\langle 10|)-(|10\rangle,Z,\langle 10|)-(|11\rangle,XZ,\langle 10|)\\
  +&(|00\rangle,1\!\!1_2,\langle 11|)+(|01\rangle,X,\langle 11|)-(|10\rangle,XZ,\langle 11|)-(|11\rangle,Z,\langle 11|)\\
  \end{split}
\en
which is the algebraic expanded formalism~(\ref{algebraic tl tele operator product state sum}) of the teleportation operator $(B\otimes 1\!\!1_2)(1\!\!1_2\otimes B)$.


\begin{thebibliography}{99}

 \bibitem{NC2011}  M.A.Nielsen and I.L. Chuang, {\it Quantum Computation and Quantum Information},
 (Cambridge University Press, Cambridge, UK, 2000 and 2011).

  \bibitem{Preskill97}  J. Preskill, {\it Lecture Notes on Quantum Computation},
  http://www.theory.caltech.edu/preskill.

 \bibitem{EPR35} A. Einstein, B. Podolsky, and N. Rosen, {\it Can Quantum-Mechanical Description of Physical Reality be Considered Complete?},
 Phys. Rev. {\bf 47}, 777-780 (1935).

  \bibitem{Bell64} J.S. Bell, {\it On the Einstein-Podolsky-Rosen
  paradox}, Physics {\bf 1} 195-200 (1964).

  \bibitem{BBCCJPW93} C.H. Bennett, G. Brassard, C. Crepeau, R.
  Jozsa, A. Peres and W.K. Wootters, {\it  Teleporting an Unknown Quantum State
  via Dual Classical and Einstein-Podolsky-Rosen Channels}, Phys. Rev. Lett. {\bf 70} 1895 (1993).

    \bibitem{Vaidman94} L. Vaidman, {\it Teleportation of Quantum States},
 Phys. Rev.  A {\bf 49}, 1473-1475 (1994).

 \bibitem{BDM00} S.L. Braunstein, G.M. D'Ariano, G.J. Milburn
 and M.F. Sacchi, {\it Universal Teleportation with a Twist}, Phys.
 Rev. Let. {\bf 84}, 3486-3489 (2000).

 \bibitem{Werner01} R. F. Werner,  {\it All Teleportation and Dense Coding
 Schemes},  J.Phys. A: Math. Theor.  {\bf 35}, 7081-7094 (2001).

   \bibitem{GC99} D. Gottesman and I. Chuang, {\it Demonstrating the Viability of Universal Quantum
   Computation Using Teleportation and Single-Qubit Operations}, Nature {\bf 402}, 390 (1999).

   \bibitem{Nielsen03} M.A. Nielsen,
 {\it Universal Quantum Computation Using Only Projective
  Measurement, Quantum Memory, and Preparation of the $0$ State},
  Phys. Lett. A \textbf{308}, 96 (2003).

 \bibitem{Leung04} D.W. Leung, {\it Quantum Computation by Measurements},
  Int. J. Quantum Inf. \textbf{2}, 33 (2004).

  \bibitem{ZP13} Y. Zhang and J-L. Pang, {\it Space-Time Topology in Teleportation-Based
 Quantum Computation}, arXiv:1309.0955 (2013).

  \bibitem{ZZ14} Y. Zhang and K. Zhang, {\it Bell Transform, Teleportation Operator and Teleportation-Based
     Quantum Computation}, arXiv:1401.7009 (2014).

 \bibitem{TL71} H.N.V. Temperley and E.H. Lieb, {\it Relations between
  the `Percolation' and `Colouring' Problem and Other Graph-Theoretical Problems
  Associated with Regular Planar Lattices: Some Exact Results for the `Percolation'
  Problem}, Proc. Roy. Soc. A {\bf 322}, 251(1971).

  \bibitem{Kauffman02} L. H. Kauffman, {\it Knots and Physics}
  (World Scientific Publishers, 2002).

 \bibitem{YBE67} C.N. Yang, {\it Some Exact Results for the Many Body Problems in One Dimension with Repulsive Delta
 Function Interaction}, Phys. Rev. Lett. {\bf 19}, 1312-1314 (1967). R.J. Baxter, {\it Partition Function of
 the Eight-Vertex Lattice Model}, Annals Phys. {\bf 70}, 193-228 (1972).  J.H.H. Perk and H. Au-Yang, {\it Yang--Baxter
 Equations}, Encyclopedia of Mathematical Physics, Vol. 5, 465-473 (Elsevier Science, Oxford, 2006).

 \bibitem{Kauffman05}
  L.H. Kauffman, {\it Teleportation Topology}. Opt. Spectrosc. {\bf 9}, 227 (2005).

 \bibitem{Zhang06}  Y. Zhang, {\it Teleportation, Braid Group and
   Temperley--Lieb Algebra},  J.Phys. A:  Math. Theor.  {\bf 39}, 11599-11622 (2006);
    Y. Zhang and L.H. Kauffman,
  {\it Topological-Like Features in Diagrammatical Quantum
  Circuits},  Quant. Inf. Proc. {\bf 6}, 477-507 (2007);
   Y. Zhang, {\it Braid Group, Temperley--Lieb Algebra,
   and Quantum Information and Computation}, AMS Contemporary Mathematics {\bf 482}, 52 (2009).

 \bibitem{Dye03}
 H. Dye, {\it Unitary Solutions to the Yang--Baxter Equation in Dimension Four},
 Quantum Inform. Process. {\bf 2}, 117-150 (2003).

 \bibitem{KL04} L.H. Kauffman and S.J. Lomonaco Jr.,  {\it Braiding Operators are Universal Quantum Gates},  New Journal
 of Physics {\bf 6}, 134 (2004).

 \bibitem{ZKG04}  Y. Zhang, L.H. Kauffman and M.L. Ge, {\it Universal Quantum Gate,
 Yang--Baxterization and Hamiltonian}. Int. J. Quant. Inform. {\bf 4}, 669-678 (2005).

 \bibitem{KL02} L.H. Kauffman and S. J. Lomonaco Jr., {\it Quantum Entanglement and Topological Entanglement},
   New Journal of Physics, {\bf 4}, 73 (2002).

  \bibitem{Barenco95b}  A. Barenco et al., {\it Elementary Gates for Quantum Computation},
   Phys. Rev. A {\bf 52}, 3457-3467 (1995).

  \bibitem{BB02} J.L. Brylinski and R. Brylinski, {\it Universal Quantum Gates},
  in {\it Mathematics of Quantum Computation}, Chapman \& Hall/CRC Press, Boca Raton, Florida, 2002 (edited by R. Brylinski and G. Chen).

 \bibitem{BMPRV00} P.O. Boykin, T. Mor, M. Pulver, V. Roychowdhury and F. Vatan,  {\it A New Universal and Fault-Tolerant Quantum Basis},
 Inf. Process. Lett, {\bf 75}, 101-107 (2000).

 \bibitem{BG11} D.J. Brod and E.F. Galv\~ao, {\it Extending Matchgates into Universal Quantum Computation}, Phys. Rev. A, {\bf 84}, 022310 (2011).

 \bibitem{RB01} R.~Raussendorf and H.J. Briegel, {\it A One-Way Quantum Computer},
 Phys. Rev. Lett. \textbf{86}, 5188 (2001).

  \bibitem{Gottesman97} D. Gottesman, {\it Stabilizer Codes and Quantum Error Correction Codes},
  Ph.D. Thesis, CalTech, Pasadena, CA, 1997.

 \bibitem{GHZ90} D.M. Greenberger, M.A. Horne, A. Shirnony, and A.Zeilinger,
    {\it Bell's Theorem Without Inequalities}, Am. J. Phys. {\bf 58}, 1131 (1990).

 \bibitem{Childs03} A.M. Childs, {\it Teleportation-based Approaches to Universal Quantum
  Computation with Single-Qubit Measurement},  Seminar at the Perimeter Institute, November 2003.

 \bibitem{BFN08} P. Bonderson, M. Freedman and C. Nayak, {\it Measurement-Only Topological Quantum Computation},
   Phys. Rev. Lett. \textbf{101}, 010501 (2008)

 \bibitem{Coecke04}  B. Coecke,  {\it The Logic of Entanglement. An Invitation}. Oxford University
 Computing Laboratory Research Report nr. PRG-RR-03-12.   An 8 page short version is at Arxiv:quant-ph/0402014.
  The full 160 page version is at ``web.comlab.ox.ac.uk/oucl/publications/tr/rr-03-12.html''.

 \bibitem{AC04} S. Abramsky and B. Coecke,   {\it A Categorical Semantics of Quantum Protocols}.
 In: Proceedings of the 19th Annual IEEE Symposium on Logic in Computer Science  (LiCS`04), IEEE
 Computer Science Press.

 \bibitem{Abramsky09} S. Abramsky, {\it Temperley--Lieb Algebra: from Knot Theory to Logica and Computation via Quantum Mechanics}, arXiv:0910.2737 (2009).

 \bibitem{NXZG2011} K. Niu, K. Xue, Q. Zhao, and M.L. Ge, {\it The Role of the $\ell_1$-Norm in Quantum Information Theory and
      Two Types of the Yang--Baxter Equation}, Journal of Physics A: Mathematical and Theoretical, {\bf 44}, 265304 (2011).
     G. Wang, K. Xue, C. Sun, C. Zhou, T. Hu, and Q. Wang, {\it Temperley--Lieb algebra, Yang--Baxterization and Universal Gate}, Quantum
     Information Processing, {\bf 9}, 699-710 (2010);
     C. Sun, G. Wang, T. Hu, C. Zhou, Q. Wang, and K. Xue, {\it The Representations of
        Temperley-Lieb Algebra and Entanglement in a Yang--Baxter System}, International Journal of Quantum Information, {\bf 7}, 1285-1293 (2009).

 \bibitem{Velez09} M. V'{e}lez and J. Ospina, {\it Generalized Temperley-Lieb Algebras and Quantum Computation},
 Mathematical Theory and Computational Practice, 5th Conference on Computability in Europe, CiE 2009, Heidelberg, Germany, July 19-24, 2009.

 \bibitem{WXSLZ15} G. Wang, K. Xue, C. Sun, B. Liu, Y. Liu, and Y. Zhang, {\it  Topological Basis Associated with B-M-W algebra: Two Spin-1/2 Realization},
   Physics Letters A, {\textbf 379}, 1-4 (2015).





 \end{thebibliography}
\end{document}